\documentclass[%
reprint,
nofootinbib,
amsmath,amssymb,
aps,
floatfix
]{revtex4-2}

\usepackage{graphicx}
\usepackage{dcolumn}
\usepackage{bm}
\usepackage{hyperref}

\usepackage{color}
\usepackage[dvipsnames]{xcolor}

\hypersetup{colorlinks=true, citecolor=MidnightBlue,linkcolor=CornflowerBlue, urlcolor=CornflowerBlue, linktocpage=true}

\usepackage[section]{placeins}

\usepackage{placeins}

\begin{document}

\preprint{APS/123-QED}

\title{UV completions, fixing the equations and nonlinearities in \(k\)-essence}

\author{Guillermo Lara}
 \email{jlaradel@sissa.it}
\author{Miguel Bezares}%
\email{mbezares@sissa.it}
\author{Enrico Barausse}
\email{barausse@sissa.it}
\affiliation{%
 SISSA, Via Bonomea 265, 34136 Trieste, Italy and INFN Sezione di Trieste\\IFPU - Institute for Fundamental Physics of the Universe, Via Beirut 2, 34014 Trieste, Italy\\
}

\date{\today}

%%%%%%%%%%%%%%%%%%%%%%%%%%%%%%%%%%%%%%%%%%%%%%%%%%%%%%%%%%%
%%%%%%%%%%%%%%%%%%%%%%%%%%%%%%%%%%%%%%%%%%%%%%%%%%%%%%%%%%%
%%%%%%%%%%%%%%%%%%%%%%%%%%%%%%%%%%%%%%%%%%%%%%%%%%%%%%%%%%%

\begin{abstract}
Scalar-tensor theories with first-derivative self interactions, known as \(k\)-essence, may provide interesting phenomenology on cosmological scales. On smaller scales, however,  initial value evolutions
(which are crucial for predicting the behavior of astrophysical systems, such as binaries of compact objects) may run into
instabilities related to the Cauchy problem becoming 
potentially ill-posed. Moreover, on local scales the 
dynamics may enter in the nonlinear regime,
which may lie beyond the range of validity of the infrared theory. Completions of \(k\)-essence in the ultraviolet, when they 
are known to exist, mitigate these problems,
as they both render  Cauchy evolutions well-posed at all times, and 
allow for checking  the relation between nonlinearities and the low energy theory's range of validity.
Here, we explore these issues explicitly by considering an ultraviolet completion to \(k\)-essence and performing vacuum 1+1 dynamical evolutions within it. The results are compared to those obtained with the low-energy theory, and with the low-energy theory suitably deformed with a phenomenological ``fixing the equations'' approach. We 
confirm
that the ultraviolet completion does not incur in any breakdown of the Cauchy problem's well-posedness, 
and 
we find that 
evolutions agree
with the results of the low-energy theory, when the system is within the regime of validity of the latter. However, we also find that the nonlinear behavior of \(k\)-essence lies (for the most part) outside this regime.

\end{abstract}

\maketitle

%%%%%%%%%%%%%%%%%%%%%%%%%%%%%%%%%%%%%%%%%%%%%%%%%%%%%%%%%%%
%%%%%%%%%%%%%%%%%%%%%%%%%%%%%%%%%%%%%%%%%%%%%%%%%%%%%%%%%%%
%%%%%%%%%%%%%%%%%%%%%%%%%%%%%%%%%%%%%%%%%%%%%%%%%%%%%%%%%%%

\section{Introduction} \label{sec: Introduction}

Six years after the first detection of gravitational waves (GWs) 
from a black hole binary coalescence
by the LIGO/Virgo Collaboration \cite{LIGOScientific:2016aoc}, General Relativity (GR) still stands as the theory that encodes our best understanding of gravity at low energies. Consistency and parametrized null tests performed with all GW observations available so far continue to show agreement with GR~\cite{TheLIGOScientific:2016src,Abbott:2018lct,LIGOScientific:2019fpa,Abbott:2020jks, LIGOScientific:2021sio}, and so do tests performed in the solar system~\cite{Will:1993hxu,Will:2014kxa} and with binary pulsars~\cite{Damour:1991rd,Kramer:2006nb,Freire:2012mg, Kramer:2021jcw}. However,
cosmological observations pointing at the existence of a ``dark sector'' (dark matter and especially dark energy) may be interpreted as a sign of a possible breakdown of GR on large scales (see e.g.~\cite{Clifton:2011jh} for a review).

This has prompted the development of effective field theories (EFTs) of dark energy, which attempt to explain the latter as a gravitational effect (caused by a deviation from GR) rather than by introducing an exotic matter component or a cosmological constant.
Restricting to scalar-tensor theories,  which postulate the existence of an additional degree of freedom (besides the metric) in the gravitational sector, EFTs of dark energy may be provided by the Horndeski class \cite{Horndeski:1974wa} (further generalizable
to degenerate higher-order scalar tensor-theories, DHOST~\cite{Gleyzes:2014dya,Langlois:2015cwa, Crisostomi:2016czh,BenAchour:2016fzp}). In this class, a prominent role is played by ``\(k\)-essence'' theories with first-derivative self interactions~\cite{Chiba:1999ka, Armendariz-Picon:2000nqq},  which are among the very few terms in the DHOST class that remain experimentally viable despite constraints from GW propagation~\cite{Monitor:2017mdv,TheLIGOScientific:2017qsa,Langlois:2017dyl,Crisostomi:2017pjs,Crisostomi:2017lbg,Dima:2017pwp,Creminelli:2018xsv, Creminelli:2019kjy, Babichev:2020tct}.

Potentially even tighter constraints may come from the {\it generation} (rather than just the propagation) of GWs~\cite{Bezares:2021yek,Bezares:2021dma}. However, obtaining predictions for GW generation is far more difficult than for  propagation, as the   nonlinear self interactions of the \(k\)-essence scalar are believed to dominate the dynamics
on the small scales characterizing compact binary systems. In fact, this expectation comes from calculations of static and quasi-static systems (such as stars), on whose scales the scalar self interactions are important and tend to suppress deviations from GR~\cite{Babichev:2009ee,terHaar:2020xxb}. This nonlinear mechanism, known as ``screening'' (of local scales from GR deviations), is common to other theories in the DHOST class (see e.g. Refs.~\cite{Vainshtein:1972sx,Babichev:2013usa,Khoury:2003rn,Hinterbichler:2010es}) and is both a blessing and a curse. On the one hand it allows \(k\)-essence to pass solar-system tests of gravity~\cite{terHaar:2020xxb,Bezares:2021yek}, but on the other hand it
renders the calculation of GW generation conceptually and practically involved~\cite{Bernard:2019fjb,Bezares:2020wkn,terHaar:2020xxb,Bezares:2021yek,Bezares:2021dma}.

In fact, because of the nonlinear scalar derivative self-interactions,  evolutions to the future of initial configurations of interest (on which  calculations of GW generation in the highly relativistic and strong-field regime of compact binaries are based) may become ``unstable'', i.e. they may depend ``discontinuously'' on the initial data 
and/or exhibit fast exponential growth.
(See e.g. Refs.~\cite{Akhoury:2011hr,Bernard:2019fjb, Kunstatter:2011ce, Gannouji:2020kas}.)
 In mathematical jargon, this corresponds to the Cauchy (i.e. initial value) problem becoming ill-posed~\cite{Hadamard10030321135}. While for astrophysically relevant initial conditions (such as neutron star binaries or gravitational collapse)
this breakdown of the Cauchy problem can be avoided by a judicious choice of gauge~\cite{Bezares:2021dma} (at least in specific \(k\)-essence theories),
for general theories/configurations this may not  always be possible. In fact, a more robust approach to ``fixing'' the Cauchy problem
is to complete \(k\)-essence to the ultraviolet (UV)~\cite{Burgess:2014lwa} (when that 
is allowed by positivity bounds~\cite{Adams:2006sv}) or to ``deform'' the 
evolution (by adding an auxiliary field that drives the dynamics to the ``real'' one on long timescales). This second approach to ``fix the equations'' was proposed by Cayuso, Ortiz and Lehner in Ref.~\cite{Cayuso_2017} (see also Refs.~\cite{Allwright:2018rut,Cayuso:2020lca}), partly inspired by dissipative hydrodynamics, and
was successfully applied to gravitational collapse in \(k\)-essence by Refs.~\cite{Bezares:2021yek,Bezares:2021dma} (where it was shown to reproduce the results obtained 
in a gauge where breakdowns of the Cauchy problem are avoided). 
On a similar note, shocks/caustics in \(k\)-essence~\cite{Bernard:2019fjb, Babichev:2016hys, Babichev:2017lrx} may also be resolved by resorting to a UV completion. In Ref.~\cite{MukohyamaUV:2020lsu}, it was illustrated that the transfer of energy to an additional (UV) degree of freedom may allow for the smoothening of shock/caustic fronts in \(k\)-essence.

In this paper, we take a step back and investigate in depth the relation between first-derivative self interactions of the scalar,  well-posedness of the Cauchy problem and  UV completions (in both the standard and ``fixing the equations'' approaches). To this purpose, we consider a \(k\)-essence model that potentially suffers from both Tricomi-type and Keldysh-type breakdowns of initial-data evolutions~\cite{Bernard:2019fjb,Bezares:2020wkn}. In more detail, the former corresponds to the equations becoming parabolic (and then elliptic) along the evolution, while the latter are caused by diverging (coordinate) characteristic speeds for the scalar mode. By 
suitably choosing the sign of the coupling of the first-derivative scalar self interactions in the action, we can then extend the 
 \(k\)-essence model to a standard \(U(1)\) symmetric UV completion~\cite{Burgess:2014lwa}. Solutions in the UV-complete theory are compared to ones
 in the low-energy \(k\)-essence theory (as long as the Cauchy problem in the latter remains well-posed) and to ones in a ``fixing the equations'' completion. We also explore the relation between the 
 regime in which the scalar self-interactions become important and the domain of validity of the low-energy EFT, finding that the two are closely connected for the example that we study.
 
 In more detail, this paper is organized as follows.
 First, in Sec.~\ref{sec: Quadratic Kessence} we review the  \(k\)-essence model that we adopt as our case study. We then introduce its standard UV-completion in Sec.~\ref{Sec: HiggsUVCompletion}, while our ``fixing the equations'' approach is introduced and applied in 
 Sec.~\ref{sec: Fixing the Equations}. We describe our numerical implementation in Sec.~\ref{sec: implementation} and present our results in Sec.~\ref{sec: results}. 
 Our findings are discussed and conclusions drawn in Sec.\ref{sec: conclusion}. 
%
% Appendices
%
In Appendix~\ref{sec: weak data example}, we present an additional example, and in Appendix~\ref{sec: Constraint propagation} we elaborate on details regarding the constraint propagation in the ``fixed" theory.
Throughout this paper, we use the metric signature \(-+++\) and work in units where $c=1$. Greek letters \(\mu, \, \nu,  \ldots\) denote spacetime indices ranging from \(0\) to \(3\), while Roman letters near the middle \(i, \, j, \ldots\) range from \(1\) to \(3\), denoting spatial indices.

%%%%%%%%%%%%%%%%%%%%%%%%%%%%%%%%%%%%%%%%%%%%%%%%%%%%%%%%%%%
%%%%%%%%%%%%%%%%%%%%%%%%%%%%%%%%%%%%%%%%%%%%%%%%%%%%%%%%%%%
%%%%%%%%%%%%%%%%%%%%%%%%%%%%%%%%%%%%%%%%%%%%%%%%%%%%%%%%%%%

\section{Quadratic \(k\)-essence } \label{sec: Quadratic Kessence}

The action of \(k\)-essence in vacuum is given by
\begin{align} \label{eq: actionKessence}
    S_\text{K} \left[g_{\mu\nu}, \pi \right]  = \int d^4 x \sqrt{-g} \left[ \dfrac{\mathcal{R}}{2\kappa}  + K\left(X\right) \right]~,
\end{align}
where $\kappa=8 \pi G$, \(\mathcal{R}\) is the Ricci scalar,  $g_{\mu\nu}$ is the spacetime metric and $K\left(X\right)$ is a function of the standard kinetic term of the scalar field $\pi(x)$, given by
$X = \nabla^{\mu} \pi \nabla_{\mu} \pi$. 
The quadratic model is defined by keeping only the leading first derivative self interaction, i.e.
\begin{align} \label{eq: KQuadratic}
    K(X) = - \dfrac{1}{2} X
    + \dfrac{\beta}{4 \Lambda^4} X^2
    + O\left(\Lambda^{-8} X^3\right)~,
\end{align}
with \(\beta \sim O(1)\) a dimensionless coupling constant and \(\Lambda\) the EFT strong coupling scale. 
Note that in the presence of matter,  screening  is present only in the \(\beta < 0 \) branch \cite{Babichev:2009ee,terHaar:2020xxb,Bezares:2020wkn}. However, positivity bounds select the branch with \(\beta > 0\) as the one consistent with  embedding in a UV theory \cite{Adams:2006sv}.

The vacuum field equations derived from action \eqref{eq: actionKessence} are given by
\begin{align} \label{eq: KsscFieldEquations}
 G_{\mu\nu} &= \kappa \, T^{(\pi)}_{\mu\nu}~,
\end{align}
where \(G_{\mu\nu}\) is the Einstein tensor and 
\begin{align}
 T^{(\pi)}_{\mu\nu} = K\left(X\right) g_{\mu\nu} - 2 K'\left(X\right) \nabla_{\mu} \pi \nabla_{\nu} \pi~, 
\end{align}
is the energy-momentum tensor of the \(k\)-essence field.
The equation for the scalar field can be written as 
\begin{align}\label{eq : KsscScalarEq}
    \gamma^{\mu\nu}\nabla_\mu \nabla_\nu \pi = 0~,
\end{align}
where
\begin{align} \label{eq: effective metric}
    \gamma^{\mu\nu} &= g^{\mu\nu} 
    + \dfrac{2K''(X)}{K'(X)}\nabla^\mu \pi \nabla^\nu \pi~,
\end{align}
is an effective metric for the scalar field. From Eq.~\eqref{eq: effective metric}, it is evident that the scalar equation \eqref{eq : KsscScalarEq} may develop shocks/caustics (e.g. discontinuities) if the scalar gradients become large, even in situations when the initial data for the scalar field is smooth~\cite{Bernard:2019fjb, Babichev:2016hys, Babichev:2017lrx}.
Additionally, other pathologies may arise if \(K'\left(X\right)\) approaches zero \cite{Kunstatter:2011ce}.

In order to study the non-linear dynamical regime, the \emph{well-posedness} of the Cauchy problem must first be assessed. 
According to Hadamard's criteria~\cite{Hadamard10030321135}, the  Cauchy initial value problem governed by  Eqs.~\eqref{eq: KsscFieldEquations} and \eqref{eq : KsscScalarEq} is well-posed if a unique solution exists and  depends continuously on the initial data.
This can be shown to occur if the associated system of equations is \emph{strongly hyperbolic} ~\cite{FriedrichRendall:10.1007/3-540-46580-4_2, Reula:1998ty}, i.e. if the system of equations can be written as a quasilinear first-order system and its principal part (consisting of the terms with the highest derivatives) has real eigenvalues and a complete set of eigenvectors~\cite{Sarbach:2012pr, Hilditch:2013sba}.
In our case, one can restrict the analysis to the scalar equation \eqref{eq : KsscScalarEq}, since the evolution equations for the metric variables [Eq.~\eqref{eq: KsscFieldEquations}] take the same form as in GR (which is well-posed~\cite{ChoquetBruhat:1952grw}) and the source terms involve only  derivatives that are lower-order than the principal part.

In the following, we will restrict to spherical symmetry, where the metric can be written in the form
\begin{align}\label{eq: metric ansatz}
    ds^2 = - \alpha^2(t, r) \, dt^2 + g_{rr}(t, r)\, dr^2 + r^2g_{\theta \theta}(t, r)\,d\Omega^2~,
\end{align}
where \(\alpha(t, r)\) is the lapse function, and \(g_{rr}(t, r)\) and \(g_{\theta \theta}(t, r)\) are the spatial components of the metric and \(d\Omega^2 = d\theta^2 + \sin^2 \left(\theta\right) d \phi^2\).
The scalar equation \eqref{eq : KsscScalarEq} can be written as a first-order system of equations of the form
\begin{align} \label{eq: FirstOrderScalarSystem}
    \partial_t \boldsymbol{U} + \mathbb{V} \, \partial_r \boldsymbol{U} = \mathcal{S}\left(\boldsymbol{U}\right),
\end{align}
where \(\boldsymbol{U} = \left(\partial_t \pi, \partial_r \pi\right)\), \(\mathcal{S}\left(\boldsymbol{U}\right)\) is a source term, and we have made use of the consistency equation \(\partial_t \partial_r \pi = \partial_r \partial_t \pi\).
The \emph{characteristic speeds}, corresponding to the eigenvalues of the characteristic matrix \(\mathbb{V}\), are given by
\begin{align} \label{eq: KsscCharacteristicSpeeds}
    V_{\pm} = - \dfrac{\gamma^{tr}}{\gamma^{tt}} 
            \pm \sqrt{ \dfrac{-\mathrm{det}\left(\gamma^{\mu\nu}\right)}{\left(\gamma^{tt}\right)^2}  }~,
\end{align}
where \(\mathrm{det}\left(\gamma^{\mu\nu}\right)\) should be understood as the determinant of the effective metric in the \((t, r)\) subspace, i.e.
\begin{align} \label{eq: effective metric determinant}
\det \left( \gamma^{\mu\nu} \right) = \gamma^{tt}\gamma^{rr} - \left(\gamma^{tr}\right)^2 ~.
\end{align}
If these speeds are real and distinct, the corresponding eigenvectors form a complete set, and thus the scalar sector is strongly hyperbolic.

Since the characteristic speeds~\eqref{eq: KsscCharacteristicSpeeds} 
depend on the effective metric (which differs in general from the spacetime 
metric $g_{\mu\nu}$),
two situations may arise that can cause a breakdown of strong hyperbolicity. 
% Tricomi
The first problem occurs when the scalar equation \eqref{eq : KsscScalarEq} changes character from hyperbolic to parabolic, i.e. when one of the eigenvalues of the effective metric [Eq.~\eqref{eq: effective metric}],
\begin{align} \label{eq: Kessence effective metric eigenvalues}
    \lambda_{\pm} = \dfrac{1}{2} \left( \gamma^{ tt} +  \gamma^{rr} \pm \sqrt{ \left(\gamma^{tt} - \gamma^{rr}\right)^2 - \left(2 \gamma^{tr} \right)^2} \right)~,
\end{align}
vanishes, implying \(\det \left(\gamma^{\mu\nu}\right) \to 0\). This referred to as a \emph{Tricomi-type} breakdown \cite{Ripley_2019a} (see also Ref.~\cite{Bernard:2019fjb})
due to its resemblance to the behavior of the Tricomi equation, 
\( \partial^2_t u (t, r) + t \, \partial^2_r u (t, r) = 0\).
The  system of evolution equations, including those for  the metric, then becomes of mixed-type, with parabolic and hyperbolic sectors~\cite{Stewart:2002vd}.
%
%Keldyish
The second problem occurs when the characteristic speeds  diverge. This referred to as a \emph{Keldyish-type} breakdown \cite{Ripley_2019a} (see also Ref.~\cite{Bernard:2019fjb}), 
in analogy with the Keldyish equation, 
\(t \,\partial^2_t u (t, r) + \partial^2_r u (t, r) = 0\).

Both problems may be solved by a suitable UV completion of the EFT. In fact, in the following we will
review a UV completion of the quadratic \(k\)-essence model given by \eqref{eq: KQuadratic} (for $\beta>0$), and show that it allows for
avoiding both Keldysh and Tricomi breakdowns of the Cauchy problem. 
Similarly, the ``fixing the equations'' approach~\cite{Cayuso_2017}
may also improve the behavior of initial-value evolutions in the branch $\beta<0$.

%%%%%%%%%%%%%%%%%%%%%%%%%%%%%%%%%%%%%%%%%%%%%%%%%%%%%%%%%%%
%%%%%%%%%%%%%%%%%%%%%%%%%%%%%%%%%%%%%%%%%%%%%%%%%%%%%%%%%%%
%%%%%%%%%%%%%%%%%%%%%%%%%%%%%%%%%%%%%%%%%%%%%%%%%%%%%%%%%%%

\subsection{\(U(1)\) UV completion} \label{Sec: HiggsUVCompletion}

The positive branch (\(\beta > 0\)) of quadratic \(k\)-essence can be obtained as the low-energy description of a UV completion 
given by the \(U(1)\)-symmetric action\footnote{To be precise, this is a \emph{partial} UV completion as it only describes the scalar degree of freedom at higher energies. A \emph{full} UV completion would also describe the gravitational degrees of freedom, e.g. in a full theory of quantum gravity.}
\begin{align} \label{eq: actionHiggs}
    S_\text{UV} \left[g_{\mu\nu}, \phi\right] =\int \mathrm{d}^4 x \, \sqrt{-g }\, \bigg[\dfrac{\mathcal{R}}{2 \kappa}  -\partial_\mu \phi^\star \partial^\mu \phi + \vphantom{]}   \nonumber\\
    \vphantom{[}- V\left(\phi^\star \phi\right)\bigg],
\end{align}
with a potential
\begin{align}
    V(\phi^\star \phi) = \dfrac{\lambda}{2}\left(\phi^\star \phi - \dfrac{v^2}{2}\right)^2,
\end{align}
where \(\phi\) is a complex scalar field (with \(\phi^\star\) its complex conjugate), \(\lambda > 0 \) is a dimensionful coupling constant and \(v\) can be interpreted as the scale of the vacuum expectation value of \(\phi\), i.e. the magnitude of \(\phi\) that minimizes \(V\left(\phi^\star \phi\right)\).

In Minkowski space, quadratic \(k\)-essence is recovered at low energies when the \(U(1)\) symmetry of action~\eqref{eq: actionHiggs} is broken spontaneously~\cite{Burgess:2014lwa}. When gravity is considered the same result holds at leading order. Indeed, by expanding \(\phi \) around the degenerate vacuum of the potential, 
\begin{align} \label{eq: VEVexpansion}
    \phi(x) = \dfrac{v}{\sqrt{2}}\left[ 1 + \rho(x) \right] e^{i \theta(x)}~,   
\end{align}
it can be seen, by substituting in action \eqref{eq: actionHiggs}, that the radial field \(\rho(x)\) acquires 
a ``mass" 
\footnote{
In our units $c=1$, the ``mass'' \(M_{\rho}\) is actually the inverse of the Compton wavelength, i.e.
the real mass is $m_\rho = M_\rho \hbar$.} 
\(M_\rho = \sqrt{\lambda} \, v \), while the phase field \(\theta(x)\) (i.e. the ``Goldstone boson''~\cite{GoldstoneSalamWeinberg:PhysRev.127.965}) remains massless. At  energies much lower than \(M_\rho\), one can use the equation of motion of the radial field, \(-\Box \rho + \left(1 + \rho \right) \partial_\mu \theta \partial^\mu \theta  + v^{-2}\partial V/\partial \rho = 0\),  to integrate it out of action \eqref{eq: actionHiggs}. More precisely, one can solve perturbatively for \(\rho\) as
\begin{align} \label{eq: rhoPertSol}
    \rho = - \dfrac{1}{M^{2}_\rho } \partial_\mu \theta \partial^\mu \theta  + \mathcal{O}\left(M^{-4}_\rho \right)~,
\end{align}
and substitute in the action \eqref{eq: actionHiggs} to obtain the effective action for the phase field  \(\theta (x)\). The latter takes the same form as Eq.~\eqref{eq: actionKessence}, i.e.
\begin{multline} \label{eq: EffectiveAction}
    S_\text{eff} \left[g_{\mu\nu}, \theta \right] =  \int d^4 x \sqrt{-g} \bigg[ \dfrac{\mathcal{R}}{2\kappa}+ \vphantom{]}\\
    \vphantom{[}+ v^2\left(- \dfrac{1}{2} \partial_\mu \theta \partial^\mu \theta 
    + \dfrac{1}{2 M_\rho^2} \left(\partial_\mu \theta \partial^\mu \theta \right)^2 \right)\bigg] +\\
    + \mathcal{O}(v^2 M_\rho^{-4}\nabla^6)~,
\end{multline}
where \(\mathcal{O}(v^2 M_\rho^{-4}\nabla^6)\) denotes higher order terms (in \(M_{\rho}^{-2}\)) with at least six   derivatives.
Therefore, this UV completion reproduces the dynamics of quadratic \(k\)-essence at leading order, and the \(k\)-essence field is 
interpreted as given by
the dimensionful ``phase'' field
\begin{align} \label{eq: UV phase field}
\pi^\text{(UV)}(x) = v \,  \theta(x)~.
\end{align} 
Direct comparison between the actions \eqref{eq: actionKessence} and \eqref{eq: EffectiveAction} yields the relation between the coupling constants in the two theories,
\begin{align} \label{eq: coupling constant matching}
    \dfrac{\beta}{2\Lambda^4} = \dfrac{1}{M_\rho^2 v^2} \geq 0~,
\end{align}
and selects the positive branch of quadratic \(k\)-essence 
(for which there is no screening mechanism in the presence of matter), 
consistently with positivity bounds \cite{Adams:2006sv}. 
At next-to-leading order, the higher order terms do not reproduce \(k\)-essence, since the UV completion introduces other six-derivative terms in addition to the cubic term appearing in Eq.~\eqref{eq: KQuadratic} -- see e.g. Ref.~\cite{Burgess:2014lwa}.

We now turn to the question of whether this UV completion admits a well-posed Cauchy problem. Since the scalar field \(\phi\) is minimally coupled to the metric, the evolution equations for the metric are
\begin{align}\label{eq: UVFieldEquations}
    G_{\mu\nu} = \kappa \, T_{\mu\nu}^{(\phi)}~,
\end{align}
where now
\begin{align}
    T_{\mu\nu}^{(\phi)} =\nabla_\mu \phi^\star \nabla_\nu \phi  & + \nabla_\mu \phi \nabla_\nu \phi^\star + \nonumber\\
    & - g_{\mu\nu} \left[\nabla^\sigma \phi \nabla_\sigma \phi + V\left(\phi^{\star} \phi\right) \right]~.
\end{align}
As before, it can be shown that the system is strongly hyperbolic~\cite{Alcubierre:1138167}. The scalar equation,
\begin{align} \label{eq: HiggsScalarEquation}
    \Box \phi - \dfrac{\partial V}{\partial \lvert \phi \rvert ^2} \phi = 0~,
\end{align}
is also manifestly strongly hyperbolic since it is a wave equation.
We split the complex scalar 
\begin{align} \label{eq: ComplexScalarSplitting}
 \phi = \phi_R + i \, \phi_I   
\end{align}
in its real and imaginary parts,  \(\phi_R\) and \(\phi_I\). Then the associated characteristic speeds are given by
\begin{align} \label{eq: HiggsCharacteristicSpeeds}
%  V^{(\phi_R)}_{\pm}= V^{(\phi_I)}_{\pm} = \pm \alpha(r) \sqrt{g^{rr}}~,   
 V^{(\phi_R)}_{\pm}= V^{(\phi_I)}_{\pm} = \pm \dfrac{\alpha}{\sqrt{g_{rr}}}~,   
\end{align}
which are always real and distinct (hence implying the existence of a complete set of eigenvectors).

%%%%%%%%%%%%%%%%%%%%%%%%%%%%%%%%%%%%%%%%%%%%%%%%%%%%%%%%%%%
%%%%%%%%%%%%%%%%%%%%%%%%%%%%%%%%%%%%%%%%%%%%%%%%%%%%%%%%%%%
%%%%%%%%%%%%%%%%%%%%%%%%%%%%%%%%%%%%%%%%%%%%%%%%%%%%%%%%%%%

\subsection{Fixing the Equations} \label{sec: Fixing the Equations}

The ``fixing the equations" approach~\cite{Cayuso_2017} (see also Refs.~\cite{Allwright:2018rut,Cayuso:2020lca}) provides a prescription to control the high frequency behavior of an EFT, which may be the cause of ill-posedness of the Cauchy problem. 
In the following, we will apply this prescription to \(k\)-essence. Unlike for the case 
of the \(U(1)\) UV completion presented in the previous section, we do not make here any assumptions on the sign
of $\beta$.

In order to deal with shocks (c.f. Sec.~\ref{sec: implementation}), it is convenient to rewrite the scalar equation~\eqref{eq : KsscScalarEq} in conservative form (as made possible by the shift-symmetry of the theory):
\begin{align} \label{eq : KsscScalarEqConservative}
    \nabla_\mu\left(K'(X)\nabla^{\mu} \pi\right) = 0~.
\end{align}
Since large gradients may occur due to the \(K'(X) \) factor,
triggering a breakdown of the Cauchy problem,
we ``fix" the scalar equation~\eqref{eq : KsscScalarEqConservative} by replacing  \(K'(X)\)  with a \emph{new} dynamical field \(\Sigma\), which in turn is prescribed to approach \(K'(X)\) by a ``driver" equation. The system of equations that we adopt (see also Ref.~\cite{Bezares:2021yek,Bezares:2021dma}) is therefore
\begin{align} \label{eq: FEScalarEquation}
    \nabla_\mu ( \Sigma \nabla^\mu \pi) = 0~,  \\
    \label{eq: FESigmaEquation}
    \tau \, \partial_t \Sigma = - \left[ \Sigma - K'\left( X \right) \right]~,
\end{align}
where \(\tau\) is a constant timescale, whose precise value controls the rate of approach of \(\Sigma\) to \(K'\left(X\right)\) and which  damps frequencies \(\omega\) in the range \(\tau^{-1} \lesssim \omega\) \cite{LLPrivateCommunication, Cayuso_2017}.
For the metric, the evolution equations remain unaltered and are given by Eq.~\eqref{eq: KsscFieldEquations}.

The characteristic speeds of the ``fixed" theory, for \(\Sigma \neq 0\), are
\begin{align} \label{eq: FECharacteristicSpeeds}
    % V^{(\text{FE})}_{\pm} & = \pm \alpha(r) \sqrt{g^{rr}}~,     
    V^{(\text{FE})}_{\pm} & = \pm \dfrac{\alpha}{ \sqrt{g_{rr}}}~,         
\end{align}
with an additional speed \( V^{(\text{FE})}_{3}  = 0\) due to the presence of the new variable \(\Sigma\).
These speeds are always real and distinct (hence implying the existence of a complete set of eigenvectors). Therefore, as long as \(\Sigma \neq 0\), the system of equations of the ``fixed" theory is strongly hyperbolic.

However, if \(\Sigma \) approaches zero during the evolution, a pathological situation occurs. 
This can be seen as follows: rewriting Eq.~\eqref{eq: FEScalarEquation} as \(\Sigma \, \Box \pi + \nabla_\mu \Sigma \, \nabla^\mu \pi = 0 \), it is evident that when \(\Sigma \to 0\) the principal part of this equation (i.e. the part consisting of the highest derivative terms) vanishes, and therefore the system~\eqref{eq: FEScalarEquation}-\eqref{eq: FESigmaEquation} changes from  second order  to  first order.

Finally, in contrast with the UV completion of Sec.~\ref{Sec: HiggsUVCompletion}, there are no restrictions on the sign of the quadratic \(k\)-essence coupling constant \(\beta\), and the ``fixing the equations" prescription can also be applied to the branch with screening (\(\beta < 0 \)).

%%%%%%%%%%%%%%%%%%%%%%%%%%%%%%%%%%%%%%%%%%%%%%%%%%%%%%%%%%%
%%%%%%%%%%%%%%%%%%%%%%%%%%%%%%%%%%%%%%%%%%%%%%%%%%%%%%%%%%%
%%%%%%%%%%%%%%%%%%%%%%%%%%%%%%%%%%%%%%%%%%%%%%%%%%%%%%%%%%%

\section{Methodology} \label{sec: implementation}

In order to explore the well-posedness of the Cauchy problem and the nonlinear dynamics in \(k\)-essence, in its \(U(1)\) UV completion and in the ``fixing the equations'' approach, the fully nonlinear equations must be considered.
In the following, we present the evolution equations in a \(1 + 1\) decomposition of the spacetime restricted to spherical symmetry
and describe the details of our numerical implementation.
First, in Sec.~\ref{sec: Evolution equations} we present the evolution equations for the scalar sector in a first-order conservative form.
We specify our working units in Sec.~\ref{sec: physical units} and then, in Sec.~\ref{sec: initial data} we describe in detail the procedure used to construct initial data. 
In Sec.~\ref{sec: evolution scheme}, we describe the numerical evolution scheme and code.
Finally, in Sec.~\ref{sec: DiagnosticQ}, we describe additional diagnostic tools needed to compare and interpret our numerical simulations.

%%%%%%%%%%%%%%%%%%%%%%%%%%%%%%%%%%%%%%%%%%%%%%%%%%%%%%%%%%%
%%%%%%%%%%%%%%%%%%%%%%%%%%%%%%%%%%%%%%%%%%%%%%%%%%%%%%%%%%%
%%%%%%%%%%%%%%%%%%%%%%%%%%%%%%%%%%%%%%%%%%%%%%%%%%%%%%%%%%%

\subsection{Evolution equations} \label{sec: Evolution equations}

We decompose the metric into space and time components by using the line element in spherical symmetry given by Eq.~\eqref{eq: metric ansatz}. 
In the \(1 + 1\) decomposition, the Einstein equations [Eq.~\eqref{eq: KsscFieldEquations} for \(k\)-essence and the ``fixed" theory and  Eq.~\eqref{eq: UVFieldEquations} for the \(U(1)\) UV completion]
can be written in first-order form [analogous to Eq.~\eqref{eq: first order scalar system}] by using the \(Z3\) formulation, which is strongly hyperbolic~\cite{Bona:2002fq,Bona:2005pp,Alic:2007ev}.
We write the evolution equations for the metric as a first order system by defining the 
variables
\begin{align}
A_{r}&=\dfrac{1}{\alpha}\partial_{r}\alpha~, & {D_{rr}}^{r}&=\dfrac{1}{2 g_{rr}}\partial_{r}g_{rr}~, &  {D_{r\theta}}^{\theta}&=\dfrac{1}{2g_{\theta\theta}}\partial_{r}g_{\theta\theta}~,
\end{align}
and the extrinsic curvature
\begin{align}
    K_{ij} = -\dfrac{1}{2} \mathcal{L}_{n} g_{ij}~,
\end{align}
where \( g_{ij} \) is the spatial metric and \( n_{\mu} = \left( -\alpha, 0 \right)   \) is the normal vector to the foliation.
We close the evolution system by prescribing  the singularity-avoidance \(1 + \mathrm{log} \) slicing condition,
\( \partial_t \mathrm{log} \alpha =  - 2 \, \mathcal{K} \), 
where the trace of the extrinsic curvature is  \( \mathcal{K} = {K_{r}}^{r} + 2{K_{\theta}}^{\theta} \)~\cite{Bona:1994dr}.   The final set of evolution fields for the \(Z3\) formulation in spherical symmetry can be found in Ref.~\cite{Valdez_Alvarado_2013}

In the following, we will also describe  the scalar equation in \(k\)-essence, in its \(U(1)\) UV completion
and in the ``fixing the equations'' approach, and write it
in first-order form.

%%%%%%%%%%%%%%%%%%%%%%%%%%%%%%%%%%%%%%%%%%%%%%%%%%%%%%%%%%%
%%%%%%%%%%%%%%%%%%%%%%%%%%%%%%%%%%%%%%%%%%%%%%%%%%%%%%%%%%%
%%%%%%%%%%%%%%%%%%%%%%%%%%%%%%%%%%%%%%%%%%%%%%%%%%%%%%%%%%%

\subsubsection{Quadratic \(k\)-essence}\label{sec: EvolEqsQuadraticKessence}

Defining the following first-order variables 
\begin{align}\label{eq: red_var}
\Phi&=\partial_{r}\pi ~,& \Pi&=-\dfrac{1}{\alpha}\partial_{t}\pi ~,
\end{align}
one can write the scalar equation~\eqref{eq : KsscScalarEq}  in first-order conservative form as
\begin{align}\label{eq: first order scalar system}
\partial_t \phi &+ \alpha \Pi  = 0 ~, \nonumber\\
\partial_t \Phi &+ \partial_r [\alpha \Pi]   = 0 ~, \nonumber\\
\partial_t \Psi &+ \partial_r F_{\Psi} = - \dfrac{2}{r} F_{\Psi}~,
\end{align}
where 
\begin{align}
\Psi & = \sqrt{g_{rr}} g_{\theta \theta} K'\left(X\right)\Pi~, \label{eq: eqPSI} \\ 
F_{\Psi} &= \dfrac{\alpha g_{\theta\theta } }{\sqrt{g_{rr} }} K'\left(X\right)\Phi~.
\end{align}
At each time step, \(\Pi\) is obtained (numerically) by solving the non-linear equation~\eqref{eq: eqPSI}.

%%%%%%%%%%%%%%%%%%%%%%%%%%%%%%%%%%%%%%%%%%%%%%%%%%%%%%%%%%%
%%%%%%%%%%%%%%%%%%%%%%%%%%%%%%%%%%%%%%%%%%%%%%%%%%%%%%%%%%%
%%%%%%%%%%%%%%%%%%%%%%%%%%%%%%%%%%%%%%%%%%%%%%%%%%%%%%%%%%%

\subsubsection{\(U(1)\) UV completion}

For the \(U(1)\) UV completion,  the scalar equation~\eqref{eq: HiggsScalarEquation} defines two real systems of equations for the real \(\phi_R\) and imaginary \(\phi_I\) parts. 
As before, we define the first-order scalar variables 
\begin{align}
    \Phi_{R, I}&=\partial_{r}\phi_{R, I} ~,& \Pi_{R, I}&=-\dfrac{1}{\alpha}\partial_{t}\phi_{R, I} ~.
\end{align}
Then, the real scalar system for \(\phi_{R, I}\) can be written as
\begin{align}%\label{eq: first order scalar system}
\partial_t \phi_{R,I} &+ \alpha \Pi_{R,I}  = 0 ~, \nonumber\\
\partial_t \Phi_{R,I} &+ \partial_r [\alpha \Pi_{R,I}]   = 0 ~, \nonumber\\
\partial_t \Pi_{R,I} &+ \partial_r\left[\dfrac{\alpha}{g_{rr}} \Phi_{R, I}\right] = S_{\Pi_{R,I}}~,
\end{align}
with source term
\begin{multline}
S_{\Pi_{R, I}} = \alpha \bigg[ \left({K_{r}}^{r} + 2 {K_{\theta}}^{\theta}\right)\Pi_{R, I} + \vphantom{]}\\
    \vphantom{[}- \dfrac{1}{g_{rr}} \left(\dfrac{2}{r} + {D_{rr}}^{r} + 2 {D_{r \theta}}^{ \theta}\right)\ \Phi_{R, I} + \dfrac{\partial V}{\partial | \phi |^2} \phi_{R, I} \bigg] ~.
\end{multline}

%%%%%%%%%%%%%%%%%%%%%%%%%%%%%%%%%%%%%%%%%%%%%%%%%%%%%%%%%%%
%%%%%%%%%%%%%%%%%%%%%%%%%%%%%%%%%%%%%%%%%%%%%%%%%%%%%%%%%%%
%%%%%%%%%%%%%%%%%%%%%%%%%%%%%%%%%%%%%%%%%%%%%%%%%%%%%%%%%%%

\subsubsection{Fixing the equations}

Finally, for the ``fixed" theory,  
the first order system of equations  can be written in the same form as in \(k\)-essence (Sec.~\ref{sec: EvolEqsQuadraticKessence}), but replacing \(K'(X) \to \Sigma\) and including the ``driver" equation~\eqref{eq: FESigmaEquation}.

%%%%%%%%%%%%%%%%%%%%%%%%%%%%%%%%%%%%%%%%%%%%%%%%%%%%%%%%%%%
%%%%%%%%%%%%%%%%%%%%%%%%%%%%%%%%%%%%%%%%%%%%%%%%%%%%%%%%%%%
%%%%%%%%%%%%%%%%%%%%%%%%%%%%%%%%%%%%%%%%%%%%%%%%%%%%%%%%%%%

\subsection{Initial data} \label{sec: initial data}

We will now describe in detail the construction of initial data in isotropic coordinates, corresponding to an initially stationary  scalar ``shell" in \(k\)-essence (Sec.~\ref{sec: IDKessence}). We will then comment on how this procedure can be generalized to the \(U(1)\) UV completion (Sec.~\ref{sec: IDHiggs}) and the ``fixed'' theory (Sec.~\ref{sec: IDFE}). 

%%%%%%%%%%%%%%%%%%%%%%%%%%%%%%%%%%%%%%%%%%%%%%%%%%%%%%%%%%%
%%%%%%%%%%%%%%%%%%%%%%%%%%%%%%%%%%%%%%%%%%%%%%%%%%%%%%%%%%%
%%%%%%%%%%%%%%%%%%%%%%%%%%%%%%%%%%%%%%%%%%%%%%%%%%%%%%%%%%%

\subsubsection{Quadratic \(k\)-essence}\label{sec: IDKessence}

On the initial slice  at time \(t = 0\), we adopt isotropic coordinates given by
\begin{align} \label{eq: initial isotropic metric}
    ds^2 = -\alpha^2(r) \, dt^2 + \psi^4(r) \left(dr^2 + r^2 \, d\Omega^2\right)~,
\end{align}
and prescribe the initial profile of the lapse function to be constant and equal to unity -- i.e. \(\alpha(r)\rvert_{t = 0} = 1 \).

In these coordinates, the Hamiltonian and momentum constraints for \(k\)-essence take the form
\begin{align} 
\dfrac{1}{r^2}\dfrac{\partial}{\partial r}\left(r^2 \dfrac{\partial \psi}{\partial r} \right) &= \dfrac{1}{4}\psi^5 {K_{\theta}}^{\theta}\left(2 {K_{r}}^{r} + {K_{\theta}}^{\theta}\right) + \nonumber\\
&\qquad +  P\left[\alpha, \psi, \partial_t \pi, \partial_r \pi, \pi \right]~, \label{eq: IsoHamConstraint}\\
\dfrac{\partial {K_{\theta}}^{\theta}}{\partial r} &= \dfrac{1}{r \psi} \left({K_{r}}^{r} - {K_{\theta}}^{\theta}\right)\left(\psi + 2 r \dfrac{\partial \psi}{\partial r} \right) +\nonumber\\
&\qquad + Q\left[\alpha, \psi, \partial_t \pi, \partial_r \pi, \pi \right]~,\label{eq: IsoMomConstraint}
\end{align}
respectively, where
\begin{align}
    P[\alpha, \psi, \partial_t \pi, \partial_r \pi, \pi] &= \dfrac{1}{4} \kappa \psi^5 \left[K\left(X\right) + 2 \Pi^2 K'\left(X\right)\right]~, \nonumber\\
    Q[\alpha, \psi, \partial_t \pi, \partial_r \pi, \pi ] &= \kappa\,\Pi \Phi K'\left(X\right)~.
\end{align}

We will consider initially stationary configurations
by imposing
%
% \(\partial_t \pi = \mathcal{K} = {K_{\theta}}^{\theta} =  0\) 
\(\partial_t \pi  = {K_{\theta}}^{\theta} = \mathcal{K} =  0\) 
for which \({K_{r}}^{r}\) and \(Q \equiv 0\). 
Therefore,  Eq.~\eqref{eq: IsoMomConstraint} is trivially satisfied and we only need to solve Eq.~\eqref{eq: IsoHamConstraint} for \(\psi\).

The initial profile for the \(k\)-essence field [the \emph{free data} in Eqs.~\eqref{eq: IsoHamConstraint}-\eqref{eq: IsoMomConstraint}] is specified as
\begin{align} \label{eq : IDKsscScalar}
    \partial_r \pi \rvert_{t= 0} &= A \exp \left[-\dfrac{\left(r-r_c\right)^2}{\sigma^2}\right] \cos\left(\dfrac{\pi}{10} r\right)~, \nonumber\\
    \partial_t \pi \rvert_{t= 0} &= 0~,
\end{align}
where \(A\) is the amplitude of the pulse, and \(r_c\) and \(\sigma\) are parameters specifying the location and root-mean-square  width of the Gaussian envelope of the pulse. Note that this form resembles the initial data used in Ref.~\cite{Bernard:2019fjb}.

We implement our initial data solver in \emph{Mathematica}~\cite{Mathematica}.
First, regularity at the origin is imposed by solving Eq.~\eqref{eq: IsoHamConstraint} perturbatively near the origin. The perturbative solution for \(\psi(r)\), which depends on one integration constant \(\psi(0)\), is then used as initial data in an outward-bound integration (in radius) starting from a small non-zero radius. Finally, using a shooting method, we fix \(\psi(0)\) by requiring that the exterior Robin boundary condition
\begin{align} \label{eq: ExteriorPsiBC}
 -1 + \psi + r \dfrac{\partial \psi}{\partial r} \Bigg\vert_{r\to \infty} = 0
\end{align}
is satisfied.
Note that
this boundary condition corresponds to imposing that \(\psi\) reduces to the asymptotically flat solution of Eq.~\eqref{eq: IsoHamConstraint}
(c.f. Birkhoff's Theorem \cite{Wald:1984rg}),
\(\psi \left(r \to \infty \right) \approx  1 + m_0 / \left(2 r \right)\), 
where \(m_0\) is the (unknown) Arnowitt-Deser-Misner (ADM)
mass.

%%%%%%%%%%%%%%%%%%%%%%%%%%%%%%%%%%%%%%%%%%%%%%%%%%%%%%%%%%%
%%%%%%%%%%%%%%%%%%%%%%%%%%%%%%%%%%%%%%%%%%%%%%%%%%%%%%%%%%%
%%%%%%%%%%%%%%%%%%%%%%%%%%%%%%%%%%%%%%%%%%%%%%%%%%%%%%%%%%%

\subsubsection{\(U(1)\) UV-completion}\label{sec: IDHiggs}

The construction of the initial data for the metric variables proceeds as in Sec.~\ref{sec: IDKessence}.
In this case, the \(P\) and \(Q\) terms in Eqs.~\eqref{eq: IsoHamConstraint}-\eqref{eq: IsoMomConstraint} are replaced by 
\begin{multline}
        P[\alpha, \psi, \partial_t \phi_{R,I}, \partial_r \phi_{R,I}, \phi_{R, I}] = -\dfrac{1}{8} \kappa \psi \big[\psi^4 \left(\Pi_{R}^2 + \Pi_{I}^2\right) + \vphantom{]} \\
        \vphantom{[}+ \Phi_{R}^2 + \Phi_{I}^2 + \psi^4 V\left(\phi^\star \phi\right)\big]~, \nonumber                
\end{multline}
\begin{align}
    Q[\alpha, \psi, \partial_t \phi_{R,I}, \partial_r \phi_{R,I}, \phi_{R, I} ] = -\dfrac{1}{2}\kappa \left(\Pi_{R} \Phi_{R} + \Pi_{I} \Phi_{I}\right)~.    
\end{align}

From the initial profile of the \(k\)-essence (phase) field [Eqs.~\eqref{eq : IDKsscScalar}], we can construct the initial configurations for the fields \(\phi_{R,I}\) by direct application of Eqs.~\eqref{eq: VEVexpansion} and \eqref{eq: rhoPertSol}.

Finally, let us comment on a subtlety regarding the initial profile for the complex scalar field. When specifying the initial configuration of the radial field \(\rho\) [Eq.~\eqref{eq: rhoPertSol}], one needs to provide also information on the configuration of the metric function \(\psi(r)\) in \(k\)-essence, which we denote by \(\psi^K(r)\). The latter is obtained from the solution of the Hamiltonian constraint [Eq.~\eqref{eq: IsoHamConstraint}]. This complicates the solution of Eq.~\eqref{eq: IsoHamConstraint} for the \(U(1)\) UV completion, since it would require the use of an interpolated function for \(\psi^K(r)\). For the cases that we consider below, \(\psi^K(r) \approx 1\). Thus, we avoid this problem by using the \textit{approximation} \(\psi^K(r) \equiv 1\) in Eq.~\eqref{eq: rhoPertSol}. We stress that the Hamiltonian constraint [Eq.~\eqref{eq: IsoHamConstraint}] in this UV completion \emph{should not} be solved by considering \(\psi^K(r) = \psi(r)\).

%%%%%%%%%%%%%%%%%%%%%%%%%%%%%%%%%%%%%%%%%%%%%%%%%%%%%%%%%%%
%%%%%%%%%%%%%%%%%%%%%%%%%%%%%%%%%%%%%%%%%%%%%%%%%%%%%%%%%%%
%%%%%%%%%%%%%%%%%%%%%%%%%%%%%%%%%%%%%%%%%%%%%%%%%%%%%%%%%%%

\subsubsection{Fixing the equations}\label{sec: IDFE}

The initial data in the ``fixing the equations" approach is prescribed in exactly the same way as in Sec.~\ref{sec: IDKessence}. The only additional information that we need to include is the initial profile of the \(\Sigma\) field, which we specify to be
\begin{align}
    \Sigma\rvert_{t= 0} & = K'(X) \lvert_{t = 0}~.
\end{align}

%%%%%%%%%%%%%%%%%%%%%%%%%%%%%%%%%%%%%%%%%%%%%%%%%%%%%%%%%%%
%%%%%%%%%%%%%%%%%%%%%%%%%%%%%%%%%%%%%%%%%%%%%%%%%%%%%%%%%%%
%%%%%%%%%%%%%%%%%%%%%%%%%%%%%%%%%%%%%%%%%%%%%%%%%%%%%%%%%%%

\subsection{Units} \label{sec: physical units}

For convenience in the numerical implementation, we will measure physical quantities with respect to the following energy, length and time units
\( E_{\Lambda} \equiv \Lambda^{-2} \kappa^{-3/2} \), \( 
  L_{\Lambda} \equiv  \Lambda^{-2} \kappa^{-1/2} \),  and \(
  T_{\Lambda} \equiv  L_\Lambda
\), 
respectively.

%%%%%%%%%%%%%%%%%%%%%%%%%%%%%%%%%%%%%%%%%%%%%%%%%%%%%%%%%%%
%%%%%%%%%%%%%%%%%%%%%%%%%%%%%%%%%%%%%%%%%%%%%%%%%%%%%%%%%%%
%%%%%%%%%%%%%%%%%%%%%%%%%%%%%%%%%%%%%%%%%%%%%%%%%%%%%%%%%%%

\subsection{Evolution scheme} \label{sec: evolution scheme}

For this paper, we extend the code of Ref.~\cite{Alic:2007ev},
which was initially written for one dimensional black hole simulations, but which
was later adapted 
in Refs.~\cite{Bernal:2009zy} to perform dynamical evolutions of boson stars, fermion-boson stars~\cite{Valdez_Alvarado_2013}, anisotropic stars~\cite{Raposo:2018rjn}, 
and also in Refs.~\cite{terHaar:2020xxb,Bezares:2020wkn,Bezares:2021yek} and in Ref.~\cite{PhysRevD.104.084017} for neutron stars in  \(k\)-essence and chameleon screening, respectively.
The metric equations are evolved using a high-resolution shock-capturing finite-difference (HRSC) scheme, described in Ref.~\cite{Alic:2007ev,Bona:2008xs}, to discretize the spacetime variables. 
This method can be interpreted as a fourth-order finite difference scheme plus third-order adaptive dissipation, where the dissipation coefficient is given by the maximum propagation speed at each grid point. 
For the scalar field sector, a more robust HRSC second-order method is employed, which is based on the Local-Lax-Friedrichs flux formula with a monotonic-centred limiter~\cite{CCC}. 
Integrations in time are carried out through the method of lines, by using a third-order accurate strong stability preserving Runge-Kutta integration scheme, with a Courant factor of \( \Delta t/ \Delta r = 0.25 \, T_\Lambda  / L_\Lambda \), such that the Courant-Friedrichs-Levy condition is satisfied.

We have used a spatial resolution of \(\Delta r = 0.01 \, L_\Lambda\)
and a spatial domain with outer boundary located at \(r = 480 \, L_\Lambda\).
We have checked that the results do not vary significantly with the position of the outer boundary or with resolution. 
For the spacetime variables, we use maximally dissipative boundary conditions, whereas
for the scalar fields we use outgoing  boundary conditions.

%%%%%%%%%%%%%%%%%%%%%%%%%%%%%%%%%%%%%%%%%%%%%%%%%%%%%%%%%%%
%%%%%%%%%%%%%%%%%%%%%%%%%%%%%%%%%%%%%%%%%%%%%%%%%%%%%%%%%%%
%%%%%%%%%%%%%%%%%%%%%%%%%%%%%%%%%%%%%%%%%%%%%%%%%%%%%%%%%%%

\subsection{Diagnostic quantities } \label{sec: DiagnosticQ}

In the \(U(1)\) UV completion, the
phase field
derivatives can be
computed
at each time step from
\begin{subequations} \label{eq: kssc reconstruction}
\begin{align}
    \partial_r \pi & = v \left(\dfrac{\phi_R \partial_r \phi_I - \phi_I \partial_r \phi_R}{\phi_R^2 + \phi_I^2}\right), \\
    \partial_t \pi & = v \left(\dfrac{\phi_R \partial_t \phi_I - \phi_I \partial_t \phi_R}{\phi_R^2 + \phi_I^2}\right). \label{eq: DtKssc}
\end{align}
\end{subequations}
The 
phase field  \(\pi(t, r)\) itself,
which at low energies is expected to reduce to the \(k\)-essence field,
can be obtained by integrating Eq.~\eqref{eq: DtKssc} along with the evolution equations. 
In the ``fixing the equations" approach, this procedure is instead not needed. 

Once the
phase field
and its derivatives are known, one can compute the
\(k\)-essence ``characteristic speeds" from Eq.~\eqref{eq: KsscCharacteristicSpeeds}. We emphasize, however, that the true characteristic speeds in the \(U(1)\) UV completion and in the ``fixed'' theory are given by Eqs.~\eqref{eq: HiggsCharacteristicSpeeds} and \eqref{eq: FECharacteristicSpeeds}, respectively. 

If an apparent horizon (defined as the outermost trapped surface) is present, its location \(r_\text{AH}\)
is given by the zeros of the expansion
of outgoing null rays [Eq.~(4.4) of Ref.~\cite{Bezares:2020wkn}]~\footnote{We correct here a typo in Eq.~(4.4) in Ref.~\cite{Bezares:2020wkn}.
}
\begin{align}
    \Theta = \dfrac{1}{\sqrt{g_{rr}}} \left(2 {D_{r\theta}}^{ \theta} + \dfrac{2}{r}\right) - 2 {K_{\theta}}^{\theta} ~.
\end{align}
Similarly, for the \(k\)-essence field, we find the location of a sound horizon (if present) by looking for the zeros of the expansion of outgoing null rays with respect to the effective metric \(\gamma^{\mu\nu}\) [Eq.~(4.5) of Ref.~\cite{Bezares:2020wkn}],
\begin{multline} \label{eq: sound horizon expansion}
    \mathcal{S} = r^2 g_{\theta \theta} \bigg[ (r {D_{r \theta}}^{\theta} + 2)^2 \gamma^{rr} + \vphantom{]}
    \\\vphantom{[}+ r {K_{\theta}}^{\theta} \alpha \left(r \gamma^{tt} {K_{\theta}}^{\theta} \alpha -2 \left(r {D_{r \theta}}^{\theta} + 2 \right)\gamma^{tr}\right) \bigg] ~.
\end{multline}

Finally, we compare the evolutions in two theories \(A\) and \(B\) by calculating a \emph{discrepancy} 
measure for a given field $\chi$ as
\begin{align} \label{eq: definition of discrepancy faithfulness}
    \mathbb{E}_{A, B}\left[\chi\right](t) = \dfrac{\left \| \chi^{(A)} - \chi^{(B)}  \right \|_\text{AH}}{\left \|  \chi^{(B)} \right \|_\text{AH} }~,
\end{align}
where the \(L_2\)-norm of a function \(\xi\) is defined as
\begin{align}
    \left \| \xi \right \|_\text{AH}^2 = \int^{\infty}_{\max_{A,B}(r_\text{AH})} \left\lvert  \xi(t, r) \right\rvert^2  \, dr~,
\end{align}
with integration domain only covering the 
exterior of the apparent horizons 
\(r_\text{AH}\) of \textit{both} theories $A$ and $B$. This measure is inspired in a similar measure introduced for Minkowski space in Ref.~\cite{Allwright:2018rut}.

%%%%%%%%%%%%%%%%%%%%%%%%%%%%%%%%%%%%%%%%%%%%%%%%%%%%%%%%%%%
%%%%%%%%%%%%%%%%%%%%%%%%%%%%%%%%%%%%%%%%%%%%%%%%%%%%%%%%%%%
%%%%%%%%%%%%%%%%%%%%%%%%%%%%%%%%%%%%%%%%%%%%%%%%%%%%%%%%%%%

\section{Results} \label{sec: results}

In this section, we will compare the dynamics of quadratic \(k\)-essence, the \(U(1)\) UV completion and the ``fixed" theory during the gravitational collapse of a scalar ``shell". 
We will first study in Sec.~\ref{sec: IRevolution} the initial stage of gravitational collapse, when the Cauchy problem in \(k\)-essence is well-posed, and confirm that the \(U(1)\) UV and the ``fixed" theory reproduce the same dynamics of quadratic \(k\)-essence. 
After a Tricomi-type breakdown of \(k\)-essence, we will continue the evolution with the \(U(1)\) UV completion and the ``fixed" theory to determine in Sec.~\ref{sec: Endstate} that the endstate of the system corresponds to that of a black hole. 
Finally, in Sec.~\ref{sec: Nonlinear regime}, we will show that the system enters the nonlinear regime and compare the dynamics 
of  the \(U(1)\) UV completion and the ``fixed" theory
within it. %
This will serve as a ``validation" test of the ``fixing the equations" approach in a setting where we have access to the UV physics, and it will also allow us to
explore the relation between the nonlinear regime and the range of validity of the EFT.
Additional comments for the case of a large coupling constant \(\beta\) are given in Sec.~\ref{sec: Miscelanea}.
In the following, we will explore the case corresponding to initial data [Eq.~\eqref{eq : IDKsscScalar}] generated with parameters \(r_c = 55 \, L_\Lambda\), \(\sigma = 1.5 \, L_\Lambda\), and \(A = 0.14 \, L_\Lambda\) and coupling constants \(\beta = 1\), \(M^2_\rho = 2  \, L^{-2}_{\Lambda}\), \(v^2 \, E^{-1}_{\Lambda} L_{\Lambda} = 1 \) and \(\tau \, T^{-1}_\Lambda = 1 \).
In Appendix~\ref{sec: weak data example}, we present an additional example with weak initial data, where no breakdown of the Cauchy problem or black hole formation occurs.

%%%%%%%%%%%%%%%%%%%%%%%%%%%%%%%%%%%%%%%%%%%%%%%%%%%%%%%%%%%
%%%%%%%%%%%%%%%%%%%%%%%%%%%%%%%%%%%%%%%%%%%%%%%%%%%%%%%%%%%
%%%%%%%%%%%%%%%%%%%%%%%%%%%%%%%%%%%%%%%%%%%%%%%%%%%%%%%%%%%

\subsection{EFT evolution and Tricomi-type breakdown} \label{sec: IRevolution}

By construction, the initial radial profile of the \(k\)-essence field \(\pi \lvert_{t = 0}\) agrees with the profiles from the \(U(1)\) UV completion phase field \(\pi^\text{(UV)}\lvert_{t = 0}\) and from the \(\pi\)-field of the ``fixed" theory. [Recall that in the \(U(1)\) UV completion, the \(k\)-essence field is described at low energies by the (dimensionful) phase mode [Eqs.~\eqref{eq: VEVexpansion}] and \eqref{eq: UV phase field} of the complex scalar \(\phi\) and needs to be computed from Eqs.~\eqref{eq: kssc reconstruction}.]
The initial data for the metric variables, obtained after solving the constraint equations, is also in agreement. 
In particular, for the \(U(1)\) UV completion, this is not a trivial statement, as the agreement in the metric occurs because the extra degree of freedom (the radial mode of the complex scalar [Eq.~\eqref{eq: VEVexpansion}]) contains a negligible fraction of the scalar energy content.
Thus, we can say that the initial data is in the regime of validity of the EFT description of \(k\)-essence.

\begin{figure}[]
 \centering
\includegraphics[width=3.4 in]{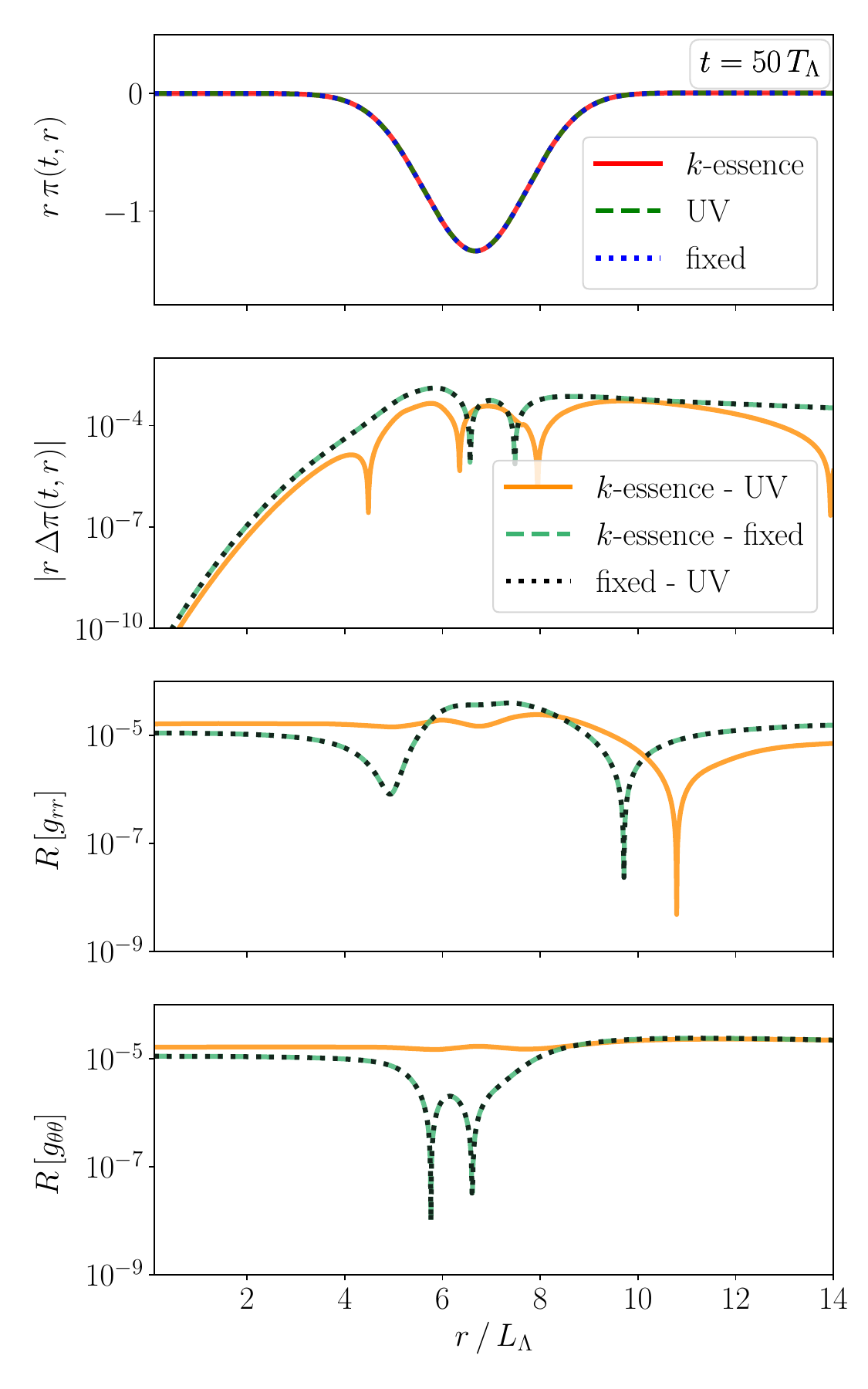}
\caption{\emph{EFT evolution.} First panel: The radial profile of the \(k\)-essence field multiplied by \(r\), at time \(t = 50 \, T_{\Lambda}\) (red solid line) compared with the
phase field of the \(U(1)\) UV completion (green dashed line) and the \(\pi\)-scalar of ``fixed" theory (blue dotted line), 
showing that they are indistinguishable from each other. Second panel: Absolute differences \( \Delta \pi  \equiv \pi^{(A)} - \pi^{(B)} \) for theories  \(A\) vs.~\(B\); namely, \(k\)-essence vs.~UV (orange solid lines), \(k\)-essence vs.~``fixed" (light green dashed lines) and ``fixed" vs.~UV (black dotted lines). Third and fourth panels: relative differences, \(R\left[g\right] \equiv \left \lvert\left( g^{(A)} - g^{(B)} \right) /g^{(B)} \right \rvert \), of the metric functions \(g = g_{rr}, g_{\theta \theta}\) for theories \(A\) vs.~\(B\).}
\label{fig: Case B: scalar profile}
\end{figure}

In the early stage of collapse in \(k\)-essence, from \(t  = 0\) to \(t \sim 55 \, T_{\Lambda}\), the scalar pulse splits into an in-going (collapsing) pulse travelling towards the origin, and  into an out-going (radiated) pulse moving towards the outer boundary of the numerical grid.  In the following, we will concentrate on the former. This stage is reproduced by the \(U(1)\) UV completion and the “fixed” theory.
In Fig.~\ref{fig: Case B: scalar profile}, in the first panel, we can observe that the \(k\)-essence scalar field at \(t \sim 50 \, T_{\Lambda}\) and \(r \sim 7\ \, L_\Lambda\) is almost indistinguishable in the \(U(1)\) UV completion and in the ``fixed" theory. 
We quantify this agreement by plotting the absolute difference of these profiles in the second panel. In the third and fourth panels, we also plot the relative difference of  \(g_{rr}\) and \(g_{\theta \theta}\), respectively, showing that the metric is also very well recovered, with a relative error of \(\lesssim 0.01 \% \).

\begin{figure}[]
 \centering
\includegraphics[width=3.4 in]{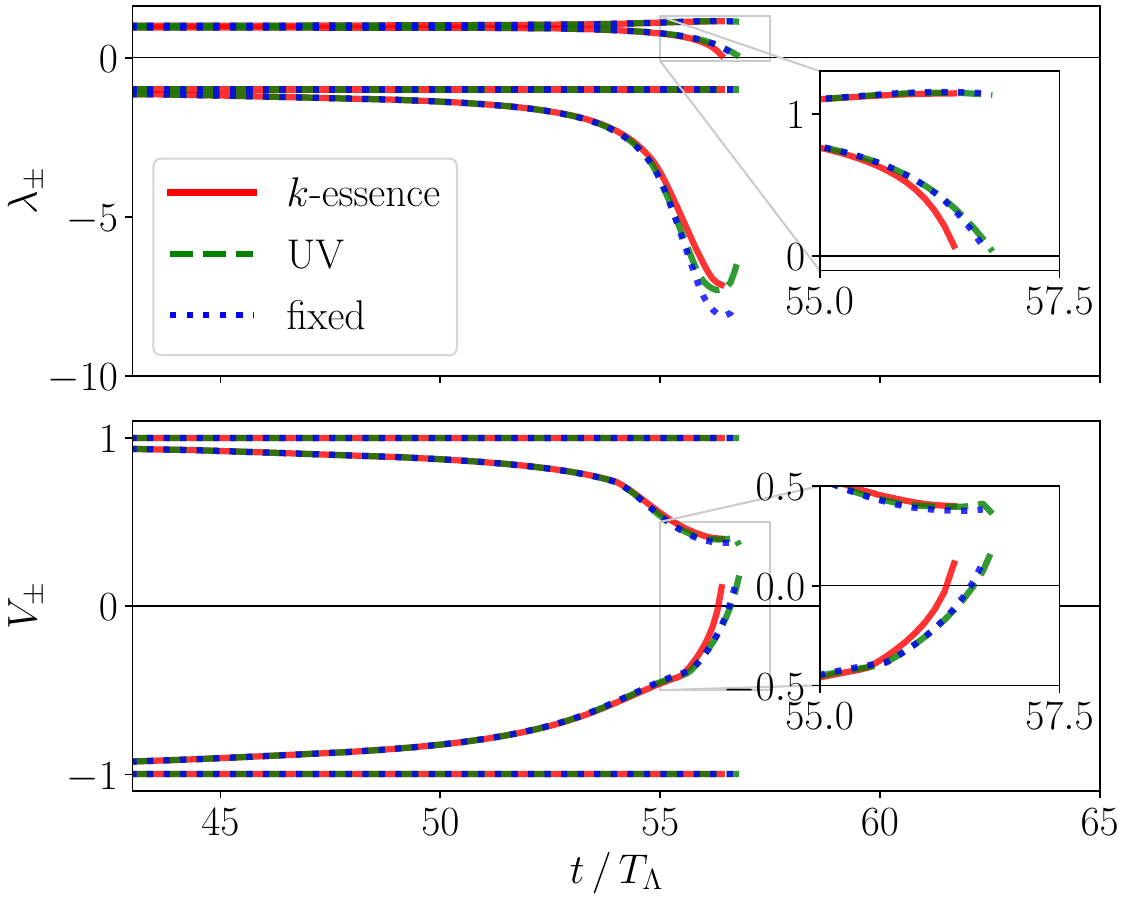}
\caption{ \emph{Character of the \(k\)-essence scalar equation. }
First panel: the minimum and maximum values of the  eigenvalues of the effective metric in \(k\)-essence (red solid line), \(U(1)\) UV completion (green dashed line) and the ``fixed" theory (blue dotted line). 
For the last two, the effective metric is not a fundamental quantity but  ``emergent" at low energies. 
From top to bottom: \(\max\left(\lambda_{+}\right)\), \(\min\left(\lambda_{+}\right)\), \(\max\left(\lambda_{-}\right)\), \(\min\left(\lambda_{-}\right)\). The Tricomi-type breakdown is signaled by \(\min \lambda_{+} \to 0\)  at \(t \sim 56.5 \, T_{\Lambda} \).
Second panel: the minimum and maximum values of the  characteristic speeds. In this panel, from top to bottom: \(\max\left(V_{+}\right)\), \(\min\left(V_{+}\right)\), \(\max\left(V_{-}\right)\), \(\min\left(V_{-}\right)\). Notice that, at \(t \sim 56.5 \, T_{\Lambda} \), \(\left \lvert\min \left(V_{+}\right) - \max \left(V_{-}\right) \right \rvert \to 0\), signaling a Tricomi-type breakdown.}
\label{fig: Case B: velocities}
\end{figure}

As the pulse approaches the origin, the \(k\)-essence scalar gradients increase. At \(t \sim 56.5 \, T_{\Lambda}\), large gradients trigger a Tricomi-type breakdown, by which the scalar equation~\eqref{eq : KsscScalarEq} transitions from hyperbolic to
parabolic and then elliptic. From the discussion in Sec.~\ref{sec: Quadratic Kessence}, this occurs when, at any point of the spatial grid, the determinant of the effective metric~\eqref{eq: effective metric} vanishes, or equivalently, when at least one eigenvalue of the latter becomes zero.
We can gain some insight by tracking the spatial maximum and minimum of the eigenvalues of the effective metric [Eq.~\eqref{eq: Kessence effective metric eigenvalues}] as a function of time, as can be seen in the first panel of Fig.~\ref{fig: Case B: velocities}.
Note that for the \(U(1)\) UV completion and the ``fixed" theory, 
the effective metric is not a fundamental but an ``emergent" quantity,
therefore, these eigenvalues have been
computed
from Eqs.~\eqref{eq: Kessence effective metric eigenvalues} and \eqref{eq: kssc reconstruction}. 
Initially, \(\lambda_{\pm} \approx \pm 1\).
As the evolution progresses, the Tricomi-type breakdown is signaled in this plot by one of the eigenvalues approaching zero. 
Specifically, we observe that \(\min(\lambda_{+}) \to 0\).
In the second panel, we plot the spatial maximum and minimum values of the characteristic speeds of \(k\)-essence [Eq.~\eqref{eq: KsscCharacteristicSpeeds}] as a function of time. 
In the early evolution, the system is clearly strongly hyperbolic since \(V_{\pm}\) are real and distinct.
As the pulse approaches the origin, first, we observe the formation of a sound horizon (roughly when \(\max \left(V_{-}\right) \approx 0\)\footnote{As mentioned earlier,
we define the location of the (apparent) sound horizon as the zero of the 
effective metric's null ray expansion \eqref{eq: sound horizon expansion}. In areal coordinates, that condition is exactly equivalent to $V_{-}=0$, and 
this equivalence carries on (albeit approximately) also in the isotropic coordinates that we utilize.}). Then, the Tricomi-type breakdown 
occurs when the characteristic speeds become equal. Indeed, we observe that \( \left\lvert \min\left(V_{+}\right) - \max\left(V_{-}\right) \right \rvert \to 0\), indicating that strong hyperbolicity is lost\footnote{This is actually a necessary and not sufficient condition for the loss of hyperbolicity, but we have checked that the effective metric also becomes degenerate when \( V_{+}=V_{-} \).}. 
Note that, as before, for the \(U(1)\) UV completion and the ``fixed" theory, the values of \(V_{\pm}\)  have been computed using Eqs.~\eqref{eq: KsscCharacteristicSpeeds} and \eqref{eq: kssc reconstruction}.

We argue that the change of character of the scalar equation~\eqref{eq : KsscScalarEq} occurs within the EFT regime since all three theories predict that the effective metric becomes degenerate (corresponding to a Tricomi transition in the low energy $k$-essence theory) at similar times.
Past 
this point,
only with the \(U(1)\) UV completion and the ``fixed" theory, for which the Cauchy problem remains well-posed, can 
the scalar and metric be evolved smoothly and the final fate of the system be predicted.

%%%%%%%%%%%%%%%%%%%%%%%%%%%%%%%%%%%%%%%%%%%%%%%%%%%%%%%%%%%
%%%%%%%%%%%%%%%%%%%%%%%%%%%%%%%%%%%%%%%%%%%%%%%%%%%%%%%%%%%
%%%%%%%%%%%%%%%%%%%%%%%%%%%%%%%%%%%%%%%%%%%%%%%%%%%%%%%%%%%

\subsection{Endstate}\label{sec: Endstate}

\begin{figure}[]
 \centering
\includegraphics[width=3.4 in]{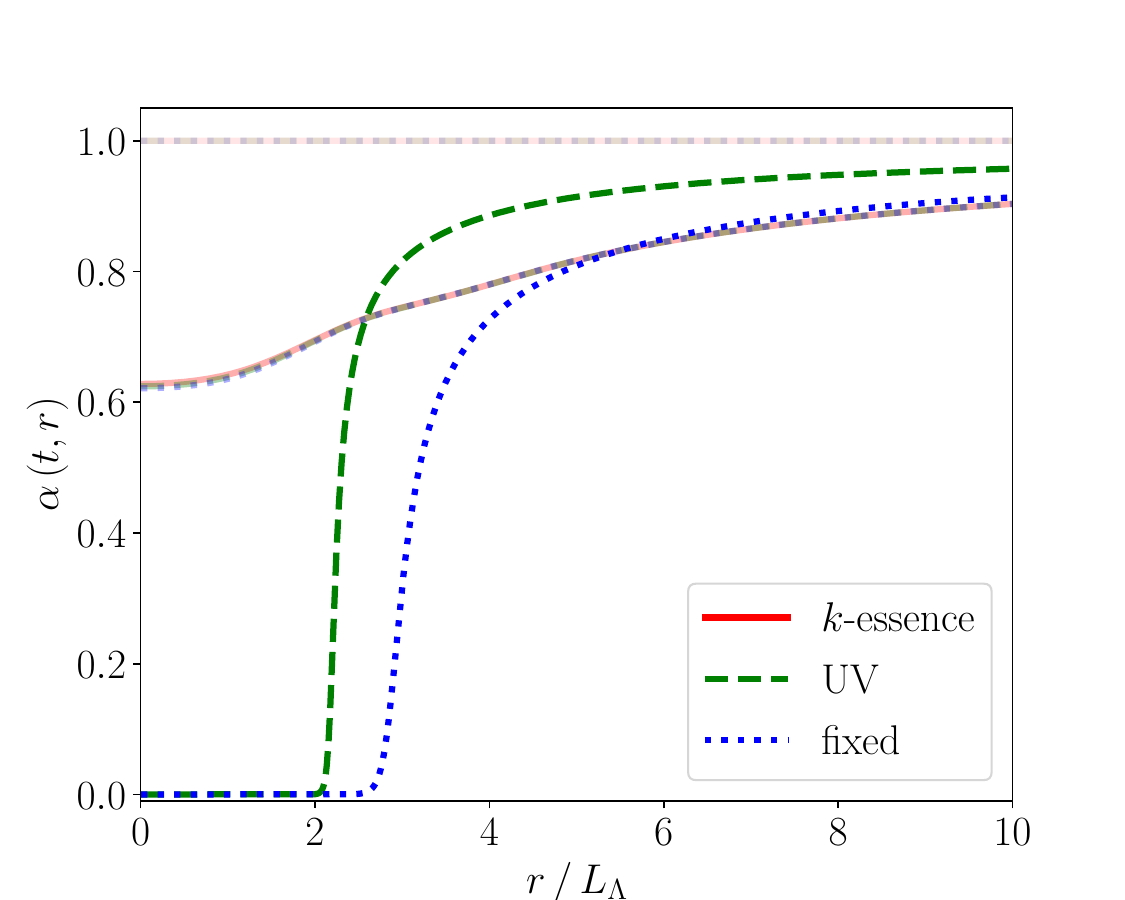}
\caption{\emph{Gravitational collapse of the pulse.} 
Radial profiles of the lapse \(\alpha\) at times \(t / T_{\Lambda} = 30 ,\, 55 , \, 80 \) in \(k\)-essence (red solid lines), \(U(1)\) UV completion (green dashed lines) and the ``fixed" theory (blue dotted lines). Increasing times are denoted by increasing intensity of the color. The lapse approaching zero near the origin is a typical effect signaling the formation of a black hole. Note that \(k\)-essence experiences a Tricomi breakdown at \(t  \sim  55.6 \, T_\Lambda\), much before any apparent horizon formation.
}
\label{fig: Case B: lapse}
\end{figure}

In both the \(U(1)\) UV completion and the ``fixed" theory, the system collapses to form a black hole. 
In Fig.~\ref{fig: Case B: lapse}, we show the lapse function approaching zero near the origin at different representative times, a typical behavior leading up to the formation of a black hole \cite{Alcubierre:1138167}.
We confirm this conclusion by identifying the appearance of an apparent horizon, which we indicate with solid vertical lines in Fig.~\ref{fig: Case B: nonlinear scalar profile}. 
For the ``fixed" theory, we have checked that the final state is a black hole when varying \(\tau / T_{\Lambda} \in \left[1,10\right] \).
This endstate remains inaccessible with the
low-energy
\(k\)-essence model, where the Tricomi-type breakdown occurs well before the lapse gets close to zero. 

With our numerical implementation (Sec.~\ref{sec: evolution scheme}), we can only track the evolution of the black hole horizon for some time after formation. This is due to the formation of steep gradients in the collapse front of the lapse \cite{Alic:2007ev}. 
Finally, in the \(U(1)\) UV completion, the final area is \(A_\text{BH} = 4 \pi R^2_\text{AH} \sim 5.2 \, L^2_{\Lambda}\), where \(R_\text{AH} \sim 0.64 \, L_{\Lambda}\) is the polar radius of the apparent horizon. In the ``fixed" theory, the initial value of the black hole area is \(A_\text{BH} \sim 7.1 \, L^2_{\Lambda}\) (\(37 \%\) larger) with \(R_\text{AH} \sim 0.75 \, L_{\Lambda}\) (\(17 \%\) larger).
However, we cannot accurately determine the final value of the area due to additional constraint violation contributions with respect to the \(U(1)\) UV completion: in the ``fixed" theory the stress-energy tensor is only strictly conserved in the limit \(\Sigma \to K'\left(X\right)\), thus, the Hamiltonian constraint time derivative is only strictly vanishing in the limit \(\Sigma \to K'\left(X\right)\). We will elaborate on this point in Appendix~\ref{sec: Constraint propagation}.

%%%%%%%%%%%%%%%%%%%%%%%%%%%%%%%%%%%%%%%%%%%%%%%%%%%%%%%%%%%
%%%%%%%%%%%%%%%%%%%%%%%%%%%%%%%%%%%%%%%%%%%%%%%%%%%%%%%%%%%
%%%%%%%%%%%%%%%%%%%%%%%%%%%%%%%%%%%%%%%%%%%%%%%%%%%%%%%%%%%

\subsection{Nonlinear vs.~UV regime } \label{sec: Nonlinear regime}

Having confirmed that the \(k\)-essence dynamics is recovered at  early times (Sec.~\ref{sec: IRevolution}) and that the evolution can be continued past the Tricomi transition to determine the final fate of the system (Sec.~\ref{sec: Endstate}), we will now proceed to compare the \(U(1)\) UV completion and the ``fixed" theory in the nonlinear regime. This will allow us to explore the relation between the latter and the range of validity of the EFT (defined by the difference between the 
\(U(1)\) UV completion and quadratic \(k\)-essence evolved within the ``fixing the equations'' approach).

To establish whether the dynamics enters the nonlinear regime, we monitor the ratio of the first \(k\)-essence self-interaction operator to the kinetic term, i.e. \(\mathrm{NL}\equiv\left | \beta \Lambda^{-4} X  \right | =  \left | 2 M^{-2}_{\rho} v^{-2} X\right |\).
As can be seen, this
 can be rewritten, using Eqs.~\eqref{eq: actionKessence} and \eqref{eq: rhoPertSol}, as
simply \(\mathrm{NL}=2 \left\lvert \rho \right\rvert\). One therefore expects the nonlinear regime [i.e. \(\mathrm{NL} \sim \mathcal{O}\left(1\right)\)] to be
closely related, if not equivalent, to the range of validity of the low energy theory,
to which the \(U(1)\) UV completion only reduces when $\rho$
becomes non-dynamical
and can be integrated out (thus implying that  \(\mathrm{NL}=2 \left\lvert \rho \right\rvert\) is small). 
We will verify this conjecture with our numerical simulations in the following.

\begin{figure}[]
 \centering
\includegraphics[width=3.4 in]{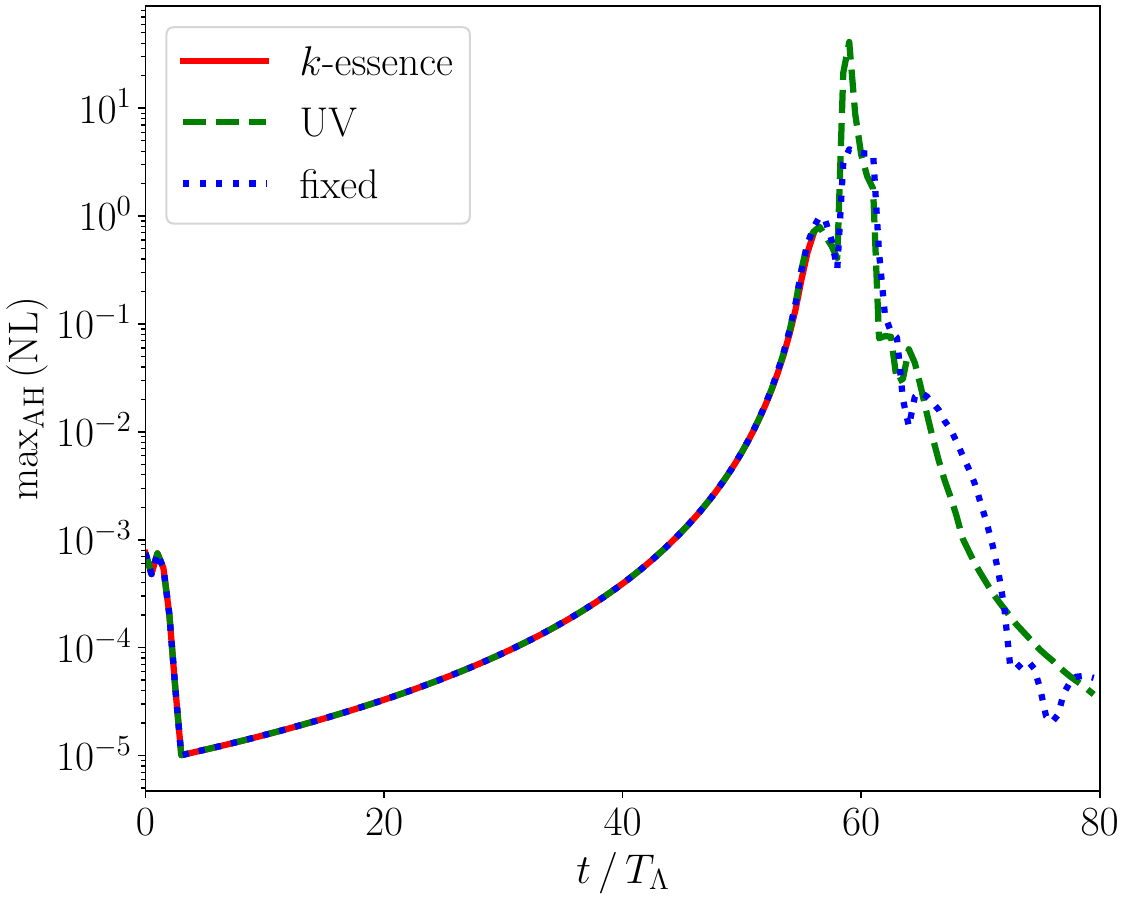}
\caption{\emph{Nonlinear regime assessment.} 
The spatial maximum of the ratio of the self-interaction term to the kinetic term for \(k\)-essence (red solid line), the \(U(1)\) UV completion (green dashed line) and the ``fixed" theory (blue dotted line). The maximum is taken in the region outside the apparent horizon, if present. During the early evolution, this measure is small (\(\lesssim 10^{-2}\)).
As the pulse approaches the origin, the system enters the nonlinear regime \(\max_\text{AH} \left(\mathrm{NL}\right) \sim \mathcal{O}(1)\), shortly after the Tricomi transition at \( t \sim  56.5 \, T_{\Lambda} \). This measure decreases in the later stage once the black hole is formed and nonlinearities are hidden behind the apparent horizon. 
}
\label{fig: Case B: max_XX}
\end{figure}

Let us first analyze when the nonlinear regime is attained. In Fig.~\ref{fig: Case B: max_XX}, we plot the spatial maximum of \(\mathrm{NL}\) as a function of time in the region outside the apparent horizon (if present). We denote this quantity as \(\max_\text{AH} \left(\mathrm{NL}\right)\). % at time \(t\).
During the early evolution, this ratio is small, signaling that the dynamics is linear. However, as the pulse approaches the origin and scalar gradients grow, both the \(U(1)\) UV completion and the ``fixed" theory enter the nonlinear regime. 
In particular, for the ``fixed" theory, the growth of the gradients within the nonlinear regime is damped in comparison to the \(U(1)\) UV completion, and we observe a milder growth in  \(\max_\text{AH} \left(\mathrm{NL}\right)\).  Recall that in the ``fixed" theory, high frequency modes are suppressed by construction.
Finally, once the black hole forms, the nonlinear regions become hidden behind the apparent horizon, and we observe a decrease in  \(\max_\text{AH} \left(\mathrm{NL}\right)\).

\begin{figure}[]
 \centering
\includegraphics[width=3.4 in]{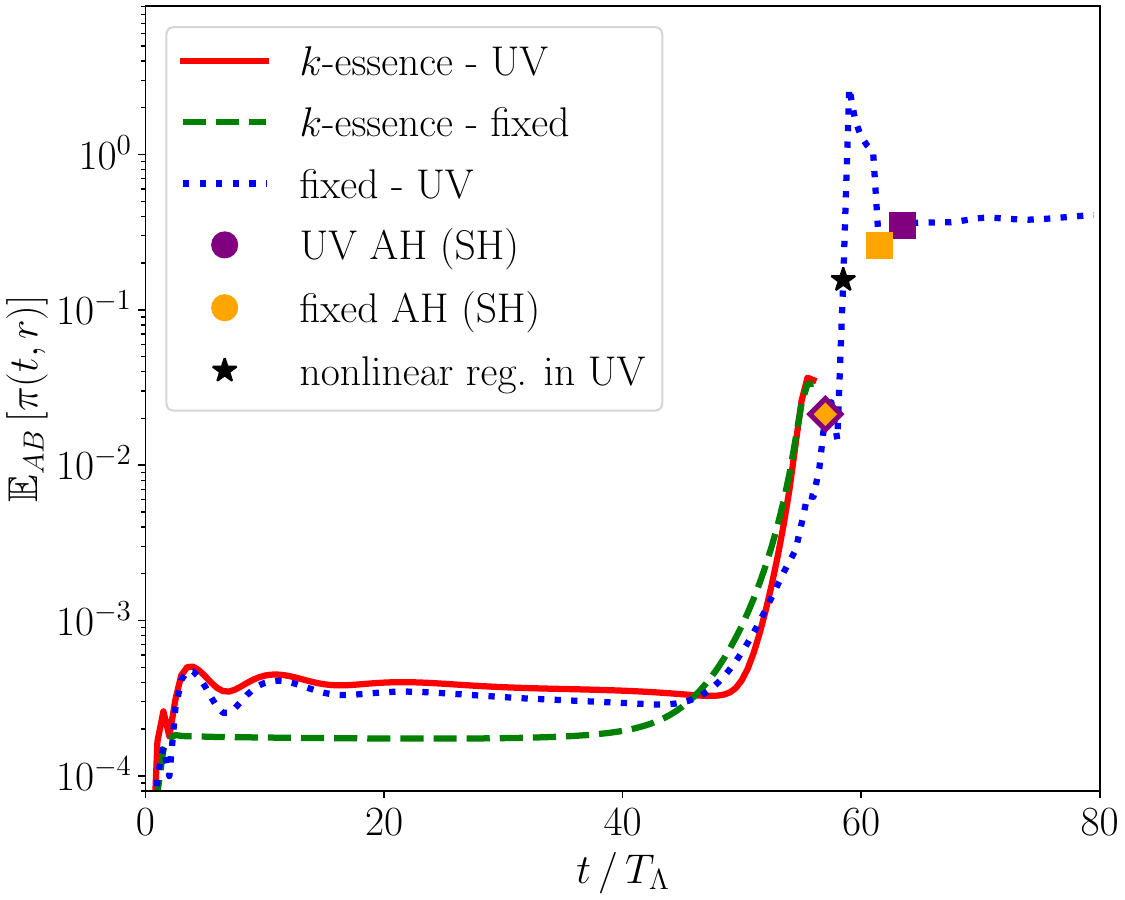}
\caption{\emph{Discrepancy of the \(k\)-essence scalar.} 
The discrepancy measure \(\mathbb{E}_{AB}\) of the \(k\)-essence scalar \(\pi\) of theories \(A\) vs.~\(B\), defined in Eq.~\eqref{eq: definition of discrepancy faithfulness}; namely, \(k\)-essence vs.~\(U(1)\) UV completion (red solid line), \(k\)-essence vs.~``fixed" theory (green dashed line), and ``fixed" theory vs.~\(U(1)\) UV completion (blue dot-dashed line). The discrepancy measures involving \(k\)-essence stop at the Cauchy breakdown of the theory. The colored diamonds and square markers, denote the (approximate) time of formation of the sound horizon (SH) and apparent horizon (AH) in each theory, respectively. Note that, the diamonds are superposed since both theories agree in the EFT regime. The black star marker denotes the approximate time where \(\max_\text{AH} \left(\mathrm{NL}\right) \approx 1\) in the \(U(1)\) UV theory --see also Fig.~\ref{fig: Case B: max_XX}. 
}
\label{fig: Case B: faithfulness pi}
\end{figure}

We now proceed to compare the 
\(k\)-essence (phase) field profiles
in the nonlinear regime. 
In Fig.~\ref{fig: Case B: faithfulness pi}, we plot the \emph{discrepancy} measure \(\mathbb{E}_{AB}\left[\pi\right]\),
between the \(k\)-essence scalar profiles of theories \(A\) and \(B\), as defined in Eq.~\eqref{eq: definition of discrepancy faithfulness}.
(Note that the plot for the discrepancy of the kinetic energy $X$ would look qualitatively similar and lead to the same conclusions.)
We  denote in colored diamonds the approximate time of formation of sound horizons, and in colored squares the approximate time of formation of apparent horizons in each theory.
We focus on the discrepancy between the ``fixed" theory (theory \(A\)) and the \(U(1)\) UV completion (theory \(B\)), plotted in blue dotted lines. 
This provides a measure of
how much the EFT and UV dynamics differ, i.e. it allows
for understanding the range of validity of the EFT.
During the early evolution the agreement is clear (\(\mathbb{E}_{AB}\left[\pi\right] < 10^{-3}\)), i.e.
the EFT is a good description of the full dynamics.
As the system enters the nonlinear regime, indicated with a black star, the discrepancy increases to \(\mathcal{O}\left(1\right)\), 
which would seem to indicate that the dynamics exits the
range of validity of the (fixed) low energy EFT.
However, the comparison of the scalar profiles is subtle and we should examine them in more detail.

\begin{figure*}[]
 \centering
\includegraphics[width=7 in]{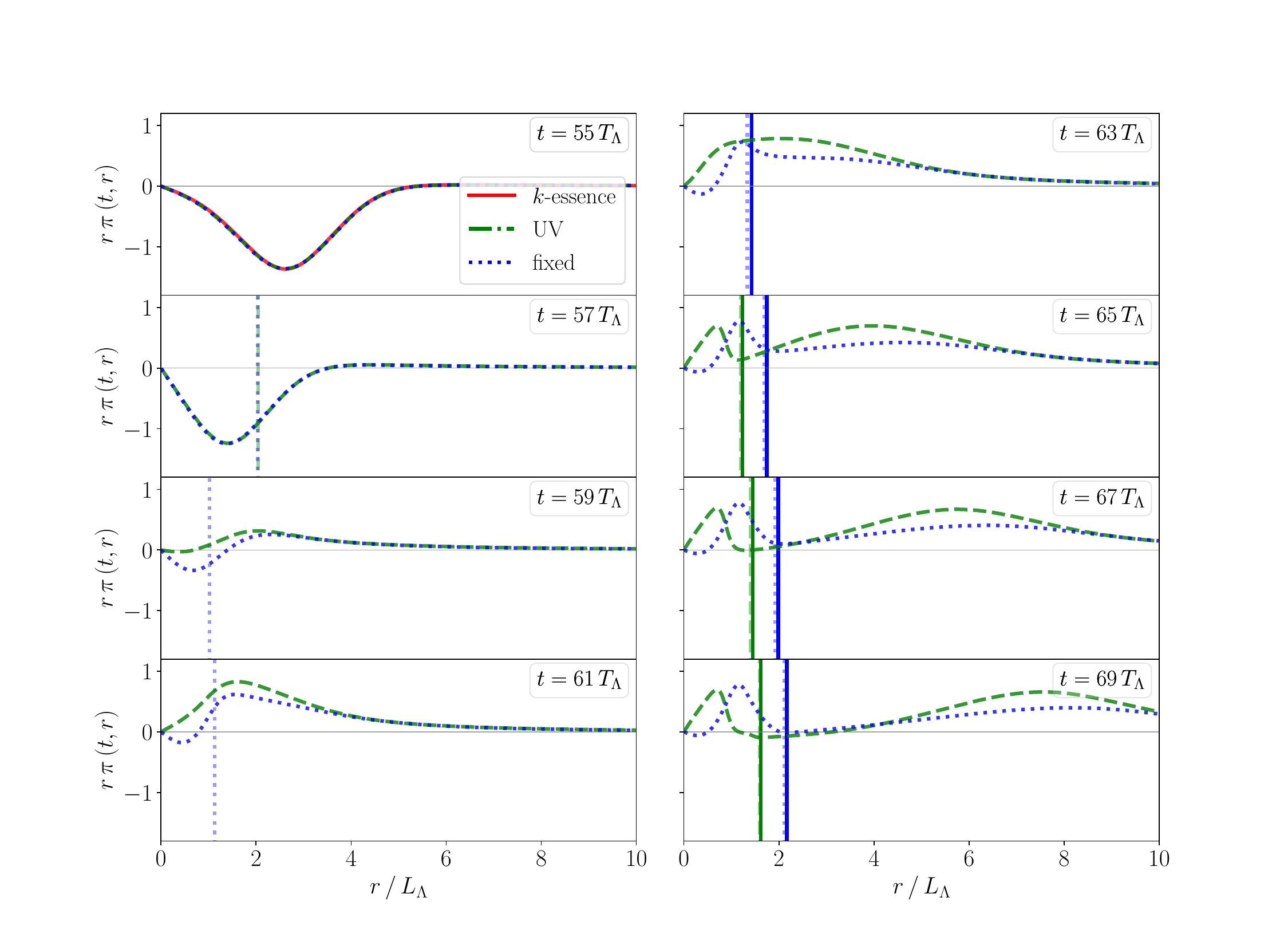}
\caption{\emph{Dynamics of the pulse in the nonlinear regime.} 
Time snapshots of the  \(k\)-essence scalar for representative times from \(t = 55 \, T_{\Lambda}\) to \(t  = 69 \, T_\Lambda\) for \(k\)-essence (red solid lines), 
the phase field of the \(U(1)\) UV completion (green dashed lines) and the \(\pi\)-scalar of the ``fixed" theory (blue dotted lines).
The profiles corresponding to the quadratic model of \(k\)-essence exist only up to the Tricomi-type breakdown of the theory (at \(t \sim 56.5 \, T_\Lambda\)), and hence, they are only shown in the first panel. The ``fixed" theory exhibits a qualitatively similar behavior to that of the \(U(1)\) UV completion. The solid vertical lines indicate the location of the apparent horizon, while the dashed and dotted vertical lines denote the location of the 
low-energy
sound horizon. The appearance and disappearance of the sound horizon between frames \(t = 57 \, T_\Lambda\) and \(t = 65 \, T_\Lambda\) occurs due to the theory entering the nonlinear/UV regime. The singularity-avoidance prescription chosen for the lapse causes the ``freezing" of the scalar profile near the origin, once the black hole forms.
}
\label{fig: Case B: nonlinear scalar profile}
\end{figure*}

In Fig.~\ref{fig: Case B: nonlinear scalar profile}, in the left panels, we show snapshots of the scalar profiles of the 
\(k\)-essence (phase) field
close to when the nonlinear regime is reached, as well as at later times. 
In the top left panel, at \(t =  55 \, T_\Lambda\), and right before the Tricomi transition, we observe that the scalar field is indistinguishable in \(k\)-essence, in the \(U(1)\) UV theory and in the ``fixed" theory. 
In the following panels, only the profiles of the last two theories are shown, as \(k\)-essence undergoes a Cauchy (Tricomi) breakdown, as mentioned earlier.
We notice that the scalar profile of the ``fixed" theory exhibits a qualitatively similar behavior of that of the \(U(1)\) UV theory.
From this figure, the \(\mathcal{O}(1)\) discrepancies in Fig.~\ref{fig: Case B: faithfulness pi} are then seen to originate mostly as a consequence of a ``lag" between the scalar profiles.
Once the black hole forms, the largest sources of discrepancy are hidden behind the apparent horizon, as can be seen for times \(t \gtrsim 61 \, T_{\Lambda}\) in Fig.~\ref{fig: Case B: faithfulness pi}. 
From the right panels of Fig.~\ref{fig: Case B: nonlinear scalar profile}, we notice that the ``fixed" theory also qualitatively follows the radiated (outgoing) scalar field of the \(U(1)\) UV completion. Note that the observed difference in amplitude is small but is magnified  by the factor \(r\). In Fig.~\ref{fig: Case B: faithfulness pi} it can be seen that the discrepancy is approximately \( \mathcal{O} \left(10^{-1}\right) \).

The observed ``lag" in Fig.~\ref{fig: Case B: nonlinear scalar profile} can be traced, at least partly, to the form of the ``driver" equation [Eq.~\eqref{eq: FESigmaEquation}] and its associated timescale \(\tau\), which controls how fast the field \(\Sigma\) relaxes to \(K'\left(X\right)\).
By decreasing (increasing) the value of \(\tau\), we can partly reduce (increase) the ``lag" in scalar profiles.
Other sources of ``delay" may be due to the slightly different evolution of the lapse in the two theories -- see Fig.~\ref{fig: Case B: lapse}. 
The latter observation illustrates that one must be careful when comparing fixed time scalar profiles from different evolutions.
To overcome these ambiguities, better measures of comparison may be defined from observables such as the scalar radiation detected by an asymptotic observer --see e.g. Ref.~\cite{Bezares:2021dma}.

Finally, we briefly comment on the low-energy sound horizons, which form prior to the formation of the black hole. 
Since physical modes in the \(U(1)\) UV completion
move along null geodesics [c.f. Eq.~\eqref{eq: HiggsScalarEquation}] and are no longer (at least in principle)
well described by the \(k\)-essence scalar equation \eqref{eq : KsscScalarEq}, the  sound horizons may lose physical meaning. 
This causes a strange behavior of the sound horizon in our simulations, as illustrated e.g. in Fig.~\ref{fig: Case B: nonlinear scalar profile}. At \(t = 57 \, T_{\Lambda}\), the sound horizon has already formed in the \(U(1)\) UV theory, and is marked by a green dashed vertical line. This horizon disappears shortly after and is not shown in subsequent frames. At \(t = 65 \, T_\Lambda\), the sound horizon instead reappears.
Again, we stress that this is probably due to the sound horizon losing physical meaning in the UV regime.

%%%%%%%%%%%%%%%%%%%%%%%%%%%%%%%%%%%%%%%%%%%%%%%%%%%%%%%%%%%
%%%%%%%%%%%%%%%%%%%%%%%%%%%%%%%%%%%%%%%%%%%%%%%%%%%%%%%%%%%
%%%%%%%%%%%%%%%%%%%%%%%%%%%%%%%%%%%%%%%%%%%%%%%%%%%%%%%%%%%

\subsection{Large coupling} \label{sec: Miscelanea}

In astrophysical settings, where 
masses and lengths are respectively of order $M_\odot$ and 
km (or larger), one typically has to employ units adapted to the system under scrutiny to simulate it, e.g. ones in which
$G=c=M_\odot=1$. In these units, 
the numerical value of the coupling constant \(\beta \Lambda^{-4} \) is extremely large~\cite{terHaar:2020xxb,Bezares:2021yek,Bezares:2021dma}. 
This coefficient is intimately connected to the scales \(M_\rho\) and \(v\) in the \(U(1)\) UV completion by Eq.~\eqref{eq: coupling constant matching}. Fixing \(v^2 =  E_{\Lambda} / L_{\Lambda}\) to avoid short wavelength oscillations in the complex scalar
\(\phi \propto \exp \left[i \pi^\text{(UV)} / v\right]\) [Eqs.~\eqref{eq: VEVexpansion} and \eqref{eq: UV phase field}],
larger values of \(\beta \Lambda^{-4}\) correspond to smaller values of \(M_\rho\). This, in turn, means a weaker suppression of higher-order terms, suppressed by powers of \(M_{\rho}^{-2} v^{-2}\). Therefore, for fixed  initial data parameters \(\{A, \sigma, r_c\}\) [Eq.~\eqref{eq : IDKsscScalar}], larger values of \(\beta \Lambda^{-4}\) will push the initial data out of the
linear regime and potentially also out of the EFT's regime of validity.
One symptom of this is an increased disagreement by the metric coefficients obtained by solving the Hamiltonian constraint [Eq.~\eqref{eq: IsoHamConstraint}] at \(t = 0\). This is due to the radial \(\rho\) field [in Eq.~\eqref{eq: VEVexpansion}] containing an increasing fraction of the scalar energy content in the \(U(1)\) UV theory, which is not accounted for in 
\(k\)-essence 
(nor in the ``fixed" theory), and resulting in ``deeper" gravitational  ``potential wells''. One way to return to the linear regime and to the EFT regime is to weaken the initial data by decreasing the amplitude \(A\) and/or choosing milder initial scalar gradients by increasing the root-mean-square width \(\sigma\).

Finally, in Sec.~\ref{sec: Fixing the Equations}, we highlighted a caveat with the strongly hyperbolic nature of the ``fixed" theory's scalar system of equations \eqref{eq: FEScalarEquation}-\eqref{eq: FESigmaEquation}. Namely,  when \(\Sigma \to 0\), the system becomes pathological.
We have performed numerical evolutions with larger values of \(\beta\) (and correspondingly smaller values of \(M_\rho\)) and observe that the \(U(1)\) UV completion evolution may drive the reconstructed value of \(K'\left(X\right)\) to zero. 
In the ``fixed" theory, \(K'\left(X\right)\) may also vanish dynamically, driving \(\Sigma\) to zero with it, and causing the code to crash. Moreover, this may happen in regions not censored by an apparent horizon (see also  Ref.~\cite{Kunstatter:2011ce}). 
This problem may be avoided in other versions of \(k\)-essence. For instance, in cubic \(k\)-essence the particular functional form of \(K\left(X\right)\) may keep \(\left \lvert K'\left(X\right) \right \rvert \geq q^2 > 0\), where \(q\) is a constant --see e.g. Refs.~\cite{Bezares:2020wkn,terHaar:2020xxb,Bezares:2021dma,Bezares:2021yek}.
Alternatively, one may look for a different way to implement the ``fixing the equations" approach.

%%%%%%%%%%%%%%%%%%%%%%%%%%%%%%%%%%%%%%%%%%%%%%%%%%%%%%%%%%%
%%%%%%%%%%%%%%%%%%%%%%%%%%%%%%%%%%%%%%%%%%%%%%%%%%%%%%%%%%%
%%%%%%%%%%%%%%%%%%%%%%%%%%%%%%%%%%%%%%%%%%%%%%%%%%%%%%%%%%%

\section{Conclusions} \label{sec: conclusion}

In this work, we have studied two general strategies to deal with the breakdown of the Cauchy problem in \(k\)-essence.
%
% \footnote{
% %
% On a similar note, shocks/caustics in \(k\)-essence theories may also be resolved by resorting to a partial UV completion. In Ref.~\cite{MukohyamaUV:2020lsu}, it was shown that the transfer of energy to an additional (UV) degree of freedom may allow for a smoothening of shock fronts in \(k\)-essence.
% %
% }
%
The first was to resort to a UV completion of the theory, which allows for an initial-value problem that remains  well-posed at all times. Unfortunately, while
this was possible for the \(k\)-essence model considered in this paper, it is not possible for generic ones, e.g. for those that possess
screening mechanisms, for which such UV completions remain unknown (if existing at all~\cite{Adams:2006sv}). 
The second strategy consisted in ``fixing the equations"~\cite{Cayuso_2017} of \(k\)-essence to control the high frequency behavior suspected of leading to the Cauchy breakdown. 
Both strategies were studied before in Minkowski space by Allwright and Lehner~\cite{Allwright:2018rut} to demonstrate their technical viability.\footnote{
See also Refs.~\cite{Kaloper:2014vqa, Reall:2021ebq} where this UV theory and its corresponding EFT description were studied without considering the coupling to gravity, and Ref.~\cite{MukohyamaUV:2020lsu}, where it is shown that shocks/caustics in \(k\)-essence may be smoothed by a suitable UV completion.   
} 
Here we have generalized them to include gravity.

By considering the specific case of
quadratic \(k\)-essence, we have shown that both approaches reproduce the EFT dynamics of \(k\)-essence up to a ``Tricomi-type'' breakdown of the Cauchy problem, where the scalar equation changes character from hyperbolic to parabolic and then elliptic. 
Furthermore,  both the UV completion and the
``fixing the equations'' approach
allow for evolving the dynamics past the Cauchy breakdown to the physical end state of
the evolution 
(in our example, the formation of a black hole). 
This should be contrasted with previous efforts to ``chart'' the space of initial data in \(k\)-essence, in order to rule out regions leading to ill-posed problems --see e.g. Ref.~\cite{Bernard:2019fjb, Bezares:2020wkn, Figueras:2020dzx, Figueras:2021abd}. With the two strategies described above, (most of) these regions need not be excluded.
In the context of compact binaries, in particular, this opens up the possibility of simulating their coalescence, allowing the study of the entire dynamics and the emission of gravitational/scalar radiation in more generic \(k\)-essence models than currently possible~\cite{Bezares:2021dma}.

Moreover, since
we have access to the high-energy regime of \(k\)-essence thanks to its UV completion,
our results for the scalar evolution provide a validation test of the ``fixing the equations" approach. It is important to stress that this approach, albeit agnostic of the details of the UV completion, qualitatively agrees with
the dynamics of the latter well into the nonlinear regime of \(k\)-essence. 
One can therefore argue that this nonlinear regime can be at least qualitatively captured by the low-energy EFT. In fact, we find that only in the high curvature/gradient region inside the black hole apparent horizon
does the  ``fixing the equations" approach significantly deviate from the UV completion evolution. This is expected, as it is in those regions that the key 
assumption of the ``fixing the equations" approach, i.e. 
the requirement that energy does not cascade into high energy modes~\cite{Cayuso_2017}, is violated. This provides hope that even the screening mechanism, which depends
crucially on the non-linear dynamics of \(k\)-essence, may be within reach of the low-energy EFT, at least qualitatively.

%%%%%%%%%%%%%%%%%%%%%%%%%%%%%%%%%%%%%%%%%%%%%%%%%%%%%%%%%%%
%%%%%%%%%%%%%%%%%%%%%%%%%%%%%%%%%%%%%%%%%%%%%%%%%%%%%%%%%%%
%%%%%%%%%%%%%%%%%%%%%%%%%%%%%%%%%%%%%%%%%%%%%%%%%%%%%%%%%%%

\FloatBarrier

\begin{acknowledgments}

It is a pleasure to thank Marco Crisostomi, Luis Lehner and Carlos Palenzuela for insightful discussions. All authors acknowledge financial support provided under the European Union's H2020 ERC Consolidator Grant ``GRavity from Astrophysical to Microscopic Scales'' grant agreement no. GRAMS-815673. This work was supported by the EU Horizon 2020 Research and
Innovation Programme under the Marie Sklodowska-Curie Grant Agreement
No. 101007855.

\end{acknowledgments}

%%%%%%%%%%%%%%%%%%%%%%%%%%%%%%%%%%%%%%%%%%%%%%%%%%%%%%%%%%%
%%%%%%%%%%%%%%%%%%%%%%%%%%%%%%%%%%%%%%%%%%%%%%%%%%%%%%%%%%%
%%%%%%%%%%%%%%%%%%%%%%%%%%%%%%%%%%%%%%%%%%%%%%%%%%%%%%%%%%%

\appendix

%%%%%%%%%%%%%%%%%%%%%%%%%%%%%%%%%%%%%%%%%%%%%%%%%%%%%%%%%%%
%%%%%%%%%%%%%%%%%%%%%%%%%%%%%%%%%%%%%%%%%%%%%%%%%%%%%%%%%%%
%%%%%%%%%%%%%%%%%%%%%%%%%%%%%%%%%%%%%%%%%%%%%%%%%%%%%%%%%%%

\section{Weak data example} \label{sec: weak data example}

In this Appendix, we show results for the case with weak initial data corresponding to parameters \(r_c = 55 \, L_{\Lambda}\), \(\sigma = 15 \, L_{\Lambda}\), and \(A = 0.02 \, L_{\Lambda}\) [Eq.~\eqref{eq : IDKsscScalar}], and the same values for the coupling constants as in the main text (Sec.~\ref{sec: results}). During the evolution, an ingoing pulse bounces off the origin and is dispersed as it propagates outwards. No apparent sound or black hole  horizons are formed.

\begin{figure}[]
 \centering
\includegraphics[width=3.4 in]{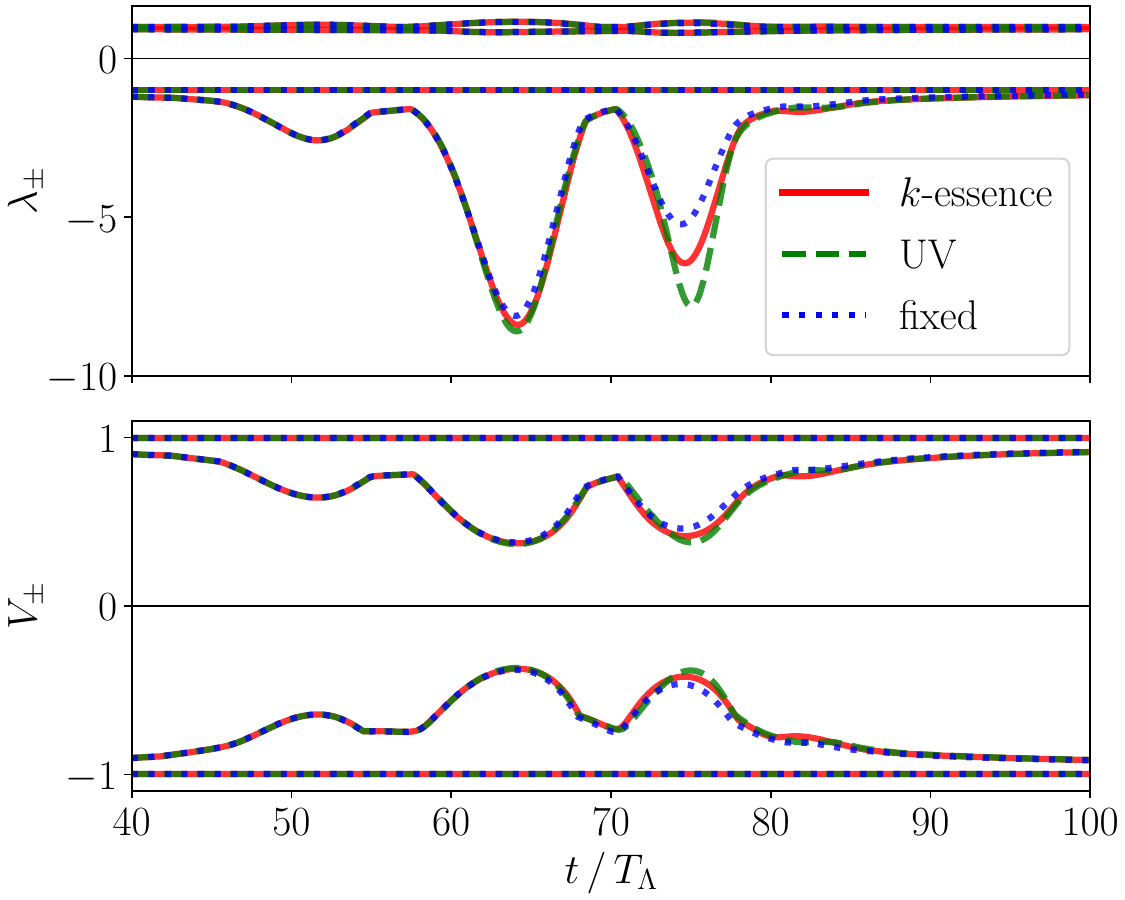}
\caption{ \emph{Character of the \(k\)-essence scalar equation (weak initial data). }
First panel: the minimum and maximum values of the  eigenvalues of the effective metric in \(k\)-essence (red solid line), the \(U(1)\) UV completion (green dashed line) and the ``fixed" theory (blue dotted line).
For the last two, the effective metric is not a fundamental quantity but  ``emergent" at low energies. 
From top to bottom: \(\max\left(\lambda_{+}\right)\), \(\min\left(\lambda_{+}\right)\), \(\max\left(\lambda_{-}\right)\), \(\min\left(\lambda_{-}\right)\).
Second panel: the minimum and maximum values of the  characteristic speeds. In this panel, from top to bottom: \(\max\left(V_{+}\right)\), \(\min\left(V_{+}\right)\), \(\max\left(V_{-}\right)\), \(\min\left(V_{-}\right)\).}
\label{fig: Case A: velocities}
\end{figure}

\begin{figure}[]
 \centering
\includegraphics[width=3.4 in]{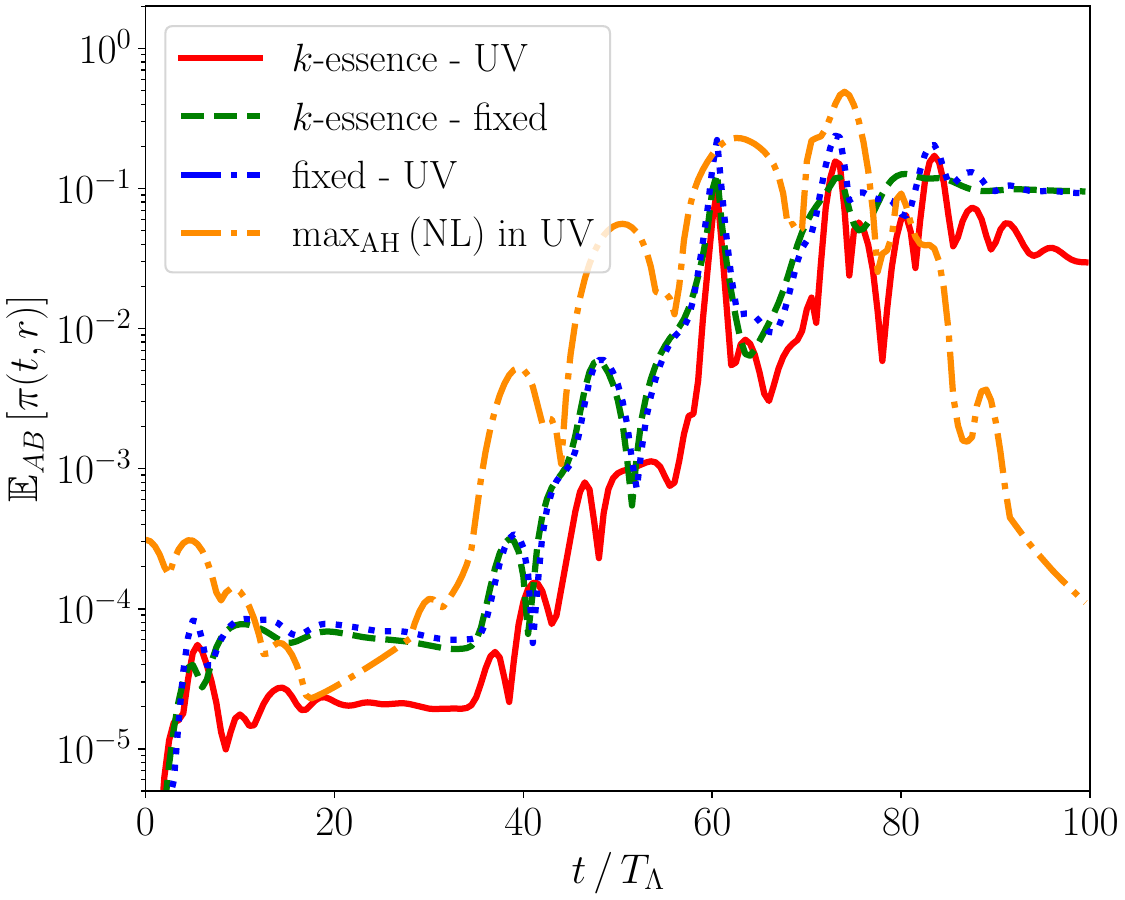}
\caption{\emph{Discrepancy of the \(k\)-essence scalar (weak initial data).} The discrepancy measure \(\mathbb{E}_{AB}\) of the \(k\)-essence scalar \(\pi\) for theories \(A\) vs.~\(B\), defined in Eq.~\eqref{eq: definition of discrepancy faithfulness}; namely, \(k\)-essence vs.~\(U(1)\) UV completion (red solid line), \(k\)-essence vs.~``fixed" theory (green dashed line), and ``fixed" theory vs.~\(U(1)\) UV completion (blue dot-dashed line). For completeness, we plot \(\max_\text{AH} \left(\mathrm{NL}\right)\) in the \(U(1)\) UV completion (orange dot-dashed line).}
\label{fig: Case A: faithfulness}
\end{figure}

\begin{figure}[]
 \centering
\includegraphics[width=3.4 in]{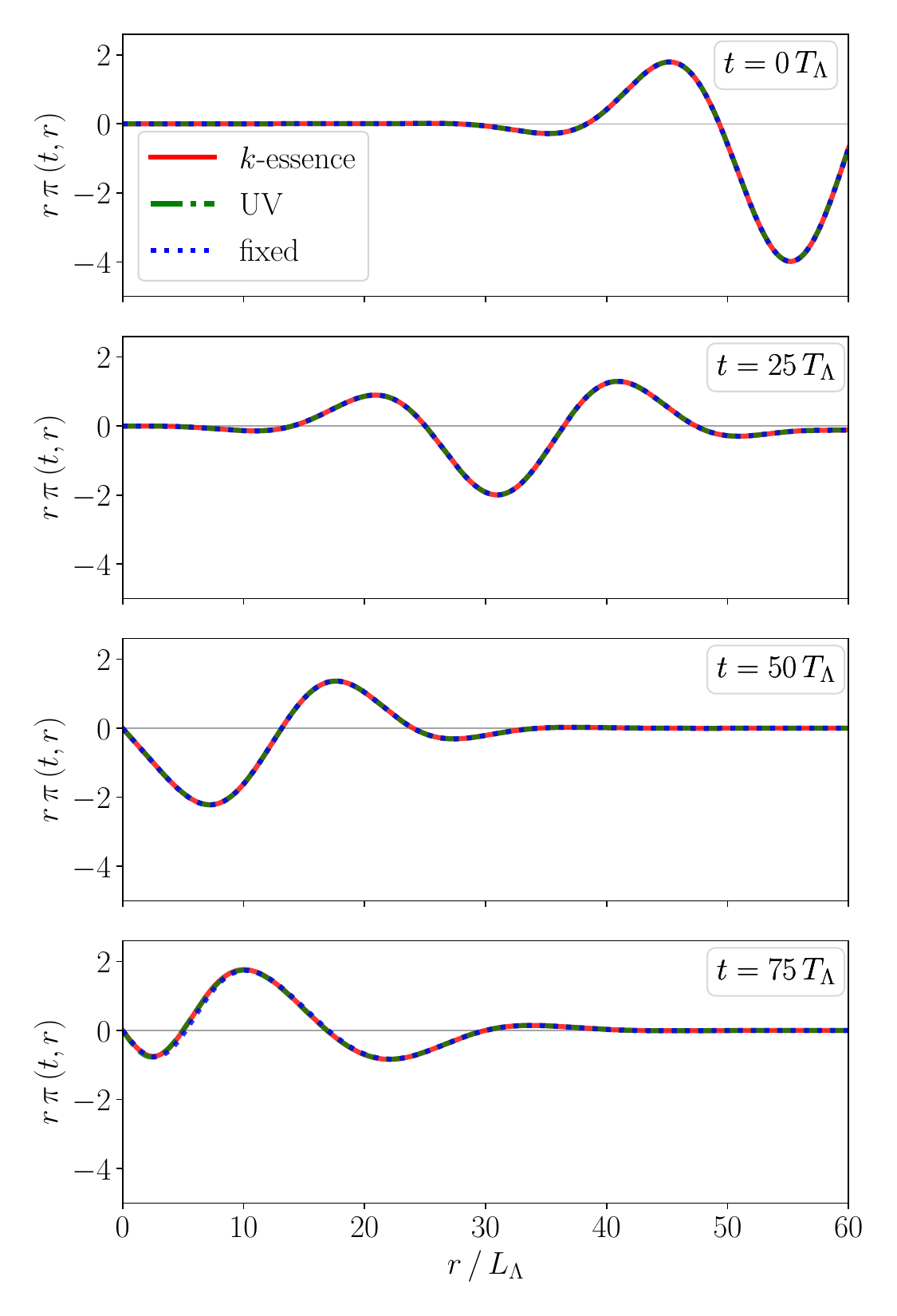}
\caption{\emph{Dynamics of the pulse in the linear/EFT regime (weak initial data).} 
Time snapshots of the  \(k\)-essence scalar for representative times from \(t = 0 \) to \(t  = 75 \, T_\Lambda\) for \(k\)-essence (red solid lines), 
the phase field of the \(U(1)\) UV completion (green dashed lines) and the \(\pi\)-scalar of the ``fixed" theory (blue dotted lines).
}
\label{fig: Case A: scalar profile}
\end{figure}

In Fig.~\ref{fig: Case A: velocities}, we show the spatial maximum and minimum values of the eigenvalues of the effective metric and of the characteristic speeds, where no Cauchy breakdown is observed. 
Consistently with the discrepancy measure \(\mathbb{E}_{AB} \left[\pi\right]\) in Fig.~\ref{fig: Case A: faithfulness}, the scalar profiles show agreement across the board in Fig.~\ref{fig: Case A: scalar profile}. For this initial data, the evolution remains in the linear/EFT regime at all times.
%

%%%%%%%%%%%%%%%%%%%%%%%%%%%%%%%%%%%%%%%%%%%%%%%%%%%%%%%%%%%
%%%%%%%%%%%%%%%%%%%%%%%%%%%%%%%%%%%%%%%%%%%%%%%%%%%%%%%%%%%
%%%%%%%%%%%%%%%%%%%%%%%%%%%%%%%%%%%%%%%%%%%%%%%%%%%%%%%%%%%

%\vspace{-10cm}
\section{Constraint propagation in the ``fixed" theory} \label{sec: Constraint propagation}

In the ``fixed" theory, the equations of motion do not automatically imply the conservation of the stress-energy tensor. Indeed, the right hand side of 
\begin{multline}
    \nabla^{\mu}T^{(\pi)}_{\mu\nu} = 2 \nabla_{\nu} \pi \nabla^{\mu}\left[\left(\Sigma - K'\left(X\right)\right)\nabla_{\mu}\pi\right] + \\
    + \text{term prop.~to Eq.~\eqref{eq: FEScalarEquation}}~
\end{multline}
is not formally zero when the equations of motion are used.
However, if the ``driver" equation [c.f. Eq.~\eqref{eq: FESigmaEquation}] is such that \(\Sigma \approx K'\left(X\right)\), an approximate conservation equation for \(T_{\mu\nu}^{(\pi)}\)  is expected, i.e. \(\nabla^{\mu}T^{(\pi)}_{\mu\nu}\approx0\).

In order to see the effect on the propagation of the constraint equations, we follow Ref.~\cite{Alcubierre:1138167} (see also Ref.~\cite{Frittelli:PhysRevD.55.5992}). 
We begin by defining the projections of Einstein equations \(E_{\mu\nu} \equiv G_{\mu\nu} - \kappa \, T^{(\pi)}_{\mu\nu}\) = 0, given by
\begin{align}
    \mathcal{H} &\equiv n^{\mu}n^{\nu} E_{\mu\nu}  ~,  \nonumber\\
    \mathcal{M}_{\mu} &\equiv -n^{\rho}{P_\mu}^{\sigma}E_{\rho\sigma} ~, \\ 
    \mathcal{E}_{\mu\nu} &\equiv {P_\mu}^{\rho}{P_\nu}^{\sigma}E_{\rho\sigma} ~, \nonumber
\end{align}
where
\(n^{\mu}\) is the vector normal to the foliation
and \({P_\mu}^{\sigma}=\left(\delta_\mu^\sigma + n_\mu n^\sigma \right)\) is the spatial projector.
Therefore, the Hamiltonian and momentum constraints can be expressed as
\(\mathcal{H} = 0\) and \(\mathcal{M}_\mu = 0\), respectively. 
The evolution equations for the metric are instead
\(\mathcal{E}_{\mu\nu} = 0\).
Finally, the evolution of the Hamiltonian and momentum constraints can be obtained from the projections of \(\nabla^{\mu}\left(G_{\mu\nu} - \kappa \, T^{(\pi)}_{\mu\nu}\right)\) and are given by, 
\begin{multline}\label{eq: HamiltonianEvolution}
    n^\nu \nabla_\nu \mathcal{H} = - D^\nu \mathcal{M}_\nu - \mathcal{E}_{\mu\nu}D^{\mu}n^{\nu} + \mathcal{L}_\mathcal{H} \left(\mathcal{H}, \mathcal{M}_\sigma \right) + \\
    + \kappa \, n^{\nu}\, \nabla^{\mu}T^{(\pi)}_{\mu\nu}~,
\end{multline}
\begin{multline}\label{eq: MomentumEvolution}
    n^\nu \nabla_\nu \mathcal{M}_\mu = - D^{\nu}\mathcal{E}_{\mu\nu} - \mathcal{E}_{\mu\nu}n^{\lambda} \nabla_{\lambda}n^{\nu} + \mathcal{L}_{\mathcal{M_\mu}}\left(\mathcal{H}, \mathcal{M}_\sigma\right) + \\
    - \kappa \, {P_\mu}^{\sigma}\nabla^{\lambda}T^{(\pi)}_{\sigma \lambda} ~,
\end{multline}
respectively, where
and \(\mathcal{L}_\mathcal{H}\) and \(\mathcal{L}_{\mathcal{M_\mu}}\) are zero for vanishing arguments, and \(D\) the spatial covariant derivative.
Thus, an approximate conservation of the constraints (\(  n^\nu \nabla_\nu \mathcal{H} \approx 0\) and \(n^\nu \nabla_\nu \mathcal{M}_\mu \approx 0\)) happens if \emph{(i)} they are satisfied initially, \emph{(ii)} we use the equations of motion, and \emph{(iii)} the driver equation ensures that \(\Sigma \approx K'\left(X\right)\) during the evolution.

In contrast, for \(k\)-essence and the \(U(1)\) UV completion, the stress energy tensor is conserved, and thus \(n^\nu \nabla_\nu \mathcal{H} = n^\nu \nabla_\nu \mathcal{M}_\mu = 0\), when the equations of motion are used.


\FloatBarrier

\bibliography{bibliography}

%apsrev4-2.bst 2019-01-14 (MD) hand-edited version of apsrev4-1.bst
%Control: key (0)
%Control: author (8) initials jnrlst
%Control: editor formatted (1) identically to author
%Control: production of article title (0) allowed
%Control: page (0) single
%Control: year (1) truncated
%Control: production of eprint (0) enabled
\begin{thebibliography}{78}%
\makeatletter
\providecommand \@ifxundefined [1]{%
 \@ifx{#1\undefined}
}%
\providecommand \@ifnum [1]{%
 \ifnum #1\expandafter \@firstoftwo
 \else \expandafter \@secondoftwo
 \fi
}%
\providecommand \@ifx [1]{%
 \ifx #1\expandafter \@firstoftwo
 \else \expandafter \@secondoftwo
 \fi
}%
\providecommand \natexlab [1]{#1}%
\providecommand \enquote  [1]{``#1''}%
\providecommand \bibnamefont  [1]{#1}%
\providecommand \bibfnamefont [1]{#1}%
\providecommand \citenamefont [1]{#1}%
\providecommand \href@noop [0]{\@secondoftwo}%
\providecommand \href [0]{\begingroup \@sanitize@url \@href}%
\providecommand \@href[1]{\@@startlink{#1}\@@href}%
\providecommand \@@href[1]{\endgroup#1\@@endlink}%
\providecommand \@sanitize@url [0]{\catcode `\\12\catcode `\$12\catcode
  `\&12\catcode `\#12\catcode `\^12\catcode `\_12\catcode `\%12\relax}%
\providecommand \@@startlink[1]{}%
\providecommand \@@endlink[0]{}%
\providecommand \url  [0]{\begingroup\@sanitize@url \@url }%
\providecommand \@url [1]{\endgroup\@href {#1}{\urlprefix }}%
\providecommand \urlprefix  [0]{URL }%
\providecommand \Eprint [0]{\href }%
\providecommand \doibase [0]{https://doi.org/}%
\providecommand \selectlanguage [0]{\@gobble}%
\providecommand \bibinfo  [0]{\@secondoftwo}%
\providecommand \bibfield  [0]{\@secondoftwo}%
\providecommand \translation [1]{[#1]}%
\providecommand \BibitemOpen [0]{}%
\providecommand \bibitemStop [0]{}%
\providecommand \bibitemNoStop [0]{.\EOS\space}%
\providecommand \EOS [0]{\spacefactor3000\relax}%
\providecommand \BibitemShut  [1]{\csname bibitem#1\endcsname}%
\let\auto@bib@innerbib\@empty
%</preamble>
\bibitem [{\citenamefont {Abbott}\ \emph
  {et~al.}(2016{\natexlab{a}})\citenamefont {Abbott} \emph
  {et~al.}}]{LIGOScientific:2016aoc}%
  \BibitemOpen
  \bibfield  {author} {\bibinfo {author} {\bibfnamefont {B.~P.}\ \bibnamefont
  {Abbott}} \emph {et~al.} (\bibinfo {collaboration} {LIGO Scientific,
  Virgo}),\ }\bibfield  {title} {\bibinfo {title} {{Observation of
  Gravitational Waves from a Binary Black Hole Merger}},\ }\href
  {https://doi.org/10.1103/PhysRevLett.116.061102} {\bibfield  {journal}
  {\bibinfo  {journal} {Phys. Rev. Lett.}\ }\textbf {\bibinfo {volume} {116}},\
  \bibinfo {pages} {061102} (\bibinfo {year} {2016}{\natexlab{a}})},\ \Eprint
  {https://arxiv.org/abs/1602.03837} {arXiv:1602.03837 [gr-qc]} \BibitemShut
  {NoStop}%
\bibitem [{\citenamefont {Abbott}\ \emph
  {et~al.}(2016{\natexlab{b}})\citenamefont {Abbott} \emph
  {et~al.}}]{TheLIGOScientific:2016src}%
  \BibitemOpen
  \bibfield  {author} {\bibinfo {author} {\bibfnamefont {B.~P.}\ \bibnamefont
  {Abbott}} \emph {et~al.} (\bibinfo {collaboration} {LIGO Scientific,
  Virgo}),\ }\bibfield  {title} {\bibinfo {title} {{Tests of general relativity
  with GW150914}},\ }\href {https://doi.org/10.1103/PhysRevLett.116.221101}
  {\bibfield  {journal} {\bibinfo  {journal} {Phys. Rev. Lett.}\ }\textbf
  {\bibinfo {volume} {116}},\ \bibinfo {pages} {221101} (\bibinfo {year}
  {2016}{\natexlab{b}})},\ \bibinfo {note} {[Erratum: Phys.Rev.Lett. 121,
  129902 (2018)]},\ \Eprint {https://arxiv.org/abs/1602.03841}
  {arXiv:1602.03841 [gr-qc]} \BibitemShut {NoStop}%
\bibitem [{\citenamefont {Abbott}\ \emph
  {et~al.}(2019{\natexlab{a}})\citenamefont {Abbott} \emph
  {et~al.}}]{Abbott:2018lct}%
  \BibitemOpen
  \bibfield  {author} {\bibinfo {author} {\bibfnamefont {B.~P.}\ \bibnamefont
  {Abbott}} \emph {et~al.} (\bibinfo {collaboration} {LIGO Scientific,
  Virgo}),\ }\bibfield  {title} {\bibinfo {title} {{Tests of General Relativity
  with GW170817}},\ }\href {https://doi.org/10.1103/PhysRevLett.123.011102}
  {\bibfield  {journal} {\bibinfo  {journal} {Phys. Rev. Lett.}\ }\textbf
  {\bibinfo {volume} {123}},\ \bibinfo {pages} {011102} (\bibinfo {year}
  {2019}{\natexlab{a}})},\ \Eprint {https://arxiv.org/abs/1811.00364}
  {arXiv:1811.00364 [gr-qc]} \BibitemShut {NoStop}%
%%CITATION = ARXIV:1811.00364;%%
\bibitem [{\citenamefont {Abbott}\ \emph
  {et~al.}(2019{\natexlab{b}})\citenamefont {Abbott} \emph
  {et~al.}}]{LIGOScientific:2019fpa}%
  \BibitemOpen
  \bibfield  {author} {\bibinfo {author} {\bibfnamefont {B.~P.}\ \bibnamefont
  {Abbott}} \emph {et~al.} (\bibinfo {collaboration} {LIGO Scientific,
  Virgo}),\ }\bibfield  {title} {\bibinfo {title} {{Tests of General Relativity
  with the Binary Black Hole Signals from the LIGO-Virgo Catalog GWTC-1}},\
  }\href {https://doi.org/10.1103/PhysRevD.100.104036} {\bibfield  {journal}
  {\bibinfo  {journal} {Phys. Rev.}\ }\textbf {\bibinfo {volume} {D100}},\
  \bibinfo {pages} {104036} (\bibinfo {year} {2019}{\natexlab{b}})},\ \Eprint
  {https://arxiv.org/abs/1903.04467} {arXiv:1903.04467 [gr-qc]} \BibitemShut
  {NoStop}%
%%CITATION = ARXIV:1903.04467;%%
\bibitem [{\citenamefont {Abbott}\ \emph
  {et~al.}(2021{\natexlab{a}})\citenamefont {Abbott} \emph
  {et~al.}}]{Abbott:2020jks}%
  \BibitemOpen
  \bibfield  {author} {\bibinfo {author} {\bibfnamefont {R.}~\bibnamefont
  {Abbott}} \emph {et~al.} (\bibinfo {collaboration} {LIGO Scientific,
  Virgo}),\ }\bibfield  {title} {\bibinfo {title} {{Tests of general relativity
  with binary black holes from the second LIGO-Virgo gravitational-wave
  transient catalog}},\ }\href {https://doi.org/10.1103/PhysRevD.103.122002}
  {\bibfield  {journal} {\bibinfo  {journal} {Phys. Rev. D}\ }\textbf {\bibinfo
  {volume} {103}},\ \bibinfo {pages} {122002} (\bibinfo {year}
  {2021}{\natexlab{a}})},\ \Eprint {https://arxiv.org/abs/2010.14529}
  {arXiv:2010.14529 [gr-qc]} \BibitemShut {NoStop}%
\bibitem [{\citenamefont {Abbott}\ \emph
  {et~al.}(2021{\natexlab{b}})\citenamefont {Abbott} \emph
  {et~al.}}]{LIGOScientific:2021sio}%
  \BibitemOpen
  \bibfield  {author} {\bibinfo {author} {\bibfnamefont {R.}~\bibnamefont
  {Abbott}} \emph {et~al.} (\bibinfo {collaboration} {LIGO Scientific, VIRGO,
  KAGRA}),\ }\bibfield  {title} {\bibinfo {title} {{Tests of General Relativity
  with GWTC-3}},\ }\href@noop {} {\  (\bibinfo {year} {2021}{\natexlab{b}})},\
  \Eprint {https://arxiv.org/abs/2112.06861} {arXiv:2112.06861 [gr-qc]}
  \BibitemShut {NoStop}%
\bibitem [{\citenamefont {Will}(1993)}]{Will:1993hxu}%
  \BibitemOpen
  \bibfield  {author} {\bibinfo {author} {\bibfnamefont {C.~M.}\ \bibnamefont
  {Will}},\ }\href {https://doi.org/10.1017/CBO9780511564246} {\emph {\bibinfo
  {title} {{Theory and Experiment in Gravitational Physics}}}}\ (\bibinfo
  {publisher} {Cambridge University Press},\ \bibinfo {address} {Cambridge},\
  \bibinfo {year} {1993})\BibitemShut {NoStop}%
\bibitem [{\citenamefont {Will}(2014)}]{Will:2014kxa}%
  \BibitemOpen
  \bibfield  {author} {\bibinfo {author} {\bibfnamefont {C.~M.}\ \bibnamefont
  {Will}},\ }\bibfield  {title} {\bibinfo {title} {{The Confrontation between
  General Relativity and Experiment}},\ }\href
  {https://doi.org/10.12942/lrr-2014-4} {\bibfield  {journal} {\bibinfo
  {journal} {Living Rev. Rel.}\ }\textbf {\bibinfo {volume} {17}},\ \bibinfo
  {pages} {4} (\bibinfo {year} {2014})},\ \Eprint
  {https://arxiv.org/abs/1403.7377} {arXiv:1403.7377 [gr-qc]} \BibitemShut
  {NoStop}%
%%CITATION = ARXIV:1403.7377;%%
\bibitem [{\citenamefont {Damour}\ and\ \citenamefont
  {Taylor}(1992)}]{Damour:1991rd}%
  \BibitemOpen
  \bibfield  {author} {\bibinfo {author} {\bibfnamefont {T.}~\bibnamefont
  {Damour}}\ and\ \bibinfo {author} {\bibfnamefont {J.~H.}\ \bibnamefont
  {Taylor}},\ }\bibfield  {title} {\bibinfo {title} {{Strong field tests of
  relativistic gravity and binary pulsars}},\ }\href
  {https://doi.org/10.1103/PhysRevD.45.1840} {\bibfield  {journal} {\bibinfo
  {journal} {Phys. Rev.}\ }\textbf {\bibinfo {volume} {D45}},\ \bibinfo {pages}
  {1840} (\bibinfo {year} {1992})}\BibitemShut {NoStop}%
%%CITATION = PHRVA,D45,1840;%%
\bibitem [{\citenamefont {Kramer}\ \emph {et~al.}(2006)\citenamefont {Kramer}
  \emph {et~al.}}]{Kramer:2006nb}%
  \BibitemOpen
  \bibfield  {author} {\bibinfo {author} {\bibfnamefont {M.}~\bibnamefont
  {Kramer}} \emph {et~al.},\ }\bibfield  {title} {\bibinfo {title} {{Tests of
  general relativity from timing the double pulsar}},\ }\href
  {https://doi.org/10.1126/science.1132305} {\bibfield  {journal} {\bibinfo
  {journal} {Science}\ }\textbf {\bibinfo {volume} {314}},\ \bibinfo {pages}
  {97} (\bibinfo {year} {2006})},\ \Eprint
  {https://arxiv.org/abs/astro-ph/0609417} {arXiv:astro-ph/0609417 [astro-ph]}
  \BibitemShut {NoStop}%
%%CITATION = ASTRO-PH/0609417;%%
\bibitem [{\citenamefont {Freire}\ \emph {et~al.}(2012)\citenamefont {Freire},
  \citenamefont {Wex}, \citenamefont {Esposito-Farese}, \citenamefont
  {Verbiest}, \citenamefont {Bailes}, \citenamefont {Jacoby}, \citenamefont
  {Kramer}, \citenamefont {Stairs}, \citenamefont {Antoniadis},\ and\
  \citenamefont {Janssen}}]{Freire:2012mg}%
  \BibitemOpen
  \bibfield  {author} {\bibinfo {author} {\bibfnamefont {P.~C.~C.}\
  \bibnamefont {Freire}}, \bibinfo {author} {\bibfnamefont {N.}~\bibnamefont
  {Wex}}, \bibinfo {author} {\bibfnamefont {G.}~\bibnamefont
  {Esposito-Farese}}, \bibinfo {author} {\bibfnamefont {J.~P.~W.}\ \bibnamefont
  {Verbiest}}, \bibinfo {author} {\bibfnamefont {M.}~\bibnamefont {Bailes}},
  \bibinfo {author} {\bibfnamefont {B.~A.}\ \bibnamefont {Jacoby}}, \bibinfo
  {author} {\bibfnamefont {M.}~\bibnamefont {Kramer}}, \bibinfo {author}
  {\bibfnamefont {I.~H.}\ \bibnamefont {Stairs}}, \bibinfo {author}
  {\bibfnamefont {J.}~\bibnamefont {Antoniadis}},\ and\ \bibinfo {author}
  {\bibfnamefont {G.~H.}\ \bibnamefont {Janssen}},\ }\bibfield  {title}
  {\bibinfo {title} {{The relativistic pulsar-white dwarf binary PSR J1738+0333
  II. The most stringent test of scalar-tensor gravity}},\ }\href
  {https://doi.org/10.1111/j.1365-2966.2012.21253.x} {\bibfield  {journal}
  {\bibinfo  {journal} {Mon. Not. Roy. Astron. Soc.}\ }\textbf {\bibinfo
  {volume} {423}},\ \bibinfo {pages} {3328} (\bibinfo {year} {2012})},\ \Eprint
  {https://arxiv.org/abs/1205.1450} {arXiv:1205.1450 [astro-ph.GA]}
  \BibitemShut {NoStop}%
%%CITATION = ARXIV:1205.1450;%%
\bibitem [{\citenamefont {Kramer}\ \emph {et~al.}(2021)\citenamefont {Kramer}
  \emph {et~al.}}]{Kramer:2021jcw}%
  \BibitemOpen
  \bibfield  {author} {\bibinfo {author} {\bibfnamefont {M.}~\bibnamefont
  {Kramer}} \emph {et~al.},\ }\bibfield  {title} {\bibinfo {title}
  {{Strong-Field Gravity Tests with the Double Pulsar}},\ }\href
  {https://doi.org/10.1103/PhysRevX.11.041050} {\bibfield  {journal} {\bibinfo
  {journal} {Phys. Rev. X}\ }\textbf {\bibinfo {volume} {11}},\ \bibinfo
  {pages} {041050} (\bibinfo {year} {2021})},\ \Eprint
  {https://arxiv.org/abs/2112.06795} {arXiv:2112.06795 [astro-ph.HE]}
  \BibitemShut {NoStop}%
\bibitem [{\citenamefont {Clifton}\ \emph {et~al.}(2012)\citenamefont
  {Clifton}, \citenamefont {Ferreira}, \citenamefont {Padilla},\ and\
  \citenamefont {Skordis}}]{Clifton:2011jh}%
  \BibitemOpen
  \bibfield  {author} {\bibinfo {author} {\bibfnamefont {T.}~\bibnamefont
  {Clifton}}, \bibinfo {author} {\bibfnamefont {P.~G.}\ \bibnamefont
  {Ferreira}}, \bibinfo {author} {\bibfnamefont {A.}~\bibnamefont {Padilla}},\
  and\ \bibinfo {author} {\bibfnamefont {C.}~\bibnamefont {Skordis}},\
  }\bibfield  {title} {\bibinfo {title} {{Modified Gravity and Cosmology}},\
  }\href {https://doi.org/10.1016/j.physrep.2012.01.001} {\bibfield  {journal}
  {\bibinfo  {journal} {Phys. Rept.}\ }\textbf {\bibinfo {volume} {513}},\
  \bibinfo {pages} {1} (\bibinfo {year} {2012})},\ \Eprint
  {https://arxiv.org/abs/1106.2476} {arXiv:1106.2476 [astro-ph.CO]}
  \BibitemShut {NoStop}%
%%CITATION = ARXIV:1106.2476;%%
\bibitem [{\citenamefont {Horndeski}(1974)}]{Horndeski:1974wa}%
  \BibitemOpen
  \bibfield  {author} {\bibinfo {author} {\bibfnamefont {G.~W.}\ \bibnamefont
  {Horndeski}},\ }\bibfield  {title} {\bibinfo {title} {{Second-order
  scalar-tensor field equations in a four-dimensional space}},\ }\href
  {https://doi.org/10.1007/BF01807638} {\bibfield  {journal} {\bibinfo
  {journal} {Int. J. Theor. Phys.}\ }\textbf {\bibinfo {volume} {10}},\
  \bibinfo {pages} {363} (\bibinfo {year} {1974})}\BibitemShut {NoStop}%
\bibitem [{\citenamefont {Gleyzes}\ \emph {et~al.}(2015)\citenamefont
  {Gleyzes}, \citenamefont {Langlois}, \citenamefont {Piazza},\ and\
  \citenamefont {Vernizzi}}]{Gleyzes:2014dya}%
  \BibitemOpen
  \bibfield  {author} {\bibinfo {author} {\bibfnamefont {J.}~\bibnamefont
  {Gleyzes}}, \bibinfo {author} {\bibfnamefont {D.}~\bibnamefont {Langlois}},
  \bibinfo {author} {\bibfnamefont {F.}~\bibnamefont {Piazza}},\ and\ \bibinfo
  {author} {\bibfnamefont {F.}~\bibnamefont {Vernizzi}},\ }\bibfield  {title}
  {\bibinfo {title} {{Healthy theories beyond Horndeski}},\ }\href
  {https://doi.org/10.1103/PhysRevLett.114.211101} {\bibfield  {journal}
  {\bibinfo  {journal} {Phys. Rev. Lett.}\ }\textbf {\bibinfo {volume} {114}},\
  \bibinfo {pages} {211101} (\bibinfo {year} {2015})},\ \Eprint
  {https://arxiv.org/abs/1404.6495} {arXiv:1404.6495 [hep-th]} \BibitemShut
  {NoStop}%
%%CITATION = ARXIV:1404.6495;%%
\bibitem [{\citenamefont {Langlois}\ and\ \citenamefont
  {Noui}(2016)}]{Langlois:2015cwa}%
  \BibitemOpen
  \bibfield  {author} {\bibinfo {author} {\bibfnamefont {D.}~\bibnamefont
  {Langlois}}\ and\ \bibinfo {author} {\bibfnamefont {K.}~\bibnamefont
  {Noui}},\ }\bibfield  {title} {\bibinfo {title} {{Degenerate higher
  derivative theories beyond Horndeski: evading the Ostrogradski
  instability}},\ }\href {https://doi.org/10.1088/1475-7516/2016/02/034}
  {\bibfield  {journal} {\bibinfo  {journal} {JCAP}\ }\textbf {\bibinfo
  {volume} {02}},\ \bibinfo {pages} {034}},\ \Eprint
  {https://arxiv.org/abs/1510.06930} {arXiv:1510.06930 [gr-qc]} \BibitemShut
  {NoStop}%
\bibitem [{\citenamefont {Crisostomi}\ \emph {et~al.}(2016)\citenamefont
  {Crisostomi}, \citenamefont {Koyama},\ and\ \citenamefont
  {Tasinato}}]{Crisostomi:2016czh}%
  \BibitemOpen
  \bibfield  {author} {\bibinfo {author} {\bibfnamefont {M.}~\bibnamefont
  {Crisostomi}}, \bibinfo {author} {\bibfnamefont {K.}~\bibnamefont {Koyama}},\
  and\ \bibinfo {author} {\bibfnamefont {G.}~\bibnamefont {Tasinato}},\
  }\bibfield  {title} {\bibinfo {title} {{Extended Scalar-Tensor Theories of
  Gravity}},\ }\href {https://doi.org/10.1088/1475-7516/2016/04/044} {\bibfield
   {journal} {\bibinfo  {journal} {JCAP}\ }\textbf {\bibinfo {volume} {1604}},\
  \bibinfo {pages} {044}},\ \Eprint {https://arxiv.org/abs/1602.03119}
  {arXiv:1602.03119 [hep-th]} \BibitemShut {NoStop}%
%%CITATION = ARXIV:1602.03119;%%
\bibitem [{\citenamefont {Ben~Achour}\ \emph {et~al.}(2016)\citenamefont
  {Ben~Achour}, \citenamefont {Crisostomi}, \citenamefont {Koyama},
  \citenamefont {Langlois}, \citenamefont {Noui},\ and\ \citenamefont
  {Tasinato}}]{BenAchour:2016fzp}%
  \BibitemOpen
  \bibfield  {author} {\bibinfo {author} {\bibfnamefont {J.}~\bibnamefont
  {Ben~Achour}}, \bibinfo {author} {\bibfnamefont {M.}~\bibnamefont
  {Crisostomi}}, \bibinfo {author} {\bibfnamefont {K.}~\bibnamefont {Koyama}},
  \bibinfo {author} {\bibfnamefont {D.}~\bibnamefont {Langlois}}, \bibinfo
  {author} {\bibfnamefont {K.}~\bibnamefont {Noui}},\ and\ \bibinfo {author}
  {\bibfnamefont {G.}~\bibnamefont {Tasinato}},\ }\bibfield  {title} {\bibinfo
  {title} {{Degenerate higher order scalar-tensor theories beyond Horndeski up
  to cubic order}},\ }\href {https://doi.org/10.1007/JHEP12(2016)100}
  {\bibfield  {journal} {\bibinfo  {journal} {JHEP}\ }\textbf {\bibinfo
  {volume} {12}},\ \bibinfo {pages} {100}},\ \Eprint
  {https://arxiv.org/abs/1608.08135} {arXiv:1608.08135 [hep-th]} \BibitemShut
  {NoStop}%
\bibitem [{\citenamefont {Chiba}\ \emph {et~al.}(2000)\citenamefont {Chiba},
  \citenamefont {Okabe},\ and\ \citenamefont {Yamaguchi}}]{Chiba:1999ka}%
  \BibitemOpen
  \bibfield  {author} {\bibinfo {author} {\bibfnamefont {T.}~\bibnamefont
  {Chiba}}, \bibinfo {author} {\bibfnamefont {T.}~\bibnamefont {Okabe}},\ and\
  \bibinfo {author} {\bibfnamefont {M.}~\bibnamefont {Yamaguchi}},\ }\bibfield
  {title} {\bibinfo {title} {{Kinetically driven quintessence}},\ }\href
  {https://doi.org/10.1103/PhysRevD.62.023511} {\bibfield  {journal} {\bibinfo
  {journal} {Phys. Rev. D}\ }\textbf {\bibinfo {volume} {62}},\ \bibinfo
  {pages} {023511} (\bibinfo {year} {2000})},\ \Eprint
  {https://arxiv.org/abs/astro-ph/9912463} {arXiv:astro-ph/9912463}
  \BibitemShut {NoStop}%
\bibitem [{\citenamefont {Armendariz-Picon}\ \emph {et~al.}(2000)\citenamefont
  {Armendariz-Picon}, \citenamefont {Mukhanov},\ and\ \citenamefont
  {Steinhardt}}]{Armendariz-Picon:2000nqq}%
  \BibitemOpen
  \bibfield  {author} {\bibinfo {author} {\bibfnamefont {C.}~\bibnamefont
  {Armendariz-Picon}}, \bibinfo {author} {\bibfnamefont {V.~F.}\ \bibnamefont
  {Mukhanov}},\ and\ \bibinfo {author} {\bibfnamefont {P.~J.}\ \bibnamefont
  {Steinhardt}},\ }\bibfield  {title} {\bibinfo {title} {{A Dynamical solution
  to the problem of a small cosmological constant and late time cosmic
  acceleration}},\ }\href {https://doi.org/10.1103/PhysRevLett.85.4438}
  {\bibfield  {journal} {\bibinfo  {journal} {Phys. Rev. Lett.}\ }\textbf
  {\bibinfo {volume} {85}},\ \bibinfo {pages} {4438} (\bibinfo {year}
  {2000})},\ \Eprint {https://arxiv.org/abs/astro-ph/0004134}
  {arXiv:astro-ph/0004134} \BibitemShut {NoStop}%
\bibitem [{\citenamefont {Abbott}\ \emph
  {et~al.}(2017{\natexlab{a}})\citenamefont {Abbott} \emph
  {et~al.}}]{Monitor:2017mdv}%
  \BibitemOpen
  \bibfield  {author} {\bibinfo {author} {\bibfnamefont {B.}~\bibnamefont
  {Abbott}} \emph {et~al.} (\bibinfo {collaboration} {LIGO Scientific, Virgo,
  Fermi-GBM, INTEGRAL}),\ }\bibfield  {title} {\bibinfo {title} {{Gravitational
  Waves and Gamma-rays from a Binary Neutron Star Merger: GW170817 and GRB
  170817A}},\ }\href {https://doi.org/10.3847/2041-8213/aa920c} {\bibfield
  {journal} {\bibinfo  {journal} {Astrophys. J. Lett.}\ }\textbf {\bibinfo
  {volume} {848}},\ \bibinfo {pages} {L13} (\bibinfo {year}
  {2017}{\natexlab{a}})},\ \Eprint {https://arxiv.org/abs/1710.05834}
  {arXiv:1710.05834 [astro-ph.HE]} \BibitemShut {NoStop}%
\bibitem [{\citenamefont {Abbott}\ \emph
  {et~al.}(2017{\natexlab{b}})\citenamefont {Abbott} \emph
  {et~al.}}]{TheLIGOScientific:2017qsa}%
  \BibitemOpen
  \bibfield  {author} {\bibinfo {author} {\bibfnamefont {B.}~\bibnamefont
  {Abbott}} \emph {et~al.} (\bibinfo {collaboration} {LIGO Scientific,
  Virgo}),\ }\bibfield  {title} {\bibinfo {title} {{GW170817: Observation of
  Gravitational Waves from a Binary Neutron Star Inspiral}},\ }\href
  {https://doi.org/10.1103/PhysRevLett.119.161101} {\bibfield  {journal}
  {\bibinfo  {journal} {Phys. Rev. Lett.}\ }\textbf {\bibinfo {volume} {119}},\
  \bibinfo {pages} {161101} (\bibinfo {year} {2017}{\natexlab{b}})},\ \Eprint
  {https://arxiv.org/abs/1710.05832} {arXiv:1710.05832 [gr-qc]} \BibitemShut
  {NoStop}%
\bibitem [{\citenamefont {Langlois}\ \emph {et~al.}(2018)\citenamefont
  {Langlois}, \citenamefont {Saito}, \citenamefont {Yamauchi},\ and\
  \citenamefont {Noui}}]{Langlois:2017dyl}%
  \BibitemOpen
  \bibfield  {author} {\bibinfo {author} {\bibfnamefont {D.}~\bibnamefont
  {Langlois}}, \bibinfo {author} {\bibfnamefont {R.}~\bibnamefont {Saito}},
  \bibinfo {author} {\bibfnamefont {D.}~\bibnamefont {Yamauchi}},\ and\
  \bibinfo {author} {\bibfnamefont {K.}~\bibnamefont {Noui}},\ }\bibfield
  {title} {\bibinfo {title} {{Scalar-tensor theories and modified gravity in
  the wake of GW170817}},\ }\href {https://doi.org/10.1103/PhysRevD.97.061501}
  {\bibfield  {journal} {\bibinfo  {journal} {Phys. Rev. D}\ }\textbf {\bibinfo
  {volume} {97}},\ \bibinfo {pages} {061501} (\bibinfo {year} {2018})},\
  \Eprint {https://arxiv.org/abs/1711.07403} {arXiv:1711.07403 [gr-qc]}
  \BibitemShut {NoStop}%
\bibitem [{\citenamefont {Crisostomi}\ and\ \citenamefont
  {Koyama}(2018{\natexlab{a}})}]{Crisostomi:2017pjs}%
  \BibitemOpen
  \bibfield  {author} {\bibinfo {author} {\bibfnamefont {M.}~\bibnamefont
  {Crisostomi}}\ and\ \bibinfo {author} {\bibfnamefont {K.}~\bibnamefont
  {Koyama}},\ }\bibfield  {title} {\bibinfo {title} {{Self-accelerating
  universe in scalar-tensor theories after GW170817}},\ }\href
  {https://doi.org/10.1103/PhysRevD.97.084004} {\bibfield  {journal} {\bibinfo
  {journal} {Phys. Rev.}\ }\textbf {\bibinfo {volume} {D97}},\ \bibinfo {pages}
  {084004} (\bibinfo {year} {2018}{\natexlab{a}})},\ \Eprint
  {https://arxiv.org/abs/1712.06556} {arXiv:1712.06556 [astro-ph.CO]}
  \BibitemShut {NoStop}%
%%CITATION = ARXIV:1712.06556;%%
\bibitem [{\citenamefont {Crisostomi}\ and\ \citenamefont
  {Koyama}(2018{\natexlab{b}})}]{Crisostomi:2017lbg}%
  \BibitemOpen
  \bibfield  {author} {\bibinfo {author} {\bibfnamefont {M.}~\bibnamefont
  {Crisostomi}}\ and\ \bibinfo {author} {\bibfnamefont {K.}~\bibnamefont
  {Koyama}},\ }\bibfield  {title} {\bibinfo {title} {{Vainshtein mechanism
  after GW170817}},\ }\href {https://doi.org/10.1103/PhysRevD.97.021301}
  {\bibfield  {journal} {\bibinfo  {journal} {Phys. Rev. D}\ }\textbf {\bibinfo
  {volume} {97}},\ \bibinfo {pages} {021301} (\bibinfo {year}
  {2018}{\natexlab{b}})},\ \Eprint {https://arxiv.org/abs/1711.06661}
  {arXiv:1711.06661 [astro-ph.CO]} \BibitemShut {NoStop}%
\bibitem [{\citenamefont {Dima}\ and\ \citenamefont
  {Vernizzi}(2018)}]{Dima:2017pwp}%
  \BibitemOpen
  \bibfield  {author} {\bibinfo {author} {\bibfnamefont {A.}~\bibnamefont
  {Dima}}\ and\ \bibinfo {author} {\bibfnamefont {F.}~\bibnamefont
  {Vernizzi}},\ }\bibfield  {title} {\bibinfo {title} {{Vainshtein Screening in
  Scalar-Tensor Theories before and after GW170817: Constraints on Theories
  beyond Horndeski}},\ }\href {https://doi.org/10.1103/PhysRevD.97.101302}
  {\bibfield  {journal} {\bibinfo  {journal} {Phys. Rev. D}\ }\textbf {\bibinfo
  {volume} {97}},\ \bibinfo {pages} {101302} (\bibinfo {year} {2018})},\
  \Eprint {https://arxiv.org/abs/1712.04731} {arXiv:1712.04731 [gr-qc]}
  \BibitemShut {NoStop}%
\bibitem [{\citenamefont {Creminelli}\ \emph {et~al.}(2018)\citenamefont
  {Creminelli}, \citenamefont {Lewandowski}, \citenamefont {Tambalo},\ and\
  \citenamefont {Vernizzi}}]{Creminelli:2018xsv}%
  \BibitemOpen
  \bibfield  {author} {\bibinfo {author} {\bibfnamefont {P.}~\bibnamefont
  {Creminelli}}, \bibinfo {author} {\bibfnamefont {M.}~\bibnamefont
  {Lewandowski}}, \bibinfo {author} {\bibfnamefont {G.}~\bibnamefont
  {Tambalo}},\ and\ \bibinfo {author} {\bibfnamefont {F.}~\bibnamefont
  {Vernizzi}},\ }\bibfield  {title} {\bibinfo {title} {{Gravitational Wave
  Decay into Dark Energy}},\ }\href
  {https://doi.org/10.1088/1475-7516/2018/12/025} {\bibfield  {journal}
  {\bibinfo  {journal} {JCAP}\ }\textbf {\bibinfo {volume} {1812}},\ \bibinfo
  {pages} {025}},\ \Eprint {https://arxiv.org/abs/1809.03484} {arXiv:1809.03484
  [astro-ph.CO]} \BibitemShut {NoStop}%
%%CITATION = ARXIV:1809.03484;%%
\bibitem [{\citenamefont {Creminelli}\ \emph {et~al.}(2020)\citenamefont
  {Creminelli}, \citenamefont {Tambalo}, \citenamefont {Vernizzi},\ and\
  \citenamefont {Yingcharoenrat}}]{Creminelli:2019kjy}%
  \BibitemOpen
  \bibfield  {author} {\bibinfo {author} {\bibfnamefont {P.}~\bibnamefont
  {Creminelli}}, \bibinfo {author} {\bibfnamefont {G.}~\bibnamefont {Tambalo}},
  \bibinfo {author} {\bibfnamefont {F.}~\bibnamefont {Vernizzi}},\ and\
  \bibinfo {author} {\bibfnamefont {V.}~\bibnamefont {Yingcharoenrat}},\
  }\bibfield  {title} {\bibinfo {title} {{Dark-Energy Instabilities induced by
  Gravitational Waves}},\ }\href
  {https://doi.org/10.1088/1475-7516/2020/05/002} {\bibfield  {journal}
  {\bibinfo  {journal} {JCAP}\ }\textbf {\bibinfo {volume} {05}},\ \bibinfo
  {pages} {002}},\ \Eprint {https://arxiv.org/abs/1910.14035} {arXiv:1910.14035
  [gr-qc]} \BibitemShut {NoStop}%
\bibitem [{\citenamefont {Babichev}(2020)}]{Babichev:2020tct}%
  \BibitemOpen
  \bibfield  {author} {\bibinfo {author} {\bibfnamefont {E.}~\bibnamefont
  {Babichev}},\ }\bibfield  {title} {\bibinfo {title} {{Emergence of ghosts in
  Horndeski theory}},\ }\href {https://doi.org/10.1007/JHEP07(2020)038}
  {\bibfield  {journal} {\bibinfo  {journal} {JHEP}\ }\textbf {\bibinfo
  {volume} {07}},\ \bibinfo {pages} {038}},\ \Eprint
  {https://arxiv.org/abs/2001.11784} {arXiv:2001.11784 [hep-th]} \BibitemShut
  {NoStop}%
%%CITATION = ARXIV:2001.11784;%%
\bibitem [{\citenamefont {Bezares}\ \emph
  {et~al.}(2021{\natexlab{a}})\citenamefont {Bezares}, \citenamefont {ter
  Haar}, \citenamefont {Crisostomi}, \citenamefont {Barausse},\ and\
  \citenamefont {Palenzuela}}]{Bezares:2021yek}%
  \BibitemOpen
  \bibfield  {author} {\bibinfo {author} {\bibfnamefont {M.}~\bibnamefont
  {Bezares}}, \bibinfo {author} {\bibfnamefont {L.}~\bibnamefont {ter Haar}},
  \bibinfo {author} {\bibfnamefont {M.}~\bibnamefont {Crisostomi}}, \bibinfo
  {author} {\bibfnamefont {E.}~\bibnamefont {Barausse}},\ and\ \bibinfo
  {author} {\bibfnamefont {C.}~\bibnamefont {Palenzuela}},\ }\bibfield  {title}
  {\bibinfo {title} {{Kinetic screening in nonlinear stellar oscillations and
  gravitational collapse}},\ }\href
  {https://doi.org/10.1103/PhysRevD.104.044022} {\bibfield  {journal} {\bibinfo
   {journal} {Phys. Rev. D}\ }\textbf {\bibinfo {volume} {104}},\ \bibinfo
  {pages} {044022} (\bibinfo {year} {2021}{\natexlab{a}})},\ \Eprint
  {https://arxiv.org/abs/2105.13992} {arXiv:2105.13992 [gr-qc]} \BibitemShut
  {NoStop}%
\bibitem [{\citenamefont {Bezares}\ \emph {et~al.}(2022)\citenamefont
  {Bezares}, \citenamefont {Aguilera-Miret}, \citenamefont {ter Haar},
  \citenamefont {Crisostomi}, \citenamefont {Palenzuela},\ and\ \citenamefont
  {Barausse}}]{Bezares:2021dma}%
  \BibitemOpen
  \bibfield  {author} {\bibinfo {author} {\bibfnamefont {M.}~\bibnamefont
  {Bezares}}, \bibinfo {author} {\bibfnamefont {R.}~\bibnamefont
  {Aguilera-Miret}}, \bibinfo {author} {\bibfnamefont {L.}~\bibnamefont {ter
  Haar}}, \bibinfo {author} {\bibfnamefont {M.}~\bibnamefont {Crisostomi}},
  \bibinfo {author} {\bibfnamefont {C.}~\bibnamefont {Palenzuela}},\ and\
  \bibinfo {author} {\bibfnamefont {E.}~\bibnamefont {Barausse}},\ }\bibfield
  {title} {\bibinfo {title} {{No Evidence of Kinetic Screening in Simulations
  of Merging Binary Neutron Stars beyond General Relativity}},\ }\href
  {https://doi.org/10.1103/PhysRevLett.128.091103} {\bibfield  {journal}
  {\bibinfo  {journal} {Phys. Rev. Lett.}\ }\textbf {\bibinfo {volume} {128}},\
  \bibinfo {pages} {091103} (\bibinfo {year} {2022})},\ \Eprint
  {https://arxiv.org/abs/2107.05648} {arXiv:2107.05648 [gr-qc]} \BibitemShut
  {NoStop}%
\bibitem [{\citenamefont {Babichev}\ \emph {et~al.}(2009)\citenamefont
  {Babichev}, \citenamefont {Deffayet},\ and\ \citenamefont
  {Ziour}}]{Babichev:2009ee}%
  \BibitemOpen
  \bibfield  {author} {\bibinfo {author} {\bibfnamefont {E.}~\bibnamefont
  {Babichev}}, \bibinfo {author} {\bibfnamefont {C.}~\bibnamefont {Deffayet}},\
  and\ \bibinfo {author} {\bibfnamefont {R.}~\bibnamefont {Ziour}},\ }\bibfield
   {title} {\bibinfo {title} {{k-Mouflage gravity}},\ }\href
  {https://doi.org/10.1142/S0218271809016107} {\bibfield  {journal} {\bibinfo
  {journal} {Int. J. Mod. Phys. D}\ }\textbf {\bibinfo {volume} {18}},\
  \bibinfo {pages} {2147} (\bibinfo {year} {2009})},\ \Eprint
  {https://arxiv.org/abs/0905.2943} {arXiv:0905.2943 [hep-th]} \BibitemShut
  {NoStop}%
\bibitem [{\citenamefont {ter Haar}\ \emph {et~al.}(2021)\citenamefont {ter
  Haar}, \citenamefont {Bezares}, \citenamefont {Crisostomi}, \citenamefont
  {Barausse},\ and\ \citenamefont {Palenzuela}}]{terHaar:2020xxb}%
  \BibitemOpen
  \bibfield  {author} {\bibinfo {author} {\bibfnamefont {L.}~\bibnamefont {ter
  Haar}}, \bibinfo {author} {\bibfnamefont {M.}~\bibnamefont {Bezares}},
  \bibinfo {author} {\bibfnamefont {M.}~\bibnamefont {Crisostomi}}, \bibinfo
  {author} {\bibfnamefont {E.}~\bibnamefont {Barausse}},\ and\ \bibinfo
  {author} {\bibfnamefont {C.}~\bibnamefont {Palenzuela}},\ }\bibfield  {title}
  {\bibinfo {title} {{Dynamics of Screening in Modified Gravity}},\ }\href
  {https://doi.org/10.1103/PhysRevLett.126.091102} {\bibfield  {journal}
  {\bibinfo  {journal} {Phys. Rev. Lett.}\ }\textbf {\bibinfo {volume} {126}},\
  \bibinfo {pages} {091102} (\bibinfo {year} {2021})},\ \Eprint
  {https://arxiv.org/abs/2009.03354} {arXiv:2009.03354 [gr-qc]} \BibitemShut
  {NoStop}%
\bibitem [{\citenamefont {Vainshtein}(1972)}]{Vainshtein:1972sx}%
  \BibitemOpen
  \bibfield  {author} {\bibinfo {author} {\bibfnamefont {A.}~\bibnamefont
  {Vainshtein}},\ }\bibfield  {title} {\bibinfo {title} {{To the problem of
  nonvanishing gravitation mass}},\ }\href
  {https://doi.org/10.1016/0370-2693(72)90147-5} {\bibfield  {journal}
  {\bibinfo  {journal} {Phys. Lett. B}\ }\textbf {\bibinfo {volume} {39}},\
  \bibinfo {pages} {393} (\bibinfo {year} {1972})}\BibitemShut {NoStop}%
\bibitem [{\citenamefont {Babichev}\ and\ \citenamefont
  {Deffayet}(2013)}]{Babichev:2013usa}%
  \BibitemOpen
  \bibfield  {author} {\bibinfo {author} {\bibfnamefont {E.}~\bibnamefont
  {Babichev}}\ and\ \bibinfo {author} {\bibfnamefont {C.}~\bibnamefont
  {Deffayet}},\ }\bibfield  {title} {\bibinfo {title} {{An introduction to the
  Vainshtein mechanism}},\ }\href
  {https://doi.org/10.1088/0264-9381/30/18/184001} {\bibfield  {journal}
  {\bibinfo  {journal} {Class. Quant. Grav.}\ }\textbf {\bibinfo {volume}
  {30}},\ \bibinfo {pages} {184001} (\bibinfo {year} {2013})},\ \Eprint
  {https://arxiv.org/abs/1304.7240} {arXiv:1304.7240 [gr-qc]} \BibitemShut
  {NoStop}%
\bibitem [{\citenamefont {Khoury}\ and\ \citenamefont
  {Weltman}(2004)}]{Khoury:2003rn}%
  \BibitemOpen
  \bibfield  {author} {\bibinfo {author} {\bibfnamefont {J.}~\bibnamefont
  {Khoury}}\ and\ \bibinfo {author} {\bibfnamefont {A.}~\bibnamefont
  {Weltman}},\ }\bibfield  {title} {\bibinfo {title} {{Chameleon cosmology}},\
  }\href {https://doi.org/10.1103/PhysRevD.69.044026} {\bibfield  {journal}
  {\bibinfo  {journal} {Phys. Rev. D}\ }\textbf {\bibinfo {volume} {69}},\
  \bibinfo {pages} {044026} (\bibinfo {year} {2004})},\ \Eprint
  {https://arxiv.org/abs/astro-ph/0309411} {arXiv:astro-ph/0309411}
  \BibitemShut {NoStop}%
\bibitem [{\citenamefont {Hinterbichler}\ and\ \citenamefont
  {Khoury}(2010)}]{Hinterbichler:2010es}%
  \BibitemOpen
  \bibfield  {author} {\bibinfo {author} {\bibfnamefont {K.}~\bibnamefont
  {Hinterbichler}}\ and\ \bibinfo {author} {\bibfnamefont {J.}~\bibnamefont
  {Khoury}},\ }\bibfield  {title} {\bibinfo {title} {{Symmetron Fields:
  Screening Long-Range Forces Through Local Symmetry Restoration}},\ }\href
  {https://doi.org/10.1103/PhysRevLett.104.231301} {\bibfield  {journal}
  {\bibinfo  {journal} {Phys. Rev. Lett.}\ }\textbf {\bibinfo {volume} {104}},\
  \bibinfo {pages} {231301} (\bibinfo {year} {2010})},\ \Eprint
  {https://arxiv.org/abs/1001.4525} {arXiv:1001.4525 [hep-th]} \BibitemShut
  {NoStop}%
\bibitem [{\citenamefont {Bernard}\ \emph {et~al.}(2019)\citenamefont
  {Bernard}, \citenamefont {Lehner},\ and\ \citenamefont
  {Luna}}]{Bernard:2019fjb}%
  \BibitemOpen
  \bibfield  {author} {\bibinfo {author} {\bibfnamefont {L.}~\bibnamefont
  {Bernard}}, \bibinfo {author} {\bibfnamefont {L.}~\bibnamefont {Lehner}},\
  and\ \bibinfo {author} {\bibfnamefont {R.}~\bibnamefont {Luna}},\ }\bibfield
  {title} {\bibinfo {title} {{Challenges to global solutions in
  Horndeski\textquoteright{}s theory}},\ }\href
  {https://doi.org/10.1103/PhysRevD.100.024011} {\bibfield  {journal} {\bibinfo
   {journal} {Phys. Rev. D}\ }\textbf {\bibinfo {volume} {100}},\ \bibinfo
  {pages} {024011} (\bibinfo {year} {2019})},\ \Eprint
  {https://arxiv.org/abs/1904.12866} {arXiv:1904.12866 [gr-qc]} \BibitemShut
  {NoStop}%
\bibitem [{\citenamefont {Bezares}\ \emph
  {et~al.}(2021{\natexlab{b}})\citenamefont {Bezares}, \citenamefont
  {Crisostomi}, \citenamefont {Palenzuela},\ and\ \citenamefont
  {Barausse}}]{Bezares:2020wkn}%
  \BibitemOpen
  \bibfield  {author} {\bibinfo {author} {\bibfnamefont {M.}~\bibnamefont
  {Bezares}}, \bibinfo {author} {\bibfnamefont {M.}~\bibnamefont {Crisostomi}},
  \bibinfo {author} {\bibfnamefont {C.}~\bibnamefont {Palenzuela}},\ and\
  \bibinfo {author} {\bibfnamefont {E.}~\bibnamefont {Barausse}},\ }\bibfield
  {title} {\bibinfo {title} {{K-dynamics: well-posed 1+1 evolutions in
  K-essence}},\ }\href {https://doi.org/10.1088/1475-7516/2021/03/072}
  {\bibfield  {journal} {\bibinfo  {journal} {JCAP}\ }\textbf {\bibinfo
  {volume} {03}},\ \bibinfo {pages} {072}},\ \Eprint
  {https://arxiv.org/abs/2008.07546} {arXiv:2008.07546 [gr-qc]} \BibitemShut
  {NoStop}%
\bibitem [{\citenamefont {Akhoury}\ \emph {et~al.}(2011)\citenamefont
  {Akhoury}, \citenamefont {Garfinkle},\ and\ \citenamefont
  {Saotome}}]{Akhoury:2011hr}%
  \BibitemOpen
  \bibfield  {author} {\bibinfo {author} {\bibfnamefont {R.}~\bibnamefont
  {Akhoury}}, \bibinfo {author} {\bibfnamefont {D.}~\bibnamefont {Garfinkle}},\
  and\ \bibinfo {author} {\bibfnamefont {R.}~\bibnamefont {Saotome}},\
  }\bibfield  {title} {\bibinfo {title} {{Gravitational collapse of
  k-essence}},\ }\href {https://doi.org/10.1007/JHEP04(2011)096} {\bibfield
  {journal} {\bibinfo  {journal} {JHEP}\ }\textbf {\bibinfo {volume} {04}},\
  \bibinfo {pages} {096}},\ \Eprint {https://arxiv.org/abs/1103.0290}
  {arXiv:1103.0290 [gr-qc]} \BibitemShut {NoStop}%
\bibitem [{\citenamefont {Leonard}\ \emph {et~al.}(2011)\citenamefont
  {Leonard}, \citenamefont {Ziprick}, \citenamefont {Kunstatter},\ and\
  \citenamefont {Mann}}]{Kunstatter:2011ce}%
  \BibitemOpen
  \bibfield  {author} {\bibinfo {author} {\bibfnamefont {C.~D.}\ \bibnamefont
  {Leonard}}, \bibinfo {author} {\bibfnamefont {J.}~\bibnamefont {Ziprick}},
  \bibinfo {author} {\bibfnamefont {G.}~\bibnamefont {Kunstatter}},\ and\
  \bibinfo {author} {\bibfnamefont {R.~B.}\ \bibnamefont {Mann}},\ }\bibfield
  {title} {\bibinfo {title} {{Gravitational collapse of K-essence Matter in
  Painlev\'e-Gullstrand coordinates}},\ }\href
  {https://doi.org/10.1007/JHEP10(2011)028} {\bibfield  {journal} {\bibinfo
  {journal} {JHEP}\ }\textbf {\bibinfo {volume} {10}},\ \bibinfo {pages}
  {028}},\ \Eprint {https://arxiv.org/abs/1106.2054} {arXiv:1106.2054 [gr-qc]}
  \BibitemShut {NoStop}%
\bibitem [{\citenamefont {Gannouji}\ and\ \citenamefont
  {Baez}(2020)}]{Gannouji:2020kas}%
  \BibitemOpen
  \bibfield  {author} {\bibinfo {author} {\bibfnamefont {R.}~\bibnamefont
  {Gannouji}}\ and\ \bibinfo {author} {\bibfnamefont {Y.~R.}\ \bibnamefont
  {Baez}},\ }\bibfield  {title} {\bibinfo {title} {{Critical collapse in
  K-essence models}},\ }\href {https://doi.org/10.1007/JHEP07(2020)132}
  {\bibfield  {journal} {\bibinfo  {journal} {JHEP}\ }\textbf {\bibinfo
  {volume} {07}},\ \bibinfo {pages} {132}},\ \Eprint
  {https://arxiv.org/abs/2003.13730} {arXiv:2003.13730 [gr-qc]} \BibitemShut
  {NoStop}%
\bibitem [{\citenamefont {Hadamard}(1902)}]{Hadamard10030321135}%
  \BibitemOpen
  \bibfield  {author} {\bibinfo {author} {\bibfnamefont {J.}~\bibnamefont
  {Hadamard}},\ }\bibfield  {title} {\bibinfo {title} {Sur les problemes aux
  derivees partielles et leur signification physique},\ }\href
  {https://ci.nii.ac.jp/naid/10030321135/en/} {\bibfield  {journal} {\bibinfo
  {journal} {Princeton university bulletin}\ ,\ \bibinfo {pages} {49}}
  (\bibinfo {year} {1902})}\BibitemShut {NoStop}%
\bibitem [{\citenamefont {Burgess}\ and\ \citenamefont
  {Williams}(2014)}]{Burgess:2014lwa}%
  \BibitemOpen
  \bibfield  {author} {\bibinfo {author} {\bibfnamefont {C.~P.}\ \bibnamefont
  {Burgess}}\ and\ \bibinfo {author} {\bibfnamefont {M.}~\bibnamefont
  {Williams}},\ }\bibfield  {title} {\bibinfo {title} {{Who You Gonna Call?
  Runaway Ghosts, Higher Derivatives and Time-Dependence in EFTs}},\ }\href
  {https://doi.org/10.1007/JHEP08(2014)074} {\bibfield  {journal} {\bibinfo
  {journal} {JHEP}\ }\textbf {\bibinfo {volume} {08}},\ \bibinfo {pages}
  {074}},\ \Eprint {https://arxiv.org/abs/1404.2236} {arXiv:1404.2236 [gr-qc]}
  \BibitemShut {NoStop}%
\bibitem [{\citenamefont {Adams}\ \emph {et~al.}(2006)\citenamefont {Adams},
  \citenamefont {Arkani-Hamed}, \citenamefont {Dubovsky}, \citenamefont
  {Nicolis},\ and\ \citenamefont {Rattazzi}}]{Adams:2006sv}%
  \BibitemOpen
  \bibfield  {author} {\bibinfo {author} {\bibfnamefont {A.}~\bibnamefont
  {Adams}}, \bibinfo {author} {\bibfnamefont {N.}~\bibnamefont {Arkani-Hamed}},
  \bibinfo {author} {\bibfnamefont {S.}~\bibnamefont {Dubovsky}}, \bibinfo
  {author} {\bibfnamefont {A.}~\bibnamefont {Nicolis}},\ and\ \bibinfo {author}
  {\bibfnamefont {R.}~\bibnamefont {Rattazzi}},\ }\bibfield  {title} {\bibinfo
  {title} {{Causality, analyticity and an IR obstruction to UV completion}},\
  }\href {https://doi.org/10.1088/1126-6708/2006/10/014} {\bibfield  {journal}
  {\bibinfo  {journal} {JHEP}\ }\textbf {\bibinfo {volume} {10}},\ \bibinfo
  {pages} {014}},\ \Eprint {https://arxiv.org/abs/hep-th/0602178}
  {arXiv:hep-th/0602178 [hep-th]} \BibitemShut {NoStop}%
%%CITATION = HEP-TH/0602178;%%
\bibitem [{\citenamefont {Cayuso}\ \emph {et~al.}(2017)\citenamefont {Cayuso},
  \citenamefont {Ortiz},\ and\ \citenamefont {Lehner}}]{Cayuso_2017}%
  \BibitemOpen
  \bibfield  {author} {\bibinfo {author} {\bibfnamefont {J.}~\bibnamefont
  {Cayuso}}, \bibinfo {author} {\bibfnamefont {N.}~\bibnamefont {Ortiz}},\ and\
  \bibinfo {author} {\bibfnamefont {L.}~\bibnamefont {Lehner}},\ }\bibfield
  {title} {\bibinfo {title} {Fixing extensions to general relativity in the
  nonlinear regime},\ }\bibfield  {journal} {\bibinfo  {journal} {Physical
  Review D}\ }\textbf {\bibinfo {volume} {96}},\ \href
  {https://doi.org/10.1103/physrevd.96.084043} {10.1103/physrevd.96.084043}
  (\bibinfo {year} {2017})\BibitemShut {NoStop}%
\bibitem [{\citenamefont {Allwright}\ and\ \citenamefont
  {Lehner}(2019)}]{Allwright:2018rut}%
  \BibitemOpen
  \bibfield  {author} {\bibinfo {author} {\bibfnamefont {G.}~\bibnamefont
  {Allwright}}\ and\ \bibinfo {author} {\bibfnamefont {L.}~\bibnamefont
  {Lehner}},\ }\bibfield  {title} {\bibinfo {title} {{Towards the nonlinear
  regime in extensions to GR: assessing possible options}},\ }\href
  {https://doi.org/10.1088/1361-6382/ab0ee1} {\bibfield  {journal} {\bibinfo
  {journal} {Class. Quant. Grav.}\ }\textbf {\bibinfo {volume} {36}},\ \bibinfo
  {pages} {084001} (\bibinfo {year} {2019})},\ \Eprint
  {https://arxiv.org/abs/1808.07897} {arXiv:1808.07897 [gr-qc]} \BibitemShut
  {NoStop}%
\bibitem [{\citenamefont {Cayuso}\ and\ \citenamefont
  {Lehner}(2020)}]{Cayuso:2020lca}%
  \BibitemOpen
  \bibfield  {author} {\bibinfo {author} {\bibfnamefont {R.}~\bibnamefont
  {Cayuso}}\ and\ \bibinfo {author} {\bibfnamefont {L.}~\bibnamefont
  {Lehner}},\ }\bibfield  {title} {\bibinfo {title} {{Nonlinear, noniterative
  treatment of EFT-motivated gravity}},\ }\href
  {https://doi.org/10.1103/PhysRevD.102.084008} {\bibfield  {journal} {\bibinfo
   {journal} {Phys. Rev. D}\ }\textbf {\bibinfo {volume} {102}},\ \bibinfo
  {pages} {084008} (\bibinfo {year} {2020})},\ \Eprint
  {https://arxiv.org/abs/2005.13720} {arXiv:2005.13720 [gr-qc]} \BibitemShut
  {NoStop}%
\bibitem [{\citenamefont {Babichev}(2016)}]{Babichev:2016hys}%
  \BibitemOpen
  \bibfield  {author} {\bibinfo {author} {\bibfnamefont {E.}~\bibnamefont
  {Babichev}},\ }\bibfield  {title} {\bibinfo {title} {{Formation of caustics
  in k-essence and Horndeski theory}},\ }\href
  {https://doi.org/10.1007/JHEP04(2016)129} {\bibfield  {journal} {\bibinfo
  {journal} {JHEP}\ }\textbf {\bibinfo {volume} {04}},\ \bibinfo {pages}
  {129}},\ \Eprint {https://arxiv.org/abs/1602.00735} {arXiv:1602.00735
  [hep-th]} \BibitemShut {NoStop}%
\bibitem [{\citenamefont {Babichev}\ and\ \citenamefont
  {Ramazanov}(2017)}]{Babichev:2017lrx}%
  \BibitemOpen
  \bibfield  {author} {\bibinfo {author} {\bibfnamefont {E.}~\bibnamefont
  {Babichev}}\ and\ \bibinfo {author} {\bibfnamefont {S.}~\bibnamefont
  {Ramazanov}},\ }\bibfield  {title} {\bibinfo {title} {{Caustic free
  completion of pressureless perfect fluid and k-essence}},\ }\href
  {https://doi.org/10.1007/JHEP08(2017)040} {\bibfield  {journal} {\bibinfo
  {journal} {JHEP}\ }\textbf {\bibinfo {volume} {08}},\ \bibinfo {pages}
  {040}},\ \Eprint {https://arxiv.org/abs/1704.03367} {arXiv:1704.03367
  [hep-th]} \BibitemShut {NoStop}%
\bibitem [{\citenamefont {Mukohyama}\ and\ \citenamefont
  {Namba}(2021)}]{MukohyamaUV:2020lsu}%
  \BibitemOpen
  \bibfield  {author} {\bibinfo {author} {\bibfnamefont {S.}~\bibnamefont
  {Mukohyama}}\ and\ \bibinfo {author} {\bibfnamefont {R.}~\bibnamefont
  {Namba}},\ }\bibfield  {title} {\bibinfo {title} {{Partial UV Completion of
  $P(X)$ from a Curved Field Space}},\ }\href
  {https://doi.org/10.1088/1475-7516/2021/02/001} {\bibfield  {journal}
  {\bibinfo  {journal} {JCAP}\ }\textbf {\bibinfo {volume} {02}},\ \bibinfo
  {pages} {001}},\ \Eprint {https://arxiv.org/abs/2010.09184} {arXiv:2010.09184
  [hep-th]} \BibitemShut {NoStop}%
\bibitem [{\citenamefont {Friedrich}\ and\ \citenamefont
  {Rendall}(2000)}]{FriedrichRendall:10.1007/3-540-46580-4_2}%
  \BibitemOpen
  \bibfield  {author} {\bibinfo {author} {\bibfnamefont {H.}~\bibnamefont
  {Friedrich}}\ and\ \bibinfo {author} {\bibfnamefont {A.}~\bibnamefont
  {Rendall}},\ }\bibfield  {title} {\bibinfo {title} {The cauchy problem for
  the einstein equations},\ }in\ \href@noop {} {\emph {\bibinfo {booktitle}
  {Einstein's Field Equations and Their Physical Implications}}},\ \bibinfo
  {editor} {edited by\ \bibinfo {editor} {\bibfnamefont {B.~G.}\ \bibnamefont
  {Schmidt}}}\ (\bibinfo  {publisher} {Springer Berlin Heidelberg},\ \bibinfo
  {address} {Berlin, Heidelberg},\ \bibinfo {year} {2000})\ pp.\ \bibinfo
  {pages} {127--223}\BibitemShut {NoStop}%
\bibitem [{\citenamefont {Reula}(1998)}]{Reula:1998ty}%
  \BibitemOpen
  \bibfield  {author} {\bibinfo {author} {\bibfnamefont {O.~A.}\ \bibnamefont
  {Reula}},\ }\bibfield  {title} {\bibinfo {title} {{Hyperbolic methods for
  Einstein's equations}},\ }\href@noop {} {\bibfield  {journal} {\bibinfo
  {journal} {Living Rev. Rel.}\ }\textbf {\bibinfo {volume} {1}},\ \bibinfo
  {pages} {3} (\bibinfo {year} {1998})}\BibitemShut {NoStop}%
\bibitem [{\citenamefont {Sarbach}\ and\ \citenamefont
  {Tiglio}(2012)}]{Sarbach:2012pr}%
  \BibitemOpen
  \bibfield  {author} {\bibinfo {author} {\bibfnamefont {O.}~\bibnamefont
  {Sarbach}}\ and\ \bibinfo {author} {\bibfnamefont {M.}~\bibnamefont
  {Tiglio}},\ }\bibfield  {title} {\bibinfo {title} {{Continuum and Discrete
  Initial-Boundary-Value Problems and Einstein's Field Equations}},\ }\href
  {https://doi.org/10.12942/lrr-2012-9} {\bibfield  {journal} {\bibinfo
  {journal} {Living Rev. Rel.}\ }\textbf {\bibinfo {volume} {15}},\ \bibinfo
  {pages} {9} (\bibinfo {year} {2012})},\ \Eprint
  {https://arxiv.org/abs/1203.6443} {arXiv:1203.6443 [gr-qc]} \BibitemShut
  {NoStop}%
\bibitem [{\citenamefont {Hilditch}(2013)}]{Hilditch:2013sba}%
  \BibitemOpen
  \bibfield  {author} {\bibinfo {author} {\bibfnamefont {D.}~\bibnamefont
  {Hilditch}},\ }\bibfield  {title} {\bibinfo {title} {{An Introduction to
  Well-posedness and Free-evolution}},\ }\href
  {https://doi.org/10.1142/S0217751X13400150} {\bibfield  {journal} {\bibinfo
  {journal} {Int. J. Mod. Phys. A}\ }\textbf {\bibinfo {volume} {28}},\
  \bibinfo {pages} {1340015} (\bibinfo {year} {2013})},\ \Eprint
  {https://arxiv.org/abs/1309.2012} {arXiv:1309.2012 [gr-qc]} \BibitemShut
  {NoStop}%
\bibitem [{\citenamefont {Foures-Bruhat}(1952)}]{ChoquetBruhat:1952grw}%
  \BibitemOpen
  \bibfield  {author} {\bibinfo {author} {\bibfnamefont {Y.}~\bibnamefont
  {Foures-Bruhat}},\ }\bibfield  {title} {\bibinfo {title} {{Theoreme
  d'existence pour certains systemes derivees partielles non lineaires}},\
  }\href {https://doi.org/10.1007/BF02392131} {\bibfield  {journal} {\bibinfo
  {journal} {Acta Mat.}\ }\textbf {\bibinfo {volume} {88}},\ \bibinfo {pages}
  {141} (\bibinfo {year} {1952})}\BibitemShut {NoStop}%
\bibitem [{\citenamefont {Ripley}\ and\ \citenamefont
  {Pretorius}(2019)}]{Ripley_2019a}%
  \BibitemOpen
  \bibfield  {author} {\bibinfo {author} {\bibfnamefont {J.~L.}\ \bibnamefont
  {Ripley}}\ and\ \bibinfo {author} {\bibfnamefont {F.}~\bibnamefont
  {Pretorius}},\ }\bibfield  {title} {\bibinfo {title} {Hyperbolicity in
  spherical gravitational collapse in a horndeski theory},\ }\bibfield
  {journal} {\bibinfo  {journal} {Physical Review D}\ }\textbf {\bibinfo
  {volume} {99}},\ \href {https://doi.org/10.1103/physrevd.99.084014}
  {10.1103/physrevd.99.084014} (\bibinfo {year} {2019})\BibitemShut {NoStop}%
\bibitem [{\citenamefont {Stewart}(2002)}]{Stewart:2002vd}%
  \BibitemOpen
  \bibfield  {author} {\bibinfo {author} {\bibfnamefont {J.~M.}\ \bibnamefont
  {Stewart}},\ }\bibfield  {title} {\bibinfo {title} {{Signature change, mixed
  problems and numerical relativity}},\ }\href
  {https://doi.org/10.1088/0264-9381/18/23/301} {\bibfield  {journal} {\bibinfo
   {journal} {Class. Quant. Grav.}\ }\textbf {\bibinfo {volume} {18}},\
  \bibinfo {pages} {4983} (\bibinfo {year} {2002})}\BibitemShut {NoStop}%
\bibitem [{\citenamefont {Goldstone}\ \emph {et~al.}(1962)\citenamefont
  {Goldstone}, \citenamefont {Salam},\ and\ \citenamefont
  {Weinberg}}]{GoldstoneSalamWeinberg:PhysRev.127.965}%
  \BibitemOpen
  \bibfield  {author} {\bibinfo {author} {\bibfnamefont {J.}~\bibnamefont
  {Goldstone}}, \bibinfo {author} {\bibfnamefont {A.}~\bibnamefont {Salam}},\
  and\ \bibinfo {author} {\bibfnamefont {S.}~\bibnamefont {Weinberg}},\
  }\bibfield  {title} {\bibinfo {title} {Broken symmetries},\ }\href
  {https://doi.org/10.1103/PhysRev.127.965} {\bibfield  {journal} {\bibinfo
  {journal} {Phys. Rev.}\ }\textbf {\bibinfo {volume} {127}},\ \bibinfo {pages}
  {965} (\bibinfo {year} {1962})}\BibitemShut {NoStop}%
\bibitem [{\citenamefont {Alcubierre}(2008)}]{Alcubierre:1138167}%
  \BibitemOpen
  \bibfield  {author} {\bibinfo {author} {\bibfnamefont {M.}~\bibnamefont
  {Alcubierre}},\ }\href
  {https://doi.org/10.1093/acprof:oso/9780199205677.001.0001} {\emph {\bibinfo
  {title} {{Introduction to 3+1 numerical relativity}}}},\ International series
  of monographs on physics\ (\bibinfo  {publisher} {Oxford Univ. Press},\
  \bibinfo {address} {Oxford},\ \bibinfo {year} {2008})\BibitemShut {NoStop}%
\bibitem [{\citenamefont {Lehner}(2021)}]{LLPrivateCommunication}%
  \BibitemOpen
  \bibfield  {author} {\bibinfo {author} {\bibfnamefont {L.}~\bibnamefont
  {Lehner}},\ }\href@noop {} {}\bibinfo {howpublished} {Private communication}
  (\bibinfo {year} {2021})\BibitemShut {NoStop}%
\bibitem [{\citenamefont {Bona}\ \emph {et~al.}(2002)\citenamefont {Bona},
  \citenamefont {Ledvinka},\ and\ \citenamefont {Palenzuela}}]{Bona:2002fq}%
  \BibitemOpen
  \bibfield  {author} {\bibinfo {author} {\bibfnamefont {C.}~\bibnamefont
  {Bona}}, \bibinfo {author} {\bibfnamefont {T.}~\bibnamefont {Ledvinka}},\
  and\ \bibinfo {author} {\bibfnamefont {C.}~\bibnamefont {Palenzuela}},\
  }\bibfield  {title} {\bibinfo {title} {{A 3+1 covariant suite of numerical
  relativity evolution systems}},\ }\href
  {https://doi.org/10.1103/PhysRevD.66.084013} {\bibfield  {journal} {\bibinfo
  {journal} {Phys. Rev. D}\ }\textbf {\bibinfo {volume} {66}},\ \bibinfo
  {pages} {084013} (\bibinfo {year} {2002})},\ \Eprint
  {https://arxiv.org/abs/gr-qc/0208087} {arXiv:gr-qc/0208087} \BibitemShut
  {NoStop}%
\bibitem [{\citenamefont {Bona}\ \emph {et~al.}(2005)\citenamefont {Bona},
  \citenamefont {Lehner},\ and\ \citenamefont
  {Palenzuela-Luque}}]{Bona:2005pp}%
  \BibitemOpen
  \bibfield  {author} {\bibinfo {author} {\bibfnamefont {C.}~\bibnamefont
  {Bona}}, \bibinfo {author} {\bibfnamefont {L.}~\bibnamefont {Lehner}},\ and\
  \bibinfo {author} {\bibfnamefont {C.}~\bibnamefont {Palenzuela-Luque}},\
  }\bibfield  {title} {\bibinfo {title} {{Geometrically motivated hyperbolic
  coordinate conditions for numerical relativity: Analysis, issues and
  implementations}},\ }\href {https://doi.org/10.1103/PhysRevD.72.104009}
  {\bibfield  {journal} {\bibinfo  {journal} {Phys. Rev. D}\ }\textbf {\bibinfo
  {volume} {72}},\ \bibinfo {pages} {104009} (\bibinfo {year} {2005})},\
  \Eprint {https://arxiv.org/abs/gr-qc/0509092} {arXiv:gr-qc/0509092}
  \BibitemShut {NoStop}%
\bibitem [{\citenamefont {Alic}\ \emph {et~al.}(2007)\citenamefont {Alic},
  \citenamefont {Bona}, \citenamefont {Bona-Casas},\ and\ \citenamefont
  {Masso}}]{Alic:2007ev}%
  \BibitemOpen
  \bibfield  {author} {\bibinfo {author} {\bibfnamefont {D.}~\bibnamefont
  {Alic}}, \bibinfo {author} {\bibfnamefont {C.}~\bibnamefont {Bona}}, \bibinfo
  {author} {\bibfnamefont {C.}~\bibnamefont {Bona-Casas}},\ and\ \bibinfo
  {author} {\bibfnamefont {J.}~\bibnamefont {Masso}},\ }\bibfield  {title}
  {\bibinfo {title} {{Efficient implementation of finite volume methods in
  Numerical Relativity}},\ }\href {https://doi.org/10.1103/PhysRevD.76.104007}
  {\bibfield  {journal} {\bibinfo  {journal} {Phys. Rev. D}\ }\textbf {\bibinfo
  {volume} {76}},\ \bibinfo {pages} {104007} (\bibinfo {year} {2007})},\
  \Eprint {https://arxiv.org/abs/0706.1189} {arXiv:0706.1189 [gr-qc]}
  \BibitemShut {NoStop}%
\bibitem [{\citenamefont {Bona}\ \emph {et~al.}(1995)\citenamefont {Bona},
  \citenamefont {Masso}, \citenamefont {Seidel},\ and\ \citenamefont
  {Stela}}]{Bona:1994dr}%
  \BibitemOpen
  \bibfield  {author} {\bibinfo {author} {\bibfnamefont {C.}~\bibnamefont
  {Bona}}, \bibinfo {author} {\bibfnamefont {J.}~\bibnamefont {Masso}},
  \bibinfo {author} {\bibfnamefont {E.}~\bibnamefont {Seidel}},\ and\ \bibinfo
  {author} {\bibfnamefont {J.}~\bibnamefont {Stela}},\ }\bibfield  {title}
  {\bibinfo {title} {{A New formalism for numerical relativity}},\ }\href
  {https://doi.org/10.1103/PhysRevLett.75.600} {\bibfield  {journal} {\bibinfo
  {journal} {Phys. Rev. Lett.}\ }\textbf {\bibinfo {volume} {75}},\ \bibinfo
  {pages} {600} (\bibinfo {year} {1995})},\ \Eprint
  {https://arxiv.org/abs/gr-qc/9412071} {arXiv:gr-qc/9412071} \BibitemShut
  {NoStop}%
\bibitem [{\citenamefont {Valdez-Alvarado}\ \emph {et~al.}(2013)\citenamefont
  {Valdez-Alvarado}, \citenamefont {Palenzuela}, \citenamefont {Alic},\ and\
  \citenamefont {Ureña-López}}]{Valdez_Alvarado_2013}%
  \BibitemOpen
  \bibfield  {author} {\bibinfo {author} {\bibfnamefont {S.}~\bibnamefont
  {Valdez-Alvarado}}, \bibinfo {author} {\bibfnamefont {C.}~\bibnamefont
  {Palenzuela}}, \bibinfo {author} {\bibfnamefont {D.}~\bibnamefont {Alic}},\
  and\ \bibinfo {author} {\bibfnamefont {L.~A.}\ \bibnamefont
  {Ureña-López}},\ }\bibfield  {title} {\bibinfo {title} {Dynamical evolution
  of fermion-boson stars},\ }\bibfield  {journal} {\bibinfo  {journal}
  {Physical Review D}\ }\textbf {\bibinfo {volume} {87}},\ \href
  {https://doi.org/10.1103/physrevd.87.084040} {10.1103/physrevd.87.084040}
  (\bibinfo {year} {2013})\BibitemShut {NoStop}%
\bibitem [{\citenamefont {Inc.}()}]{Mathematica}%
  \BibitemOpen
  \bibfield  {author} {\bibinfo {author} {\bibfnamefont {W.~R.}\ \bibnamefont
  {Inc.}},\ }\href@noop {} {\bibinfo {title} {Mathematica, {V}ersion 12.2}},\
  \bibinfo {note} {champaign, IL, 2020}\BibitemShut {NoStop}%
\bibitem [{\citenamefont {Wald}(1984)}]{Wald:1984rg}%
  \BibitemOpen
  \bibfield  {author} {\bibinfo {author} {\bibfnamefont {R.~M.}\ \bibnamefont
  {Wald}},\ }\href {https://doi.org/10.7208/chicago/9780226870373.001.0001}
  {\emph {\bibinfo {title} {{General Relativity}}}}\ (\bibinfo  {publisher}
  {Chicago Univ. Pr.},\ \bibinfo {address} {Chicago, USA},\ \bibinfo {year}
  {1984})\BibitemShut {NoStop}%
\bibitem [{\citenamefont {Bernal}\ \emph {et~al.}(2010)\citenamefont {Bernal},
  \citenamefont {Barranco}, \citenamefont {Alic},\ and\ \citenamefont
  {Palenzuela}}]{Bernal:2009zy}%
  \BibitemOpen
  \bibfield  {author} {\bibinfo {author} {\bibfnamefont {A.}~\bibnamefont
  {Bernal}}, \bibinfo {author} {\bibfnamefont {J.}~\bibnamefont {Barranco}},
  \bibinfo {author} {\bibfnamefont {D.}~\bibnamefont {Alic}},\ and\ \bibinfo
  {author} {\bibfnamefont {C.}~\bibnamefont {Palenzuela}},\ }\bibfield  {title}
  {\bibinfo {title} {{Multi-state Boson Stars}},\ }\href
  {https://doi.org/10.1103/PhysRevD.81.044031} {\bibfield  {journal} {\bibinfo
  {journal} {Phys. Rev.}\ }\textbf {\bibinfo {volume} {D81}},\ \bibinfo {pages}
  {044031} (\bibinfo {year} {2010})},\ \Eprint
  {https://arxiv.org/abs/0908.2435} {arXiv:0908.2435 [gr-qc]} \BibitemShut
  {NoStop}%
%%CITATION = ARXIV:0908.2435;%%
\bibitem [{\citenamefont {Raposo}\ \emph {et~al.}(2019)\citenamefont {Raposo},
  \citenamefont {Pani}, \citenamefont {Bezares}, \citenamefont {Palenzuela},\
  and\ \citenamefont {Cardoso}}]{Raposo:2018rjn}%
  \BibitemOpen
  \bibfield  {author} {\bibinfo {author} {\bibfnamefont {G.}~\bibnamefont
  {Raposo}}, \bibinfo {author} {\bibfnamefont {P.}~\bibnamefont {Pani}},
  \bibinfo {author} {\bibfnamefont {M.}~\bibnamefont {Bezares}}, \bibinfo
  {author} {\bibfnamefont {C.}~\bibnamefont {Palenzuela}},\ and\ \bibinfo
  {author} {\bibfnamefont {V.}~\bibnamefont {Cardoso}},\ }\bibfield  {title}
  {\bibinfo {title} {{Anisotropic stars as ultracompact objects in General
  Relativity}},\ }\href {https://doi.org/10.1103/PhysRevD.99.104072} {\bibfield
   {journal} {\bibinfo  {journal} {Phys. Rev.}\ }\textbf {\bibinfo {volume}
  {D99}},\ \bibinfo {pages} {104072} (\bibinfo {year} {2019})},\ \Eprint
  {https://arxiv.org/abs/1811.07917} {arXiv:1811.07917 [gr-qc]} \BibitemShut
  {NoStop}%
%%CITATION = ARXIV:1811.07917;%%
\bibitem [{\citenamefont {Dima}\ \emph {et~al.}(2021)\citenamefont {Dima},
  \citenamefont {Bezares},\ and\ \citenamefont
  {Barausse}}]{PhysRevD.104.084017}%
  \BibitemOpen
  \bibfield  {author} {\bibinfo {author} {\bibfnamefont {A.}~\bibnamefont
  {Dima}}, \bibinfo {author} {\bibfnamefont {M.}~\bibnamefont {Bezares}},\ and\
  \bibinfo {author} {\bibfnamefont {E.}~\bibnamefont {Barausse}},\ }\bibfield
  {title} {\bibinfo {title} {Dynamical chameleon neutron stars: Stability,
  radial oscillations, and scalar radiation in spherical symmetry},\ }\href
  {https://doi.org/10.1103/PhysRevD.104.084017} {\bibfield  {journal} {\bibinfo
   {journal} {Phys. Rev. D}\ }\textbf {\bibinfo {volume} {104}},\ \bibinfo
  {pages} {084017} (\bibinfo {year} {2021})}\BibitemShut {NoStop}%
\bibitem [{\citenamefont {Bona}\ \emph {et~al.}(2009)\citenamefont {Bona},
  \citenamefont {Bona-Casas},\ and\ \citenamefont {Terradas}}]{Bona:2008xs}%
  \BibitemOpen
  \bibfield  {author} {\bibinfo {author} {\bibfnamefont {C.}~\bibnamefont
  {Bona}}, \bibinfo {author} {\bibfnamefont {C.}~\bibnamefont {Bona-Casas}},\
  and\ \bibinfo {author} {\bibfnamefont {J.}~\bibnamefont {Terradas}},\
  }\bibfield  {title} {\bibinfo {title} {{Linear high-resolution schemes for
  hyperbolic conservation laws: TVB numerical evidence}},\ }\href
  {https://doi.org/10.1016/j.jcp.2008.12.010} {\bibfield  {journal} {\bibinfo
  {journal} {J. Comput. Phys.}\ }\textbf {\bibinfo {volume} {228}},\ \bibinfo
  {pages} {2266} (\bibinfo {year} {2009})},\ \Eprint
  {https://arxiv.org/abs/0810.2185} {arXiv:0810.2185 [gr-qc]} \BibitemShut
  {NoStop}%
\bibitem [{\citenamefont {{Bona}}\ \emph {et~al.}(2009)\citenamefont {{Bona}},
  \citenamefont {{Palenzuela-Luque}},\ and\ \citenamefont
  {{Bona-Casas}}}]{CCC}%
  \BibitemOpen
  \bibinfo {editor} {\bibfnamefont {C.}~\bibnamefont {{Bona}}}, \bibinfo
  {editor} {\bibfnamefont {C.}~\bibnamefont {{Palenzuela-Luque}}},\ and\
  \bibinfo {editor} {\bibfnamefont {C.}~\bibnamefont {{Bona-Casas}}},\ eds.,\
  \href {https://doi.org/10.1007/978-3-642-01164-1} {\emph {\bibinfo {title}
  {Elements of Numerical Relativity and Relativistic Hydrodynamics}}},\
  \bibinfo {series} {Lecture Notes in Physics, Berlin Springer Verlag}, Vol.\
  \bibinfo {volume} {783}\ (\bibinfo {year} {2009})\BibitemShut {NoStop}%
\bibitem [{\citenamefont {Kaloper}\ \emph {et~al.}(2015)\citenamefont
  {Kaloper}, \citenamefont {Padilla}, \citenamefont {Saffin},\ and\
  \citenamefont {Stefanyszyn}}]{Kaloper:2014vqa}%
  \BibitemOpen
  \bibfield  {author} {\bibinfo {author} {\bibfnamefont {N.}~\bibnamefont
  {Kaloper}}, \bibinfo {author} {\bibfnamefont {A.}~\bibnamefont {Padilla}},
  \bibinfo {author} {\bibfnamefont {P.}~\bibnamefont {Saffin}},\ and\ \bibinfo
  {author} {\bibfnamefont {D.}~\bibnamefont {Stefanyszyn}},\ }\bibfield
  {title} {\bibinfo {title} {{Unitarity and the Vainshtein Mechanism}},\ }\href
  {https://doi.org/10.1103/PhysRevD.91.045017} {\bibfield  {journal} {\bibinfo
  {journal} {Phys. Rev. D}\ }\textbf {\bibinfo {volume} {91}},\ \bibinfo
  {pages} {045017} (\bibinfo {year} {2015})},\ \Eprint
  {https://arxiv.org/abs/1409.3243} {arXiv:1409.3243 [hep-th]} \BibitemShut
  {NoStop}%
\bibitem [{\citenamefont {Reall}\ and\ \citenamefont
  {Warnick}(2021)}]{Reall:2021ebq}%
  \BibitemOpen
  \bibfield  {author} {\bibinfo {author} {\bibfnamefont {H.~S.}\ \bibnamefont
  {Reall}}\ and\ \bibinfo {author} {\bibfnamefont {C.~M.}\ \bibnamefont
  {Warnick}},\ }\bibfield  {title} {\bibinfo {title} {{Effective field theory
  and classical equations of motion}},\ }\href@noop {} {\  (\bibinfo {year}
  {2021})},\ \Eprint {https://arxiv.org/abs/2105.12028} {arXiv:2105.12028
  [hep-th]} \BibitemShut {NoStop}%
\bibitem [{\citenamefont {Figueras}\ and\ \citenamefont
  {Fran\c{c}a}(2020)}]{Figueras:2020dzx}%
  \BibitemOpen
  \bibfield  {author} {\bibinfo {author} {\bibfnamefont {P.}~\bibnamefont
  {Figueras}}\ and\ \bibinfo {author} {\bibfnamefont {T.}~\bibnamefont
  {Fran\c{c}a}},\ }\bibfield  {title} {\bibinfo {title} {{Gravitational
  Collapse in Cubic Horndeski Theories}},\ }\href
  {https://doi.org/10.1088/1361-6382/abb693} {\bibfield  {journal} {\bibinfo
  {journal} {Class. Quant. Grav.}\ }\textbf {\bibinfo {volume} {37}},\ \bibinfo
  {pages} {225009} (\bibinfo {year} {2020})},\ \Eprint
  {https://arxiv.org/abs/2006.09414} {arXiv:2006.09414 [gr-qc]} \BibitemShut
  {NoStop}%
\bibitem [{\citenamefont {Figueras}\ and\ \citenamefont
  {Fran\c{c}a}(2021)}]{Figueras:2021abd}%
  \BibitemOpen
  \bibfield  {author} {\bibinfo {author} {\bibfnamefont {P.}~\bibnamefont
  {Figueras}}\ and\ \bibinfo {author} {\bibfnamefont {T.}~\bibnamefont
  {Fran\c{c}a}},\ }\bibfield  {title} {\bibinfo {title} {{Black Hole Binaries
  in Cubic Horndeski Theories}},\ }\href@noop {} {\  (\bibinfo {year}
  {2021})},\ \Eprint {https://arxiv.org/abs/2112.15529} {arXiv:2112.15529
  [gr-qc]} \BibitemShut {NoStop}%
\bibitem [{\citenamefont {Frittelli}(1997)}]{Frittelli:PhysRevD.55.5992}%
  \BibitemOpen
  \bibfield  {author} {\bibinfo {author} {\bibfnamefont {S.}~\bibnamefont
  {Frittelli}},\ }\bibfield  {title} {\bibinfo {title} {Note on the propagation
  of the constraints in standard 3+1 general relativity},\ }\href
  {https://doi.org/10.1103/PhysRevD.55.5992} {\bibfield  {journal} {\bibinfo
  {journal} {Phys. Rev. D}\ }\textbf {\bibinfo {volume} {55}},\ \bibinfo
  {pages} {5992} (\bibinfo {year} {1997})}\BibitemShut {NoStop}%
\end{thebibliography}%

\end{document}